\documentclass{aa}
\usepackage[varg]{txfonts}
\usepackage{enumerate}
\usepackage{hyperref}
\usepackage{graphicx}% Include figure files
\usepackage{color}
\usepackage[dvipsnames]{xcolor}
\usepackage{natbib}
\usepackage{longtable}
\usepackage{caption}
\usepackage{amssymb}
\usepackage{overpic}
\usepackage{float}
\usepackage{pdfpages}
\usepackage{enumitem}
\usepackage[export]{adjustbox}
\usepackage{rotating}
\usepackage{txfonts}
\usepackage{color}
\usepackage[dvipsnames]{xcolor}
\usepackage{subcaption}
\usepackage{float}
\usepackage{gensymb}

\def \kms {\ifmmode  \,\rm km\,s^{-1} \else $\,\rm km\,s^{-1}  $ \fi }
\def \Lsun {\ifmmode L_{\odot} \else $L_{\odot}$ \fi}
\def \Msun {\ifmmode M_{\odot} \else $M_{\odot}$\fi}

\begin{document}

\title{A refined mass distribution of the cluster MACS~J0416.1$-$2403
  from a new large set of spectroscopic multiply lensed sources}

\author{G.~B.~Caminha       \inst{\ref{unife}}\thanks{e-mail address: \href{mailto:gbcaminha@fe.infn.it}{gbcaminha@fe.infn.it}} \and
        C.~Grillo           \inst{\ref{unimilano},\,\ref{dark}}                           \and
        P.~Rosati           \inst{\ref{unife}}                          \and
        I.~Balestra         \inst{\ref{obs_munich},\,\ref{inaftrieste}} \and
        A.~Mercurio         \inst{\ref{inafcapo}}                       \and
        E.~Vanzella         \inst{\ref{inafbologna}}                    \and
        A.~Biviano          \inst{\ref{inaftrieste}}                    \and
        K.~I.~Caputi        \inst{\ref{Kapteyn}}                        \and
        C.~Delgado-Correal  \inst{\ref{unife}}                          \and
        W.~Karman           \inst{\ref{Kapteyn}}                        \and
        M.~Lombardi         \inst{\ref{unimilano}}                      \and
        M.~Meneghetti       \inst{\ref{inafbologna},\,\ref{infnbologna}}\and
        B.~Sartoris         \inst{\ref{unitrieste}}                     \and
        P.~Tozzi            \inst{\ref{inafflorence}}                   
        }
\institute{
Dipartimento di Fisica e Scienze della Terra, Universit\`a degli Studi di Ferrara, Via Saragat 1, I-44122 Ferrara, Italy\label{unife}\and
Dipartimento di Fisica, Universit\`a  degli Studi di Milano, via Celoria 16, I-20133 Milano, Italy\label{unimilano} \and
Dark Cosmology Centre, Niels Bohr Institute, University of Copenhagen, Juliane Maries Vej 30, DK-2100 Copenhagen, Denmark\label{dark}\and
University Observatory Munich, Scheinerstrasse 1, 81679 Munich, Germany\label{obs_munich}\and
INAF - Osservatorio Astronomico di Trieste, via G. B. Tiepolo 11, I-34143, Trieste, Italy\label{inaftrieste}\and
INAF - Osservatorio Astronomico di Capodimonte, Via Moiariello 16, I-80131 Napoli, Italy\label{inafcapo} \and
INAF - Osservatorio Astronomico di Bologna, Via Ranzani 1, I- 40127 Bologna, Italy\label{inafbologna}\and
Kapteyn Astronomical Institute, University of Groningen, Postbus 800, 9700 AV Groningen, The Netherlands \label{Kapteyn} \and
INFN - Sezione di Bologna, viale Berti Pichat 6/2, 40127 Bologna, Italy\label{infnbologna} \and
Dipartimento di Fisica, Universit\`a  degli Studi di Trieste, via G. B. Tiepolo 11, I-34143 Trieste, Italy\label{unitrieste}\and
INAF - Osservatorio Astrofisico di Arcetri, Largo E. Fermi, I-50125, Firenze, Italy\label{inafflorence}
}

\abstract { We report the spectroscopic confirmation of 22 new
  multiply lensed sources behind the {\it Hubble Frontier Field} (HFF)
  galaxy cluster MACS~J0416.1$-$2403 (MACS~0416), using archival data from the
  Multi Unit Spectroscopic Explorer (MUSE) on the VLT.  Combining with
  previous spectroscopic measurements of 15 other multiply imaged
  sources, we have obtained a sample of 102 secure multiple images with
  measured redshifts, the largest to date in a single strong lensing
  system. The newly confirmed sources are largely low-luminosity
  Lyman-$\alpha$ emitters with redshift in the range $[3.08-6.15]$.
  With such a large number of secure constraints, and a significantly
  improved sample of galaxy members in the cluster core, we have
  improved our previous strong lensing model and obtained a robust
  determination of the projected total mass distribution of MACS 0416.
  We find evidence of three cored dark-matter halos, adding to the
  known complexity of this merging system.  The total mass density
  profile, as well as the sub-halo population, are found to be 
  in good agreement with previous works. 
  We update and make public the redshift catalog of MACS 0416 from our previous spectroscopic campaign with the new MUSE redshifts. We also release lensing maps (convergence, shear, magnification)
  in the standard HFF format.
}


\titlerunning{A refined mass distribution of the cluster MACS~J0416.1-2403}

\authorrunning{G.~B.~Caminha et al.}

\maketitle

\section{Introduction}

The use of gravitational lensing by galaxy clusters has intensified in
recent years and has led to significant progress in our understanding
of the mass distribution in clusters, as well as to the discovery of
some of the most distant galaxies \citep[e.g.,][]{2013ApJ...762...32C, 2014ApJ...795..126B} thanks to the magnification of selected cluster
lenses. Key to this progress has been the combination of homogeneous
multi-band surveys of a sizeable number of massive clusters with the
Hubble Space Telescope (HST), primarily with the Cluster Lensing And Supernova survey with Hubble \citep[CLASH,][]{2012ApJS..199...25P},
with wide-field imaging \citep[e.g.,][]{2014ApJ...795..163U, 2016ApJ...821..116U} and
spectroscopic follow-up work from the ground and space. Studies with HST have
inevitably focused on the cluster cores, where a variety of strong
lensing models have been developed to cope with the increasing data
quality and to deliver the precision needed to determine the physical
properties of background lensed galaxies (such as stellar masses,
sizes and star formation rates), which critically depend on the
magnification measurement across the cluster cores. Following the
CLASH project, which has provided a panchromatic, relatively shallow
imaging of 25 massive clusters, the Hubble Frontier Fields (HFF)
program \citep{2016arXiv160506567L} has recently targeted six clusters
(three in common with CLASH) to much greater depth ($\sim2$ mag) in seven
optical and near-IR bands with the ACS and WFC3 cameras.  This has
provided a very rich legacy data set to investigate the best
methodologies to infer mass distributions of the inner ($R\lesssim
300$ kpc) regions of galaxy clusters, and is stimulating a transition
to precision strong lensing modeling with parametric
\citep[e.g.,][]{2014MNRAS.444..268R, 2015MNRAS.452.1437J, M2016,2016ApJ...819..114K} and non-parametric lens models \citep[e.g.,][]{2014ApJ...797...98L, 2016MNRAS.459.3447D, 2015ApJ...811...29W, 2016arXiv160300505H}.

Spectroscopic follow-up information on a large number of multiply
lensed sources is critical to achieve high-precision cluster mass
reconstruction through strong lensing modeling.  Early works heavily relied
on photometric redshifts or color information to identify multiple
images. While this method has been shown to be adequate for determining 
robust mass density profiles \citep[e.g.,][]{2015ApJ...801...44Z}, it is prone to systematics due to possible
misidentifications of multiple images and degeneracies between angular
diameter distances and the cluster mass distribution. This typically
leads to root-mean-square offsets ($\Delta_{\rm rms}$) between the observed and lens
model-predicted positions of $\Delta_{\rm rms} \gtrsim 1\arcsec$ \citep[see][for the CLASH sample]{2015ApJ...801...44Z}.
Using extensive redshift measurements for both cluster member galaxies
and background lensed galaxies, high-fidelity mass maps can be
obtained with $\Delta_{\rm rms}\approx 0\arcsec.3$, as shown for example, in
the study of the HFF clusters MACS~J0416.3$-$2403
(hereafter MACS~0416) \citep[][hereafter Gr15]{2015ApJ...800...38G} and MACS~J1149.5$+$2223 with the
sucessful prediction of the lensed supernova Refsdal
\citep{2016ApJ...817...60T, 2016ApJ...822...78G}.

Exploiting these new high-quality spectroscopic data sets in clusters
that are relatively free from other intervening line-of-sight
structures, strong lensing modeling even becomes sensitive to the
adopted cosmology \citep[][hereafter Ca16]{2016A&A...587A..80C}.
In addition, new large spectroscopic samples of cluster member
galaxies over a sufficiently wide area allow the cluster total mass to be
derived based on galaxy dynamics \citep[e.g.,][]{2013A&A...558A...1B}.
This provides an independent, complementary probe of the cluster mass
out to large radii, which, when combined with high-quality weak-lensing
determinations, can in principle be used to infer dark-matter
properties \citep{2014ApJ...783L..11S} or to test modified theories of gravity
\citep{2016JCAP...04..023P}.

The combination of photometric and spectroscopic data now available
for MACS~0416, from extensive HST and VLT observations, makes it one
of the best data sets with which to investigate the dark-matter distribution in
the central region of a massive merging cluster through strong lensing
techniques and to unveil high-redshift magnified galaxies owing to
its large magnification area.  The high-precision strong lensing model
of MACS~0416 presented by Gr15
was based on CLASH imaging data and spectroscopic information obtained
as part of the CLASH-VLT survey, presented in  \citet{2015arXiv151102522B}.

MACS~0416 is a massive and X-ray luminous \citep[$M_{200} \approx 0.9 \times 10^{15} \rm M_{\odot}$ and $L_X \approx 10^{45}$ erg s$^{-1}$,][]{2015arXiv151102522B} galaxy cluster at $z=
0.396$, originally selected as one of the five clusters with high
magnification in the CLASH sample.  This system was readily identified
as a merger, given its unrelaxed X-ray morphology and the observed
projected separation ($\sim 200$ kpc) of the two brightest cluster
galaxies (BCGs) \citep[see][]{2012MNRAS.420.2120M}. \citet{2013ApJ...762L..30Z}
performed the first strong lensing analysis using the available CLASH
HST photometry, which revealed a quite elongated projected mass
distribution in the cluster core ($\sim 250$ kpc). In subsequent
works \citet{2014MNRAS.443.1549J,2015MNRAS.446.4132J} combined weak and strong lensing analyses,
detecting two main central mass concentrations. When comparing their
mass reconstruction with shallow Chandra observations, they were not
able to unambiguously discern between a pre-collisional or
post-collisional merger.

The CLASH-VLT spectroscopic sample of about 800 cluster member
galaxies out to $\sim 4$ Mpc has recently allowed a detailed dynamical
and phase-space distribution analyses, which revealed a very complex
structure in the cluster core \citep{2015arXiv151102522B}. The most
likely scenario, supported also by deep X-ray Chandra observations and
VLA radio data, suggests a merger composed of two main subclusters
observed in a pre-collisional phase.

In this work, we present a further improved strong lensing model of
MACS~0416, which exploits a new unprecendeted sample of more than 100
spectroscopically confirmed multiple images (corresponding to 37
multiply imaged sources) and $\sim\! 200$ cluster member galaxies in
the cluster core. In Section 2, we describe the MUSE spectroscopic
data set, the data reduction procedure and the method used for redshift
measurements. In Section 3, we describe the strong lensing model and
discuss the results of our strong lensing analysis.  In Section 4, we
summarize our conclusions.

Throughout this article, we adopt a flat $\rm \Lambda CDM$ cosmology
with $\Omega_m = 0.3$ and $H_0 = 70\, {\rm km/s/Mpc}$.  In this
cosmology, $1''$ corresponds to a physical scale of $5.34\, {\rm kpc}$
at the cluster redshift ($z_{lens} = 0.396$). All magnitudes are given
in the AB system.

\begin{figure*}
  \centering
   \includegraphics[width = 1.0\textwidth]{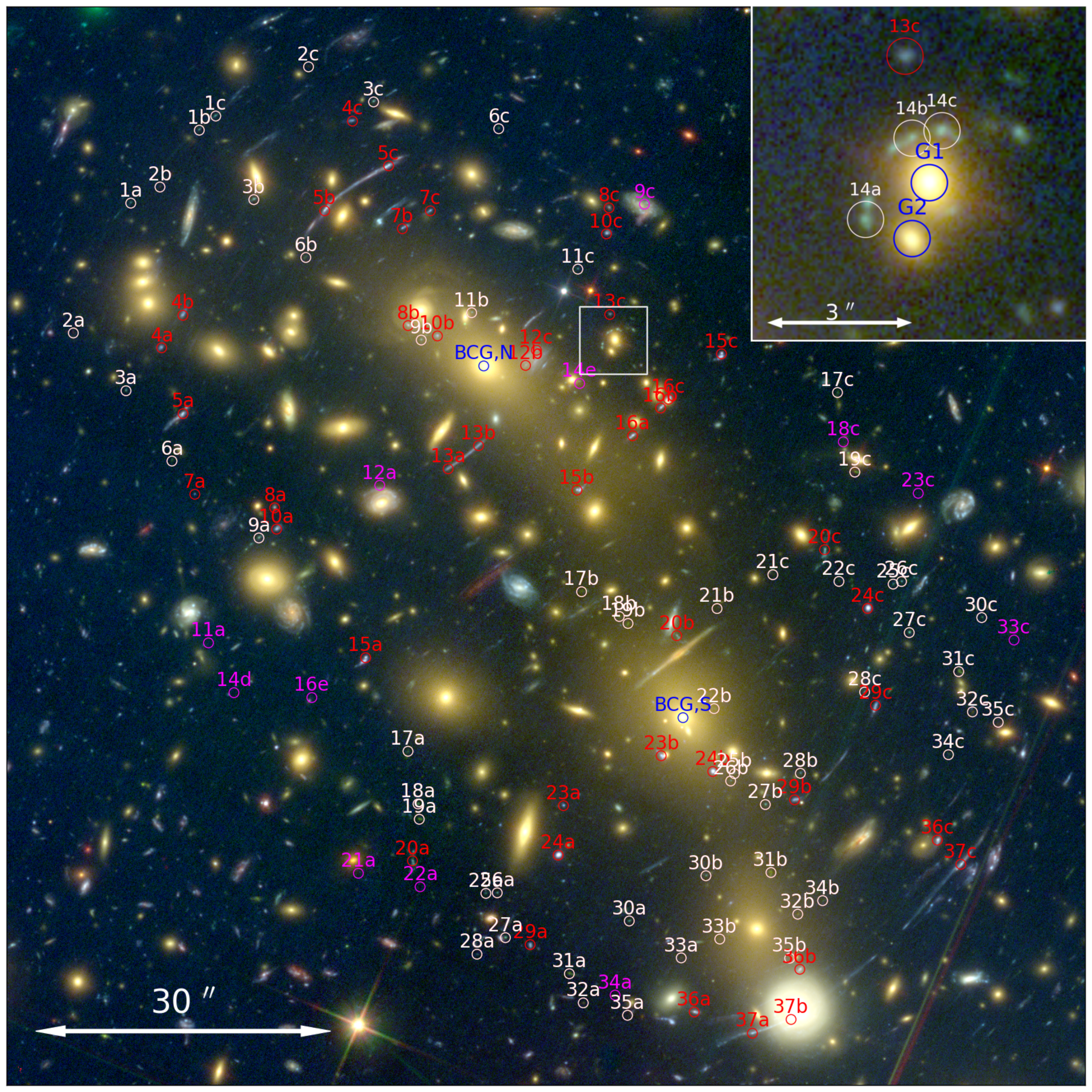}
   \caption{Color composite image of MACS~0416 from Hubble Frontier
     Fields data. Blue, green and red channels are the combination of
     filters F435W, F606W+F814W and F105W+F125W+F140W+F160W,
     respectively. White circles mark the positions of the 59
     multiple images belonging to 22 families with new spectroscopic
     confirmation in this work, while red circles show multiple images
     previously known in spectroscopic families.  Magenta circles show
     the model-predicted positions of multiple images not included in our model, lacking secure identifications (see Table \ref{tab:multiple_images}). The inset
     is a blow-up of the region around family 14, around two galaxy cluster members, G1 and G2, with total mass density profile parameters free to vary in our model (see Section \ref{sec:sl_modelling}). The blue circles indicate the positions of the
     BCGs (BCG,N and BCG,S).}
  \label{fig:arcs}
\end{figure*}

\section{Data}

In this work, we take advantage of the enhanced imaging data from the
HFF campaign \citep{2016arXiv160506567L} and significantly augment the
CLASH-VLT wide-field spectroscopic campaign of MACS 0416
\citep{2015arXiv151102522B} with a large number of new
spectroscopic redshifts obtained with the MUSE integral-field
spectrograph at the VLT, over the central area of 2 $\rm
arcmin^2$. The latter has led us to identify more than three times the
number of secure multiple images with spectroscopic redshift used in
Gr15 and two times when comparing with \citet{2016arXiv160300505H}, as well as to define a highly complete and pure sample of
cluster members. Details on MUSE data reduction and analysis are given
in Section \ref{sec:muse_data}.

\subsection{MUSE observations and data reduction}
\label{sec:muse_data}

\begin{figure}
  \centering
   \includegraphics[width = 1.0\columnwidth]{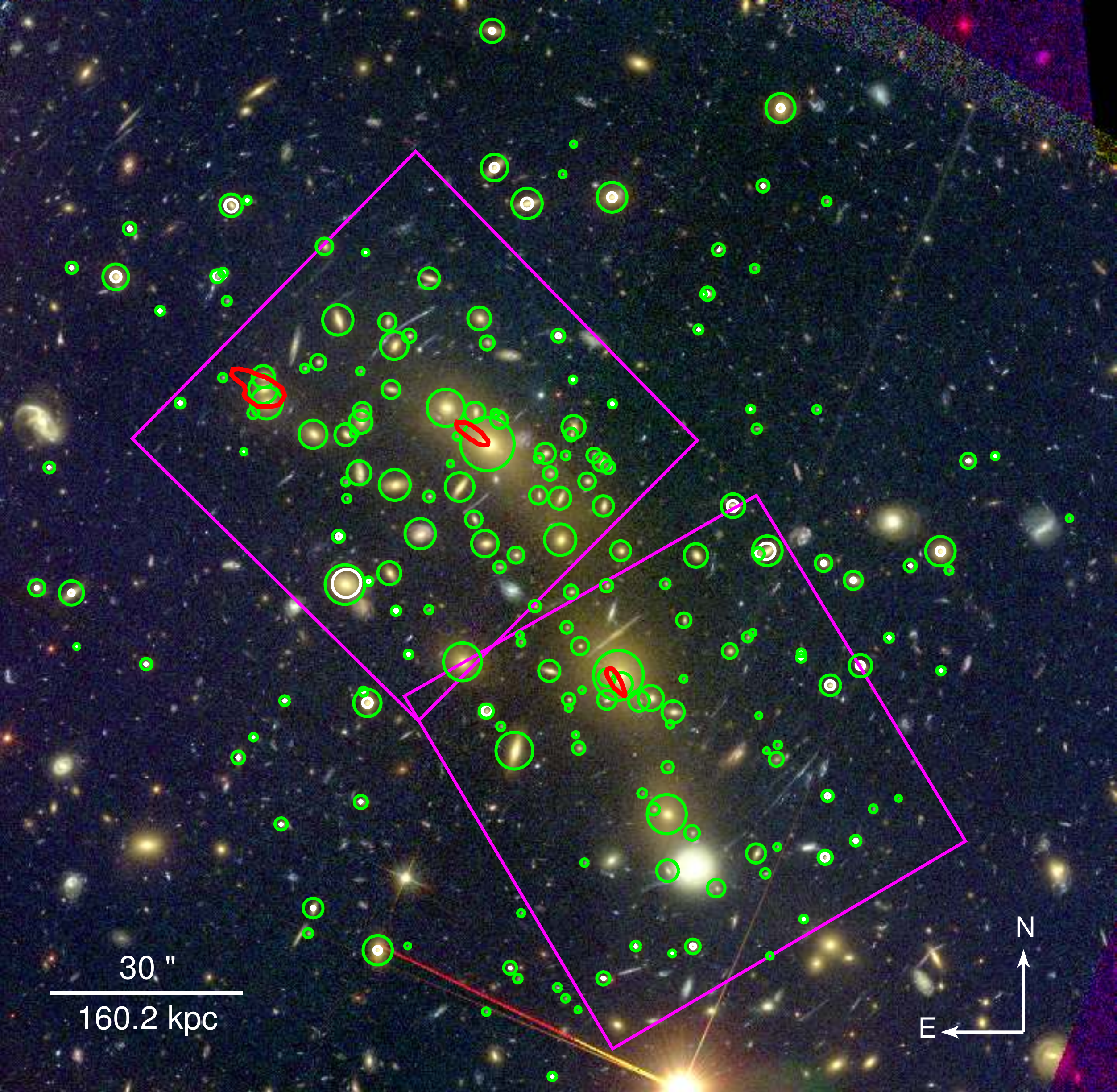}
   \caption{Color image of MACS~0416 from the Hubble Frontier Fields
      data (using the ACS filters F435W, F606W and F814W for the blue, green and red channels respectively) with the two overlaid MUSE
     pointings (magenta boxes), each $\approx 1\arcmin$ across. The green
     circles indicate the 193 selected galaxy cluster members (75\% of which
     spectroscopically confirmed), with the radii proportional to the values of 
     $\sigma_v$ obtained from the best-fitting lensing model and using
     Eq. \ref{eq:mass_to_light}. The three red contours show the 95\%
     confidence level of the dark-matter halo centers included in
     the lensing model.
    }
   \label{fig:members}
\end{figure}

We used archival MUSE \citep{2012Msngr.147....4B} data from two different programs which covered
the North-East (NE) and South-West (SW) regions of MACS~0416.  The
footprints of these two MUSE pointings are shown in Figure
\ref{fig:members} (see the magenta squares), overlaid onto the HST
color image of the cluster.  The NE region was observed within a GTO
program (ID 094.A-0115B, PI: J. Richard) in November 2014, for a
total of two hours split into four exposures.  Significantly deeper
observations in the SW region of the cluster were carried out by the
program ID 094.A-0525(A) (PI: F.E. Bauer).  The latter includes 58
exposures of approximately 11 minutes each, executed over the period
October 2014 -- February 2015.  In both programs, each exposure was
offset by fractions of arcseconds and rotated by 90 degrees to improve
sky subtraction.  The seeing conditions of the NE pointing were very
good, $\approx 0\arcsec.5$, based on the DIMM monitor at Paranal (the
lack of bright stars in both pointings did not allow us to directly
measure the seeing on MUSE data).  Most of the exposures of the SW
pointing, 30 out of 58, were taken in seeing conditions $< 1\arcsec$,
whereas the others have FWHM$\simeq 1\arcsec-1\arcsec.3$.
Moreover, a visual inspection of all SW
exposures from the stacked data-cubes did not show evidence of
significant variations in observational conditions. Only one exposure
was discarded due to the presence of a satellite track, leading to a
total of 11 hours of exposure time in the SW pointing.

We used the MUSE reduction pipeline version 1.2.1 to process the raw
calibration and science exposures of each single night, and to combine the data into the two final data-cubes.
During this process we applied all the standard calibration procedures (bias and flat field corrections, wavelength and flux calibration, etc) provided by the pipeline.
We then combined the observing blocks of the NE and SW observations
(taking into account the offset of each observation) into two final
data-cubes.  Different configurations of the MUSE pipeline recipes
were experimented to improve the quality of the final data-cubes,
particularly the sky subtraction, with no significant differences
however. The final WCS adjustment was made matching compact sources
detected with SExtractor \citep{1996A&AS..117..393B} in the broad-band
images of the two final data-cubes with the corresponding objects in
the HFF catalog for the filter F606W.  As a final post-processing
step to minimize the sky residuals, we applied the Zurich Atmosphere
Purge \citep[ZAP,][]{2016MNRAS.458.3210S} tool using SExtractor
segmentation maps to define sky regions.

The two final data-cubes have a spatial pixel scale of $0\arcsec.2$, a
spectral coverage from 4750 $\AA$ to 9350 $\AA$, with a dispersion of
$1.25\, \AA$/pixel and a fairly constant spectral resolution of $\approx
2.4\,\AA$ over the entire spectral range.  We noticed that after using
different configurations of the MUSE pipeline and applying the ZAP
tool, an overall improvement was achieved in the sky subtraction even though artefacts in the
background at specific wavelengths still remain particularly in the SW
pointing, due to residual instrumental signatures and sky subtraction.
Nonetheless, the quality of the reduced data-cubes allows the spectroscopic identifications of approximately one hundred
sources in each pointing, reaching very faint levels as described
below.

We notice that despite the significant longer exposure, the signal-to-noise of spectra in the SW pointing does not scale according to expectations, resulting only in a moderately larger depth when compared to the NE pointing.
We attribute this difference to the significantly better seeing of the NE pointing ($0\arcsec.5$ versus ~1\arcsec) and the large number of short exposures used in the observations of the SW pointing,  which due to residual systematics in the background subtraction, did not yield the expected depth in the coadded datacube.

\subsection{Spectra extraction and redshift measurements}

We describe here the strategy and methodology to extract spectra
and measure redshifts for all detectable objects in the MUSE fields,
specifically cluster members and multiple images of background lensed sources,
which are critical inputs of our strong lensing model.

To maximize the completeness of spectroscopic identifications in the
two MUSE data-cubes we proceeded in two steps. Firstly, we used the
ASTRODEEP Frontier Fields catalog by \citet{2016A&A...590A..31C},
and blindly extracted spectra at each object position from the MUSE data-cubes, whose world coordinate system is aligned with
the HFF within $\sim\! 0\arcsec.1$. Spectra were extracted within $0\arcsec.8$-radius circular apertures, which provide a good
compromise in the effort to maximize signal-to-noise and  minimize
source confusion. The ASTRODEEP HFF catalog reaches a 90\%
completeness limit at $mag_{\rm F160W} \approx 27.25$ for disk-like
galaxies. Objects which are flagged as possible spurious detections in
the ASTRODEEP catalog were not considered.  In total, we extracted 716 and 699
spectra in the SW and NE MUSE pointings, respectively.

Two team members used the software EZ \citep{2010PASP..122..827G} and SpecPro
\citep{2011PASP..123..638M} to measure independently redshifts,
assigning quality flags following the scheme described in
\citet{2016A&A...587A..80C} and \citet{2015arXiv151102522B},
that is 3$=$secure, 2$=$likely, 1$=$not-reliable and 9$=$ based on a
single emission line.  Since the MUSE spectral resolution allows us to
distinguish the shape or doublet nature of narrow emission lines
(Lyman-$\alpha$ and \ion{O}{II}, for instance), the redshifts with
quality flag equal to 9 are considered very reliable.  With this procedure,
we measured $\approx 300$ reliable redshifts with quality flag greater than
1. Approximately one-third of the objects show emission lines in their
spectra.

As an additional step, we visually inspected the original and
continuum-subtracted cubes.  Continuum subtraction was obtained at
each wavelength frame by considering two windows, ten spectral pixels
($=12.5\,\AA$) wide, in the blue and red side of each frame, separated
by ten spectral pixels.  The continuum is estimated from the mean of
the median counts in each of these two regions and then subtracted
from each wavelength slice. The inspection of such a continuum
subtracted data-cube allowed us to identify faint emission lines of
sources close to bright galaxies.  In this way, we identified 14
additional faint sources, mostly Lyman-$\alpha$ emitters, with very faint
or non-detectable counterpart in the HFF images.

In summary, the analysis of the available MUSE observations of MACS~0416
led us to extend the previous redshift catalog from the CLASH-VLT survey, published in
\citet{2015arXiv151102522B}, which now contains  301 MUSE based 
redshifts, of which 208 are new. With this paper, we also
electronically release the updated version of the redshift catalog, combining the VIMOS and MUSE
observations,
which now contains approximately 4600 objects with redshifts (quality flag greater than 1).
A direct comparison between MUSE and \mbox{VIMOS} spectra can be done with Figure \ref{fig:specs} and Figure 2 of Gr15.

  The GLASS survey \citep{2015ApJ...812..114T} has provided 
redshift measurements of 170 sources in an area of $\approx 4\, \rm
arcmin^2$ around the cluster core. We have found that 
 103 redshifts from the CLASH-VLT and MUSE data sets are in common with GLASS.
 Within the MUSE field of view, we were
not able to measure redshifts for ten GLASS sources, of which six are
close to the edge of the MUSE observations.  The remaining four have
redshifts below $z=2.9$, the lower limit of a Lyman-$\alpha$ emission to
appear in the MUSE spectrum.

Regarding the identification of multiple images for the strong lensing
model described below, in this work we confirm 11 of the 15
spectroscopic multiple-image families from previous studies
\citep{2014MNRAS.444..268R,2014ApJ...797...48J, 2014MNRAS.443.1549J, 2015ApJ...800...38G, 2016arXiv160300505H} and
measure secure redshifts for additional 56 multiple images, belonging
to 21 new families, thus more than doubling the number of multiple
images with spectroscopic redshift known to date for MACS~0416.  We
also measure the redshift of additional five multiple images belonging to known
spectroscopic families but with no previous spectroscopic
confirmation.  Finally, we update the redshift of family 1 to
$z=3.238$,  previously reported in the GLASS catalog at $z=2.19$ with quality flag ``probable'' in their definition. We
therefore spectroscopically determine with MUSE a total of 21+1 new multiply 
lensed systems.  All except one of
these 37 families have at least two images with measured redshift.
All these multiple images are indicated in Figure~\ref{fig:arcs},
while spectra and image cutouts are shown in Figure~\ref{fig:specs}.
Many of these lensed sources have faint magnitudes for ordinary
ground-based spectroscopic work, ranging from $mag_{F814W}=24$ down to
$\approx 29$ (see Table \ref{tab:multiple_images}) and redshifts
$z\gtrsim 3.08$ (reflecting the visibility of the Lyman-$\alpha$ in the
MUSE window). These sources are primarily low-luminosity Lyman-$\alpha$
emitters, whose spectro-photometric properties can be used to
constrain physical properties of low-mass galaxies which are considered
to be the main candidates for reionization \citep{2014MNRAS.442.2560W,2014ApJ...788..121K}. 
Moreover, the rest-frame equivalent widths of these lensed Lyman-$\alpha$ emitters range between $\approx 10\,\AA$ to $\approx 120\,\AA$ and have extremely low luminosities ($L \approx \rm 10^{40}\, L_{\odot} - 10^{42}\, L_{\odot}$), which are comparable to those measured in the sample of lensed LAEs in \citet{2016arXiv160601471K}.

When combined with previous measurements, from the CLASH-VLT and GLASS
surveys, the multiple-image systems span a redshift range from
$z\approx 0.94$ up to $z\approx 6.15$. The highest redshift source is a remarkable giant arc with three multiple images
(2a, 2b and 2c), as revealed by the MUSE data-cube at the $8686\,\AA$.
A very interesting multiply imaged system is System 9, which consists
of a complex, double-peaked, extended \mbox{Lyman-$\alpha$} emission in which
three faint galaxies, detected on the HST image, are embedded. This
system has been studied in detail in \citet{2016arXiv160703112V}.
 We also note that two families (IDs 21 and 35) do not have 
significant counterparts in the deep HFF images (see Table
\ref{tab:multiple_images}), showing the remarkable ability that MUSE
has in identifying very faint emission line objects.

\begin{figure}
  \centering
   \includegraphics[width = 1.0\columnwidth]{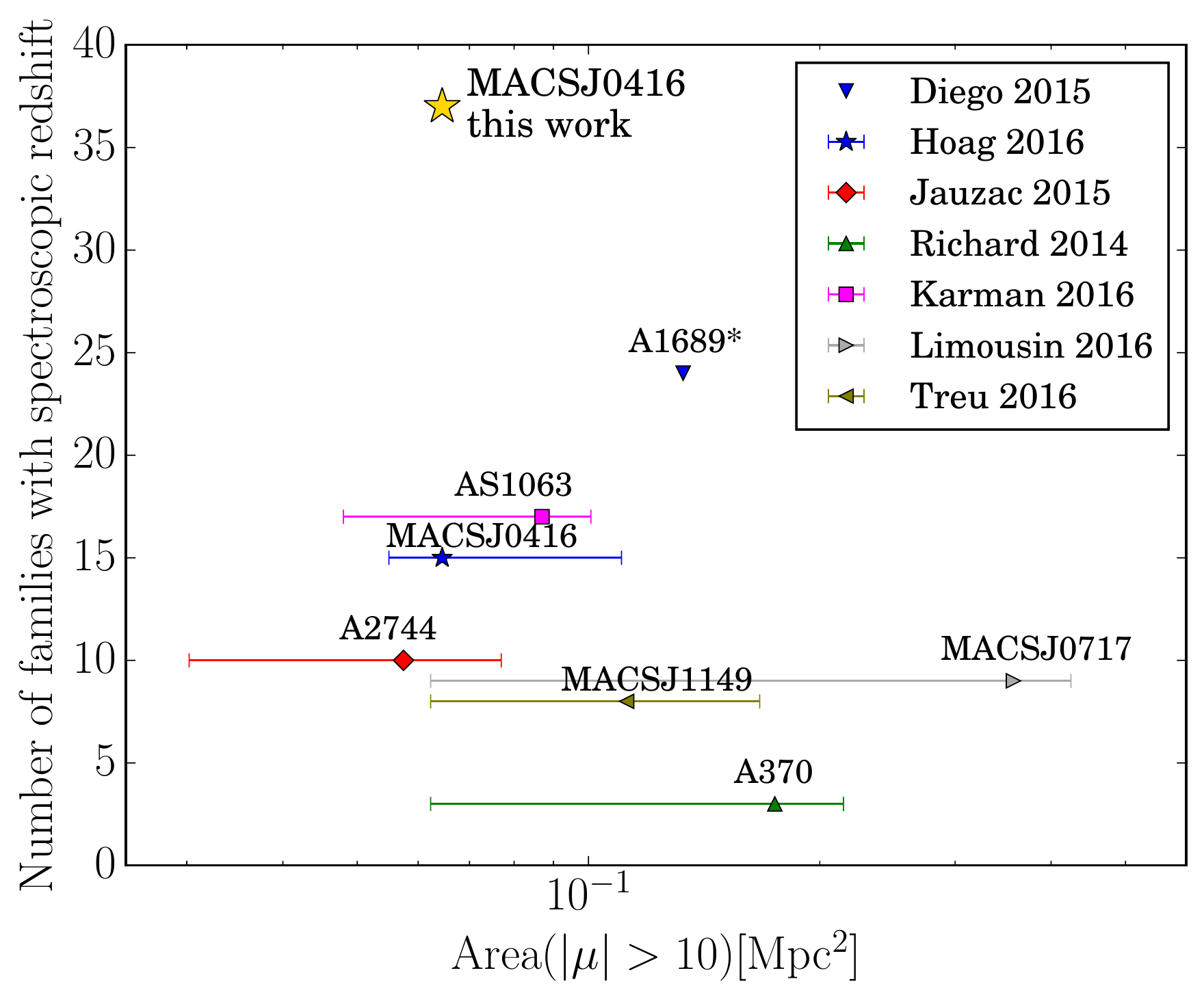}
   \caption{Number of spectroscopically confirmed families of multiple
     images identified to date, as a function of the physical area on the lens plane
     with absolute magnification greater than 10, for a
     source at  $z=4$. In addition to the HFF clusters, we also
     include Abell~1689. The error bars of the magnification area reflect the
     variance (ninetieth percentile) of the most recent HFF
     strong lensing models.}
   \label{fig:mag_specz}
\end{figure}

  In Figure~\ref{fig:mag_specz}, we compare the number of
  spectroscopically confirmed multiple-image families and physical
  magnification areas for all HFF clusters, as well as for Abell 1689,
  which has long been a reference cluster for high-quality lensing
  studies \citep[in this case, we computed the magnification area using the public lens model in][]{2007ApJ...668..643L}.
The error bars show the variance of the results of the different strong lensing models that can be found on the HFF webpage\footnote{\url{https://archive.stsci.edu/prepds/frontier/lensmodels/}}.
For Abell~1689, only one strong lensing model is publicly available, so no error bars are provided.
These two parameters enter a figure of merit for the
  quality of the reconstructed mass distribution of a cluster, since the fidelity
  of the mass map strictly depends on the number of multiple images
  {\sl with spectroscopic redshift}, while the spatial resolution of
  this map depends on the number of strong lensing constraints for a
  given extent of the magnification area.

In Table \ref{tab:areas}, we quote strong lensing constraints and
references for the HFF clusters and Abell 1689. We show the progress in
the identification of multiple-image families for MACS~0416, from
Gr15 (using \mbox{VLT/VIMOS}), to \citet{2016arXiv160300505H} (using
HST/GLASS), to this work (with \mbox{VLT/MUSE}).  The 21 additional
spectroscopic multiply lensed sources presented here make MACS~0416
arguably the best-studied strong lensing cluster to date.

\begin{table}[h]
\centering
\caption{Comparison of strong lensing constraints, that is the number of multiple image families with spectroscopic confirmation ($\rm N_{spec}$), for the the best-studied strong lensing clusters to date.}
    \begin{tabular}{l c c c c} \hline \hline
Cluster & $\rm N_{spec}$ & Reference\\
\hline
MACS~0416        & 37 & this work \\
MACS~0416        & 15 & \citet{2016arXiv160300505H} \\
MACS~0416        & 8  & \citet{2015ApJ...800...38G} \\
\hline
Abell 1689       & 24   & \citet{2015MNRAS.446..683D} \\
Abell S1063      & 17   & \citet{2016arXiv160601471K} \\
MACS~J1149.5+2223& 8   & \citet{2016ApJ...817...60T} \\
Abell 2744       & 10   & \citet{2015MNRAS.452.1437J} \\
MACS~J0717.5+3745& 9   & \citet{M2016} \\
Abell 370        & 3    & \citet{2010MNRAS.402L..44R} \\
\hline\hline
    \end{tabular}
\label{tab:areas}
\end{table}

The MUSE data also allow us to extend and check the purity of the
catalog of 175 cluster members presented in Gr15.
In that work, 12 CLASH photometric bands were used to define the distribution in color space of spectroscopic members from the CLASH-VLT campaign, thereby assigning a probability to all other galaxies to be a member based on their N-dimensional color.
We have already emphasized that the completeness and purity of the sample of sub-halos
associated to cluster members play an important role in the quality of
the strong lensing model.

\begin{figure}
  \centering
   \includegraphics[width = 1.0\columnwidth]{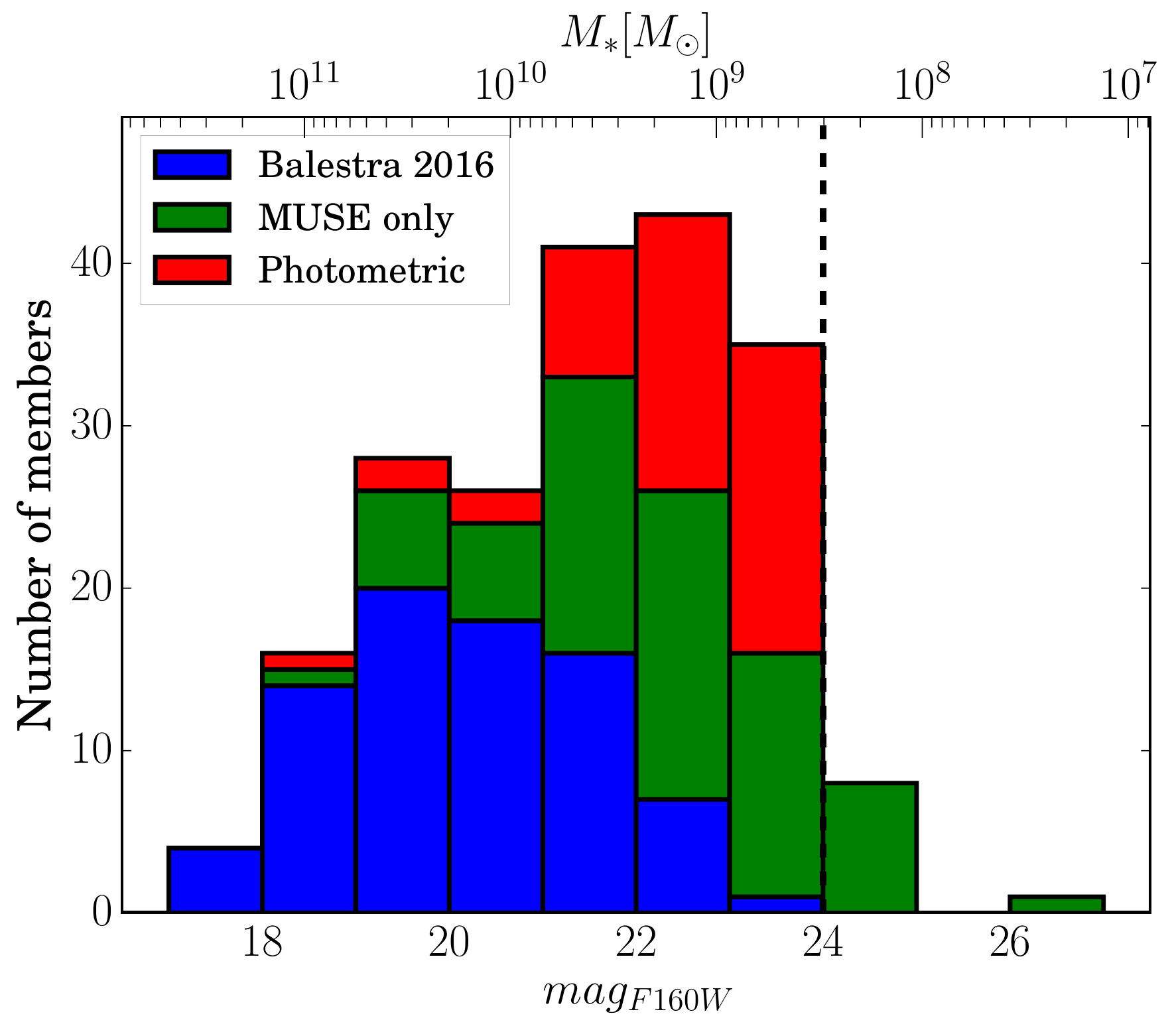}
   \caption{Stacked distribution of F160W magnitudes of galaxy members in the
     core of MACS~0416. Spectroscopically confirmed members from
     CLASH-VLT \citep{2015arXiv151102522B} are shown in blue (80), the
     newly identified members by MUSE in green (73) and the remaining
     photometrically selected galaxies with $mag_{F160W}<24$ from Gr15
     in red (193 in total). The stellar mass on the top axis is
     computed from $\log(M_{*}/M_{\odot}) = 18.541 - 0.416 \times
     mag_{F160W}$ from Gr15. }
   \label{fig:members_hist}
\end{figure}

Interestingly, we find that only four galaxies were misidentified as
galaxy members in Gr15, and add 22 new spectroscopic members brighter
than $mag_{F160W} = 24$. The latter is the limiting magnitude adopted
in Gr15 to define the photometric sample within the HST/WFC3 FoV
($\approx 5.5 \,\rm arcmin^2$). Following Gr15, we used the redshift
range $[0.382-0.410]$ to define membership. This interval,
corresponding to approximately three times the cluster velocity dispersion
in the rest frame, is somewhat wider than the velocity range of members
selected with kinematic methods in \citet{2015arXiv151102522B}.  In
Figure~\ref{fig:members_hist}, we show the magnitude distribution of
the cluster members, highlighting the improvement enabled by the MUSE
data, which provides also nine members fainter than $mag_{F160W} = 24$.  This
is not surprising, considering that MUSE observations outperform VIMOS
spectroscopy (with 1-2 hr exposure) when measuring redshifts of faint
early-type galaxies, and that the Gr15 cluster member catalog was
constructed maximizing purity over completeness.  Thus, the new sample
of cluster members inlcudes 193 galaxies, of which 144 (75\%) have
measured spectroscopic redshifts.  Finally, we note that all cluster
members with measured GLASS redshifts are confirmed by MUSE.

\section{Strong lensing modeling}
\label{sec:sl_modelling}

We use the positions of the new large set of multiple images
described above to study the mass distribution of MACS~0416 with
the strong lensing technique described in Ca16
(see Section 3 of that paper for details and equations). We briefly
summarize here the main characteristics of our strong lensing model.
We use the public software \emph{lenstool}
\citep{1996ApJ...471..643K, 2007NJPh....9..447J} to reconstruct the
projected total mass distribution in a parametric form, by minimizing the
distances between the model-predicted and observed positions of
the multiple images.

As emphasized in Gr15 and Ca16, we only rely on secure multiple images
with spectroscopic redshift, since additional {\sl photometric}
multiply lensed systems lead to systematic uncertainties in the model,
due to possible misidentification of multiple images, and introduce
degeneracies in the mass distribution, due to uncertain angular
diameter distances associated with photometric or unknown redshfits. 
We select the new multiple images firstly from our spectroscopic catalog, identifying the sources with very similar redshifts.
Only a few sources are found to be at the same redshift but are not multiple images.
These sources likely belong to a proto-cluster or group of galaxies behind MACS~0416.
Notice that we did not use our previous strong lensing model of Gr15 to select the new multiple image families.
Our selection strongly relies on the spectroscopic identifications.

In order not to bias our model, we do not include any family without spectroscopic confirmation and do not exclude spectroscopically confirmed multiple images that cannot be reproduced well by our model. 
Moreover, we do not include multiple images when the identification is not secure, either because no detectable images or spectroscopic emission is found, or the image association is not unique (see magenta circles in Figure~\ref{fig:arcs}).
Specifically, we find more than
  one candidate image at the location where we expected to find the
  multiple images 9c, 11a, 14d, 16e, 18c and 23c, making their
  identification uncertain. In the case of multiple image 12a, its
  strong distortion due to a galaxy member does not allow us to
  identify the luminosity peak corresponding to the images 12b and
  12c. Moreover, we do not find clear counterparts of 21a, 22a, 33c
  and 34a in the HFF imaging. On the other hand, although we do not
  have spectroscopic confirmation of the multiple images 17b, 17c and
  28a, their color, morphology and parity are in very good agreement
  with other images with measured redshifts, resulting in secure
  multiple-image systems.  Interestingly, the families 21 and 35 show
  very clear Lyman-$\alpha$ emission but no evident counterparts in the
  HFF imaging.  For these two families, we consider multiple-image positions at the peaks of the
  emission in the MUSE data. In summary, we build a secure set of 102
  multiple images (belonging to 37 families) with spectroscopic redshifts, to
  reconstruct the total mass distribution of MACS~0416.  The coordinates,
  redshifts, magnitudes, previous literature information of the
  multiple images in our final set are presented in Table
  \ref{tab:multiple_images}.

  We adopt an uncertainty on the observed multiple
  image positions ($\sigma^{obs}$) of $0\arcsec.5$, which takes into
  account possible perturbations form line-of-sight structures, as discussed in
  Gr15 and Ca16 and close to the theoretical expectations
  \citep{2010Sci...329..924J, 2012MNRAS.420L..18H}. We show below
  that this value yields a reduced $\chi^2$ value very close to one.
To compute the posterior probability distributions of the model parameters, hence their statistical errors and correlations, \emph{lenstool} uses a Bayesian Markov Chain Monte Carlo (MCMC) technique.

\subsection{Mass model components}
The overall total mass distribution of the cluster is modeled with a smooth component, made of one
or more halos which represent the dominant dark matter, the hot
gas and the intra-cluster light, and a clumpy sub-halo population
traced by the member galaxies.

Each smooth component is parametrized with a pseudo-isothermal
elliptical mass distribution \citep[PIEMD;][]{1993ApJ...417..450K}.
This model is characterized by the values of an effective velocity dispersion
($\sigma_v$), core radius ($r_{core}$), ellipticity
($\varepsilon$) and position angle ($\theta$).
The ellipticity is defined as $\varepsilon \equiv (a^2 - b^2)/(a^2 + b^2)$, where $a$ and $b$ are the semi-major and minor axis,
respectively.
Since the distribution of dark matter is not necessarily associated to an
observable counterpart, the center of this component ($x$
and $y$) is also a free parameter in the model.

As for the clumpy component, we exploit the new highly complete sample
of cluster members described above, attaching a halo to each of the
193 member galaxies. Each halo is modeled with a circular pseudo-isothermal mass distribution
\citep[dPIE;][]{2007arXiv0710.5636E, 2010A&A...524A..94S}. The
position of each dPIE is fixed at the luminosity center of member galaxies,
while the values of effective velocity dispersion $\sigma^{gals}_{v,i}$ and
truncation radius $r^{gals}_{cut, i}$ are free parameters.  Following
Gr15 anf Ca16, we scale these parameters with the observed luminosity in the filter $F160W$
of each galaxy, using:

\begin{equation}
  \sigma^{gals}_{v,i} = \sigma^{gals}_v\left( \frac{L_i}{L_{0}} \right)^{0.35}\,\, {\rm and}\,\,\, r^{gals}_{cut,i} = r^{gals}_{cut}\left( \frac{L_i}{L_{0}} \right)^{0.5},
\label{eq:mass_to_light}
\end{equation}
where $L_0$ is a reference luminosity which we choose to coincide with that of the northern BCG (BCG,N in Figure \ref{fig:arcs}, $mag_{F160W} = 17.02$).
We are thus left with only two parameters ($\sigma^{gals}_v$ and
$r^{gals}_{cut}$) describing the overall sub-halo population.  These
specific relations yield a shallow dependence of the galaxy total
mass-to-light ratio with their luminosity, that is $M_{total}/L
\propto L^{0.2}$. Such a ``tilted'' scaling relation is known from studies on the Fundamental Plane \citep{1987nngp.proc..175F, 1992ApJ...399..462B} of early-type galaxies and has been
shown to better reproduce the observed positions of multiple images in
previous high-precision lensing models, such as those for MACS~0416 itself
(Gr15) and MACS~J1149.5$+$2223
\citep{2016ApJ...822...78G}, when predicting the reappearance of SN Refsdal \citep{2016ApJ...819L...8K}.

The inset in Figure \ref{fig:arcs} shows a galaxy-scale
  lensing system (family 14) embedded in the cluster potential. Since the
  cluster galaxies G1 and G2 are the main contributors to the creation of the
  multiple images of this system, the total mass density parameters of these two galaxies are left free to
  vary in the modeling.  For the more luminous galaxy G1, we consider an elliptical profile and optimise also its values of ellipticity and position angle. We therefore
  have six extra free parameters describing G1 and G2:
  ($\sigma_v^{G1}$, $r_{cut}^{G1}$, $\varepsilon_{G1}$, $\theta_{G1}$)
  and ($\sigma_v^{G2}$, $r_{cut}^{G2}$).

  Similarly to previous works \citep[Gr15,][]{2014ApJ...797...48J,
    2014MNRAS.444..268R,2016ApJ...819..114K}, we take into account
  the lensing contribution of a foreground galaxy at $z=0.112$ located in
  the South-West region of MACS~0416, very close to family 37 (see
  Figure \ref{fig:arcs}). To do that, we include an extra dPIE mass component at the galaxy
  position (RA = 04:16:06.82 and DEC=$-$24:05:08.4) and cluster redshift, making its
  ``effective'' parameters $\sigma_v^{fore}$ and $r_{cut}^{fore}$ also free to vary.
This is a first simple approximation to a correct multi-plane lensing model, which is not included yet in the \emph{lenstool} software.

\subsection{Results}

Following Gr15, we first use two halos for the smooth total mass
component of MACS 0416, a complex merging cluster, as clearly
indicated by the distribution of cluster galaxies in two main clumps
around the BCG North and South.
This model has a total of 22 free parameters describing the cluster total mass distribution and can reproduce the observed positions of the multiple images with values of root-mean-square offset ($\Delta_{\rm rms}$) of $0\arcsec.82$ and minimum $\chi^{2}$ of 275.
In this case, the number of degrees of freedom (i.e., the number of observables minus the number of free parameters) is 108, resulting in a reduced $\chi^2$ of $2.55$, significantly higher than what was found in previous strong lensing studies on this cluster.

By inspecting the spatial distribution
of the offsets between the observed and model-predicted positions of
the multiple images, we notice larger
offsets in the region NE of the northern BCG. We then add a third
extended spherical pseudo-isothermal halo with four free parameters,
namely $\sigma_{v3}, r_{core,3}$ and its center ($x_3,y_3$) free to
vary across the entire FoV.
We find that the addition of this third
halo reduces significantly the $\Delta_{\rm rms}$ and $\chi^2$ values to $0\arcsec.59$ and $143$, respectively, and that the best-fitting center of this extra component is very close to a relatively minor clump in the
galaxy distribution (see Figure \ref{fig:members}).
Interestingly,
such a clump was not selected as an overdensity in the phase-space
analysis of cluster galaxies over the entire cluster by
\citet{2015arXiv151102522B}, however its center is in good
agreement with a peak in the convergence map obtained by
\citet{2016arXiv160300505H} (the offset is only $\approx 5\arcsec.5$, see their Figure 5), who performed a free
form reconstruction of the total mass distribution of MACS~0416,
combining both weak and strong lensing.

  In summary, the cluster total mass distribution adopted here has 26 free
  parameters: 1) 16 describing the three smooth dark-matter components; 2) six
  describing the galaxy-scale lensing system; 3) two for the prominent foreground
  galaxy; and 4) two describing the mass-luminosity scaling relation of
  galaxy members.  Overall, the number of constraints from the
  positions of the 102 multiple images associated to 37 sources is 130
  ($102 \times 2-37 \times 2$), and therefore the number of degrees of
  freedom is 104 ($130-26$). Notably, these constraints are well
  distributed across the entire central region of MACS~0416, spanning
  a wide redshift range $(0.94-6.15)$, which is key to the
  fidelity of the reconstructed cluster total mass distribution and break the mass-sheet degeneracy \citep{1985ApJ...289L...1F,1997A&A...318..687S}.

\begin{figure}
  \centering
    \includegraphics[width = 1.0\columnwidth]{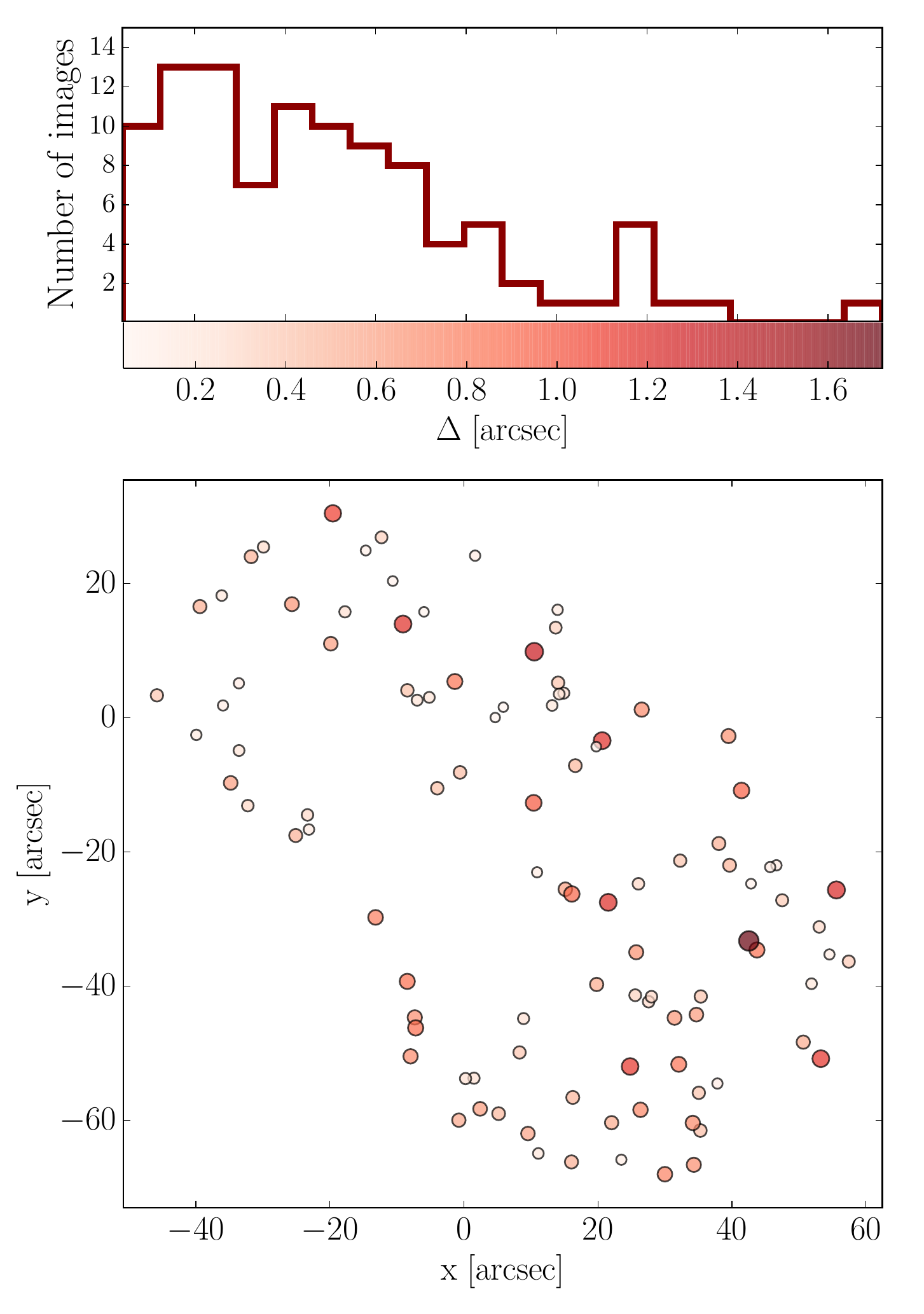}
    \caption{The top panel shows the distribution of the absolute
      value of the offsets $\Delta$ between the observed and predicted (from our best-fitting model) positions of
      the multiple images. The bottom map shows the spatial distribution of
      these offsets. The circles indicate the positions of the observed multiple images
      (relatively to the northern BCG), with colors and sizes scaling
      with their positional offsets.}
   \label{fig:offsetmap}
\end{figure}

\begin{table}[h]
    \caption{Median values and confidence levels of the cluster total mass distribution parameters from the MCMC analysis of the strong lensing model. Coordinates are relative to the position of the BCG,N  (RA=04:16:09.15 and Dec=$-$24:04:03.0)}
    \begin{tabular}{c c c c c} \hline \hline
    ~ & Median              & $68\%$ CL & $95\%$ CL & $99.7\%$ CL \\ \hline
 \multicolumn{5}{l}{Halo around BCG-North (relative position = $(0\arcsec, 0\arcsec)$)} \\ \hline
$x_1(\arcsec)$ & $-2.4$ & $_{-0.6}^{+0.7}$ & $_{-1.2}^{+1.4}$ & $_{-1.8}^{+2.2}$ \\
$y_1(\arcsec)$ & $1.8$ & $_{-0.5}^{+0.4}$ & $_{-1.1}^{+0.8}$ & $_{-1.8}^{+1.2}$ \\
$\varepsilon_1$ & $0.85$ & $_{-0.01}^{+0.01}$ & $_{-0.03}^{+0.03}$ & $_{-0.04}^{+0.04}$ \\
$\theta_1(\rm deg)$ & $143.9$ & $_{-0.8}^{+0.8}$ & $_{-1.7}^{+1.7}$ & $_{-2.5}^{+2.4}$ \\
$r_{core,1}(\arcsec)$ & $6.3$ & $_{-0.7}^{+0.7}$ & $_{-1.3}^{+1.3}$ & $_{-1.9}^{+1.9}$ \\
$\sigma_{v1}(\rm km/s)$ & $707$ & $_{-28}^{+26}$ & $_{-56}^{+50}$ & $_{-83}^{+79}$ \\
\hline
\multicolumn{5}{l}{Halo around BCG-South (relative position = $(20\arcsec.3, -35\arcsec.8)$)} \\ \hline
$x_2(\arcsec)$ & $19.6$ & $_{-0.2}^{+0.3}$ & $_{-0.5}^{+0.8}$ & $_{-0.7}^{+1.3}$ \\
$y_2(\arcsec)$ & $-36.4$ & $_{-0.5}^{+0.4}$ & $_{-1.2}^{+0.8}$ & $_{-1.8}^{+1.2}$ \\
$\varepsilon_2$ & $0.77$ & $_{-0.01}^{+0.01}$ & $_{-0.03}^{+0.02}$ & $_{-0.04}^{+0.03}$ \\
$\theta_2(\rm deg)$ & $125.6$ & $_{-0.4}^{+0.4}$ & $_{-0.8}^{+0.8}$ & $_{-1.1}^{+1.2}$ \\
$r_{core,2}(\arcsec)$ & $12.5$ & $_{-0.5}^{+0.6}$ & $_{-1.1}^{+1.1}$ & $_{-1.6}^{+1.6}$ \\
$\sigma_{v2}(\rm km/s)$ & $1102$ & $_{-17}^{+16}$ & $_{-33}^{+32}$ & $_{-48}^{+47}$ \\
\hline
\multicolumn{5}{l}{Third spherical halo} \\ \hline
$x_3(\arcsec)$ & $-34.4$ & $_{-1.0}^{+0.8}$ & $_{-2.5}^{+1.4}$ & $_{-4.2}^{+2.1}$ \\
$y_3(\arcsec)$ & $7.9$ & $_{-0.6}^{+0.7}$ & $_{-1.1}^{+1.9}$ & $_{-1.7}^{+3.0}$ \\
$r_{core,3}(\arcsec)$ & $6.4$ & $_{-2.0}^{+2.2}$ & $_{-4.0}^{+4.7}$ & $_{-5.9}^{+6.8}$ \\
$\sigma_{v3}(\rm km/s)$ & $434$ & $_{-52}^{+58}$ & $_{-101}^{+121}$ & $_{-145}^{+170}$ \\
\hline
\multicolumn{5}{l}{Galaxy scale system 14} \\ \hline
$\varepsilon_{G1}$ & $0.3$ & $_{-0.2}^{+0.2}$ & $_{-0.3}^{+0.3}$ & $_{-0.3}^{+0.3}$ \\
$\theta_{G1}(\rm deg)$ & $115$ & $_{-75}^{+39}$ & $_{-111}^{+60}$ & $_{-115}^{+65}$ \\
$\sigma_{v}^{G1}(\rm km/s)$ & $143$ & $_{-11}^{+15}$ & $_{-22}^{+34}$ & $_{-32}^{+40}$ \\
$r_{cut}^{G1}(\arcsec)$ & $8.1$ & $_{-4.7}^{+4.6}$ & $_{-6.8}^{+6.5}$ & $_{-7.5}^{+6.9}$ \\
$\sigma_{v}^{G2}(\rm km/s)$ & $48$ & $_{-24}^{+29}$ & $_{-34}^{+46}$ & $_{-36}^{+49}$ \\
$r_{cut}^{G2}(\arcsec)$ & $5.2$ & $_{-3.6}^{+3.4}$ & $_{-4.9}^{+4.6}$ & $_{-5.1}^{+4.8}$ \\
%$\sigma_{v}^{fore}$ & $221.30$ & $_{-16.34}^{+15.99}$ & $_{-33.15}^{+31.15}$ & $_{-51.07}^{+43.55}$ \\
%$r_{cut}^{fore}$ & $67.98$ & $_{-23.65}^{+21.74}$ & $_{-40.38}^{+30.42}$ & $_{-51.46}^{+31.93}$ \\
\hline
\multicolumn{5}{l}{Sub-halo population} \\ \hline
$r^{gals}_{cut}(\arcsec)$ & $10.5$ & $_{-2.4}^{+2.7}$ & $_{-4.6}^{+6.2}$ & $_{-6.2}^{+10.3}$ \\
$\sigma_{v}^{gals}(\rm km/s)$ & $251$ & $_{-14}^{+14}$ & $_{-26}^{+31}$ & $_{-40}^{+48}$ \\
\hline\hline
    \end{tabular}
\label{tab:bf_parameters}
\end{table}

In Table \ref{tab:bf_parameters}, we show the median values of the
parameters of our strong lensing models with their intervals at the
$68\%$, $95\%$ and $99.7\%$ confidence levels (CL).  The
coordinates are relative to the position of the northern BCG and the
angles are counted counterclockwise from the horizontal axis.  The
final reduced $\chi^2$ value is close to one ($= 1.37$), while the rms
offset is $\Delta_{\rm rms} = 0\arcsec.59$ ($\Delta_{\rm median} = 0\arcsec.5$, see Figure~\ref{fig:offsetmap}).

The best-fitting centers of the three diffuse halos are shown as red
contours (corresponding to a 95\% significance level) in
Figure~\ref{fig:members}. In Gr15, we discussed the apparent offest in
the projected distance between the centers of the two main halos and
the corresponding BCGs. We cautioned that while an offset of
$9\arcsec.3$ and $5\arcsec.8$, relative to BCG,N and BCG,S
respectively, was statistically significant, systematics inherent in
the parametric form of the lens model, as well as projection effects, made
it difficult to claim such an offset, which has often been used to constrain the
collisionless nature of DM \citep{2011MNRAS.415..448W,
  2015MNRAS.452L..54K}, between the DM and stellar component.
Interestingly, our new model, which is based on $\approx3$-times the
number of multiple images of Gr15, albeit with an extra halo
component, leads to a projected distance of only $3\arcsec.0_{-2.6}^{+2.1}$ and
$1\arcsec.0_{-1.4}^{+0.5}$ (99.7 \% CL) of the two main halos from the BCG North and South,
respectively. The statistical significance of the offset between each
DM halo and its hosting BCG is therefore reduced when compared with
Gr15.  We defer a further analysis on this issue to a future paper,
where the velocity dispersions of the BCGs are used to alleviate the
degeneracy between the parameters (centers and scale) of the DM halos
and those associated to the BCGs.

The center of the third dark matter halo ($x_3$ and $y_3$) is on the top of a clump of three galaxy members in the the north-east region of the cluster (see Figure \ref{fig:members}) and its mass is significantly smaller than that of the other two halos.
However, the possibility of having a vanishing mass (i.e., $\sigma_{v3}=0$) is excluded by the posterior distribution computed from the MCMC, statistically confirming the existence of this halo.
Finally, within a circle with radius of $\approx 15 \arcsec$ from this halo there is no evidence of a background or foreground structure in our redshift measurements and in the HST imaging, thus excluding a possible existence in this region of a significant perturber not belonging to MACS~0416.

Regarding the galaxy-scale lensing system 14, it is worth noticing
that although the value of the position angle of G1 is not well
constrained (see Table \ref{tab:bf_parameters}), its median value is
in very good agreement with that of the light distribution,
$\theta_{G1}^{light}=119^\circ$, measured with SExtractor.  On the
other hand, the ellipticity is unconstrained due to degeneracies with
the smooth dark-matter halos. Moreover, the values of effective
velocity dispersion from the best-fitting scaling relation of the
cluster members (using Eq. \ref{eq:mass_to_light} and the MCMC chain used to compute the
values in Table \ref{tab:bf_parameters}) are $108^{+21}_{-17} \rm km/s$
and $59^{+11}_{-9.4} \rm km/s$ (99.7\% CL) for the luminosities of G1 and
G2, respectively.  Interestingly, they agree, within the statistical
errors, with the values optimized separately for $\sigma_v^{G1}$ and
$\sigma_v^{G2}$ (see Table \ref{tab:bf_parameters}), thus indicating
that the total intrinsic scatter in the total mass-to-light ratios of the
sub-halo population does not seem to play an important role in the
cluster strong lensing modeling.

\begin{figure*}
  \centering
   \includegraphics[width = 1.\columnwidth]{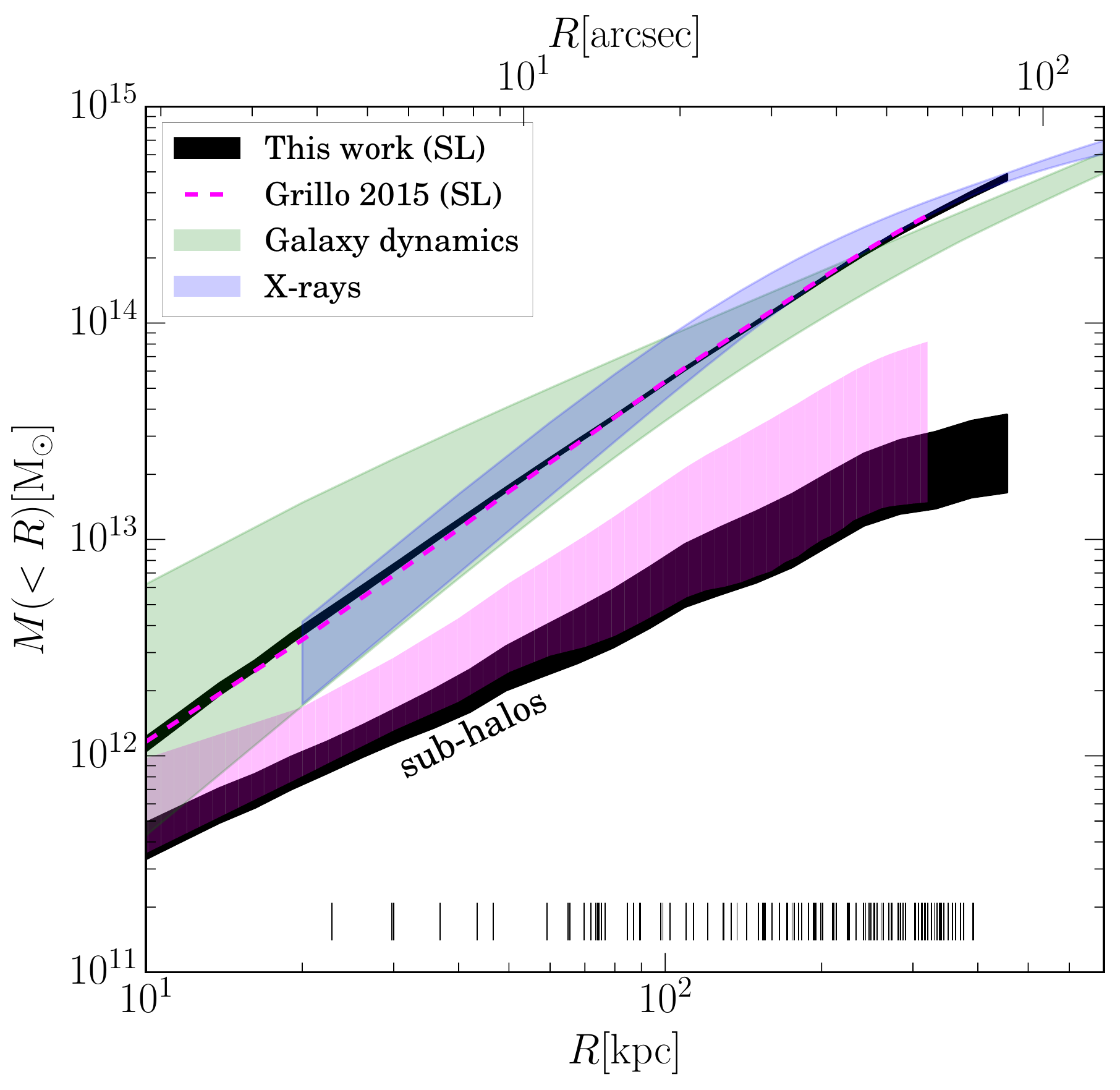}
   \includegraphics[width = 1.\columnwidth]{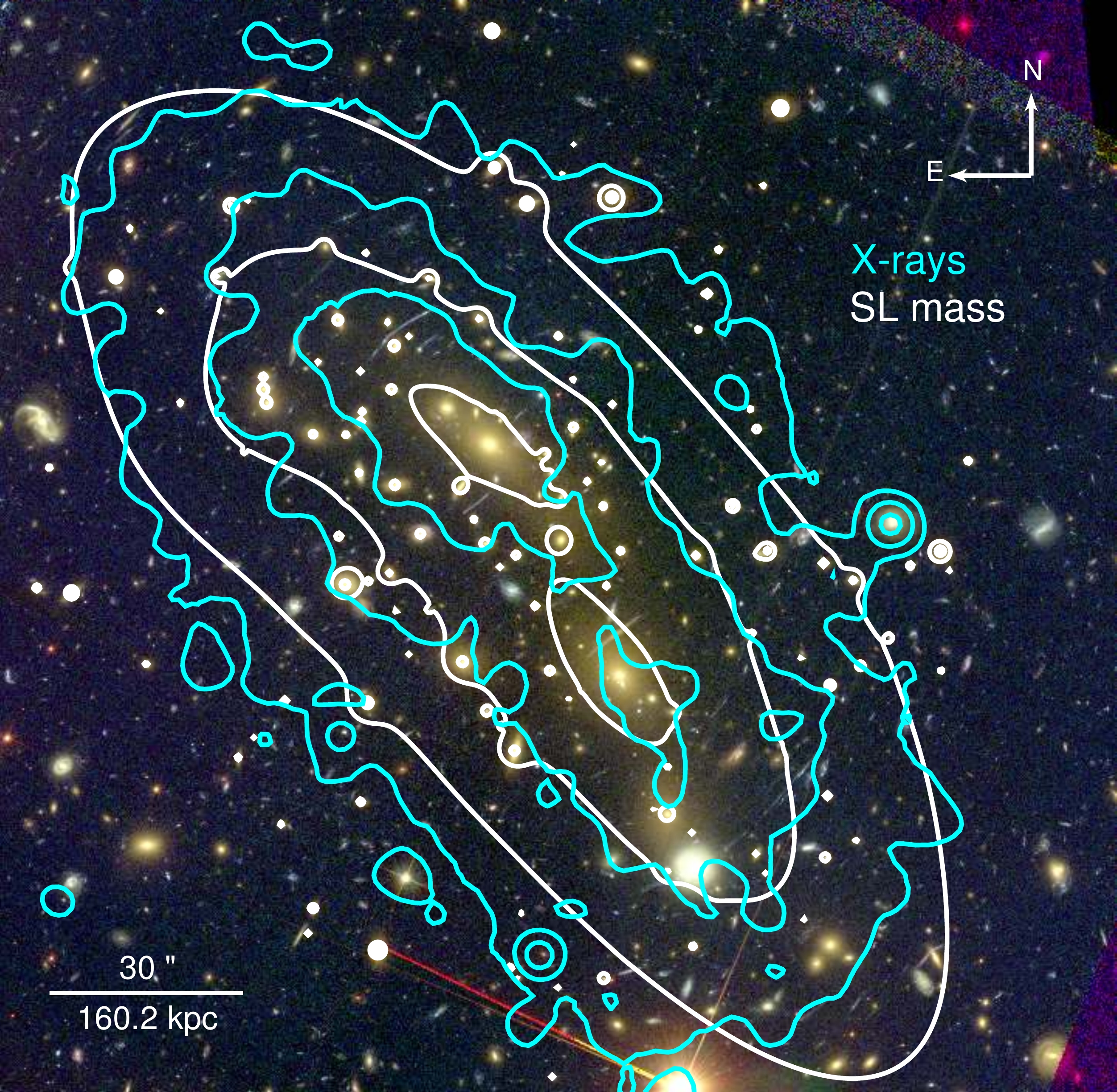}
   \caption{In the left panel we cumulative projected mass distribution of MACS~0416 from
     our new strong lensing model (total mass and sub-halo component),
     relative to the position of the BCG,N. The black regions
     correspond to the 99.7\% confidence level, while the magenta
     line and shaded area show the best-fitting profiles from our previous model
     presented in Gr15. The green and blue regions are the $1\,\sigma$
     determination of the mass profile from the dynamical analysis of
     $\sim\! 800$ cluster galaxies and the X-ray emission, respectively, reported in
     \citet{2015arXiv151102522B}. The total mass profiles associated
     to the sub-halo population of cluster galaxies from the lensing
     model is also shown. The vertical lines show the projected radial distances
     of the multiple images used in this work. In the right panel, we show the projected total mass iso-contours ($[\rm 0.6, \;1,\; 2] \times 10 ^{15}\,M_{\odot}Mpc^{-2}$ in white), and \emph{Chandra} X-ray contours (0.5-2 KeV) overlaid on the HFF-ACS color image.}
  \label{fig:mass}
\end{figure*}

In Figure~\ref{fig:mass}, we compare the cumulative projected total mass profile resulting from our best-fitting strong lensing model with that
obtained from the X-ray emission and dynamical analysis of $\sim\!  800$ cluster
galaxies. Interestingly, the latter is also best described by a softened (i.e., cored) isothermal
sphere model, as discussed in \citet{2015arXiv151102522B}.
The new strong lensing total mass
profile is perfectly consistent with that of Gr15, who used $\sim\!
1/3$ of the multiple images and the GLEE \citep{2010A&A...524A..94S, 2012ApJ...750...10S} software for the modeling.
Although the sub-halo total mass distribution in our model is lower than in Gr15, both measurements agree within the statistical errors (99.7\% CL).
This difference in mainly related to the inclusion of the third dark matter halo and the extra constraints in our strong lensing model.
In the right panel, we also compare the projected total mass and the X-ray surface brightness distribution.
As discussed in \citet{2015arXiv151102522B} the close resemblance of the total mass and gas component, adds further evidence to a pre-merger scenario for MACS~0416.

\begin{figure}
  \centering
  \hspace{-1.5pt}\includegraphics[clip=True, width = 0.49\columnwidth, frame=1pt]{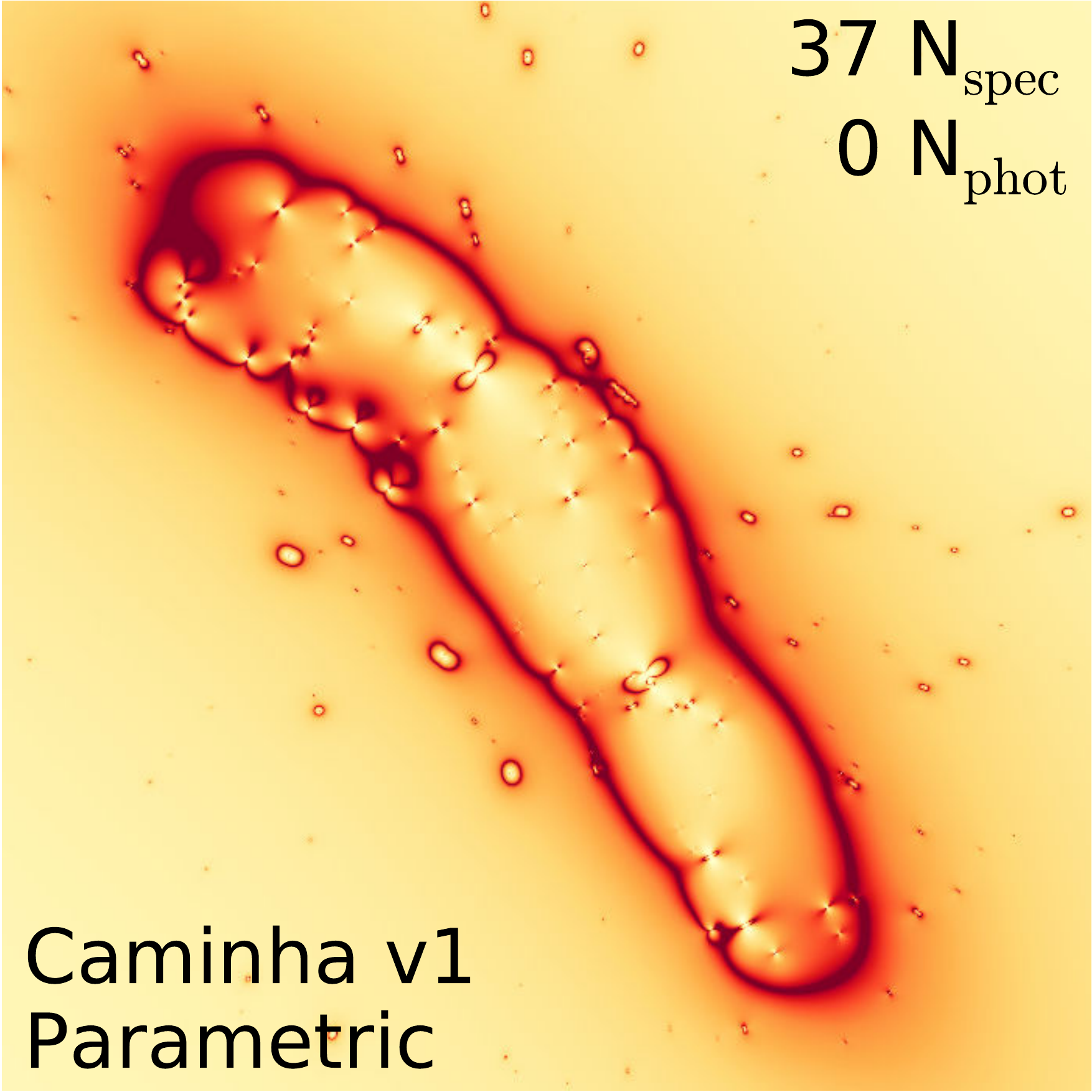}
  \includegraphics[clip=True, width = 0.49\columnwidth]{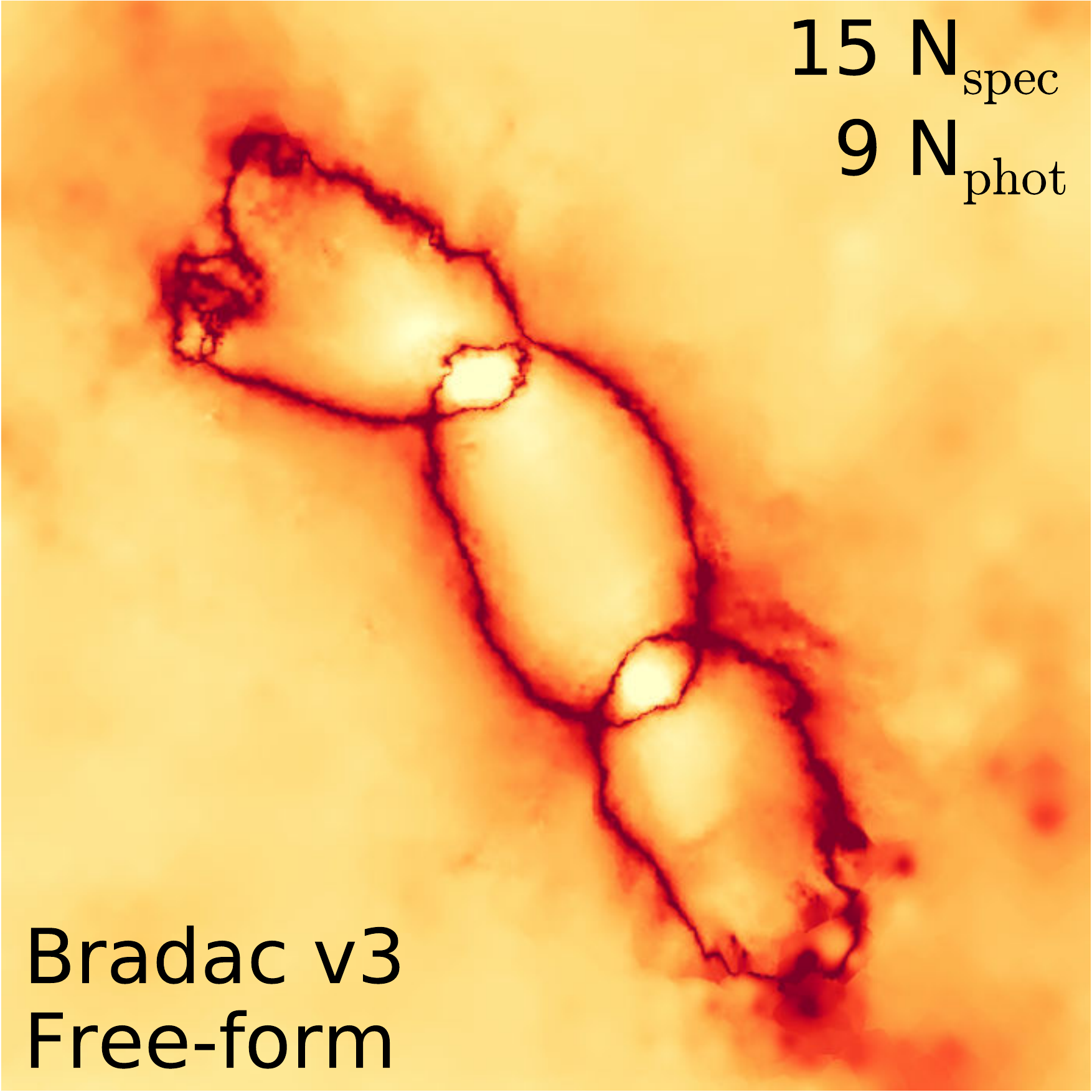}

  \includegraphics[clip=True, width = 0.49\columnwidth]{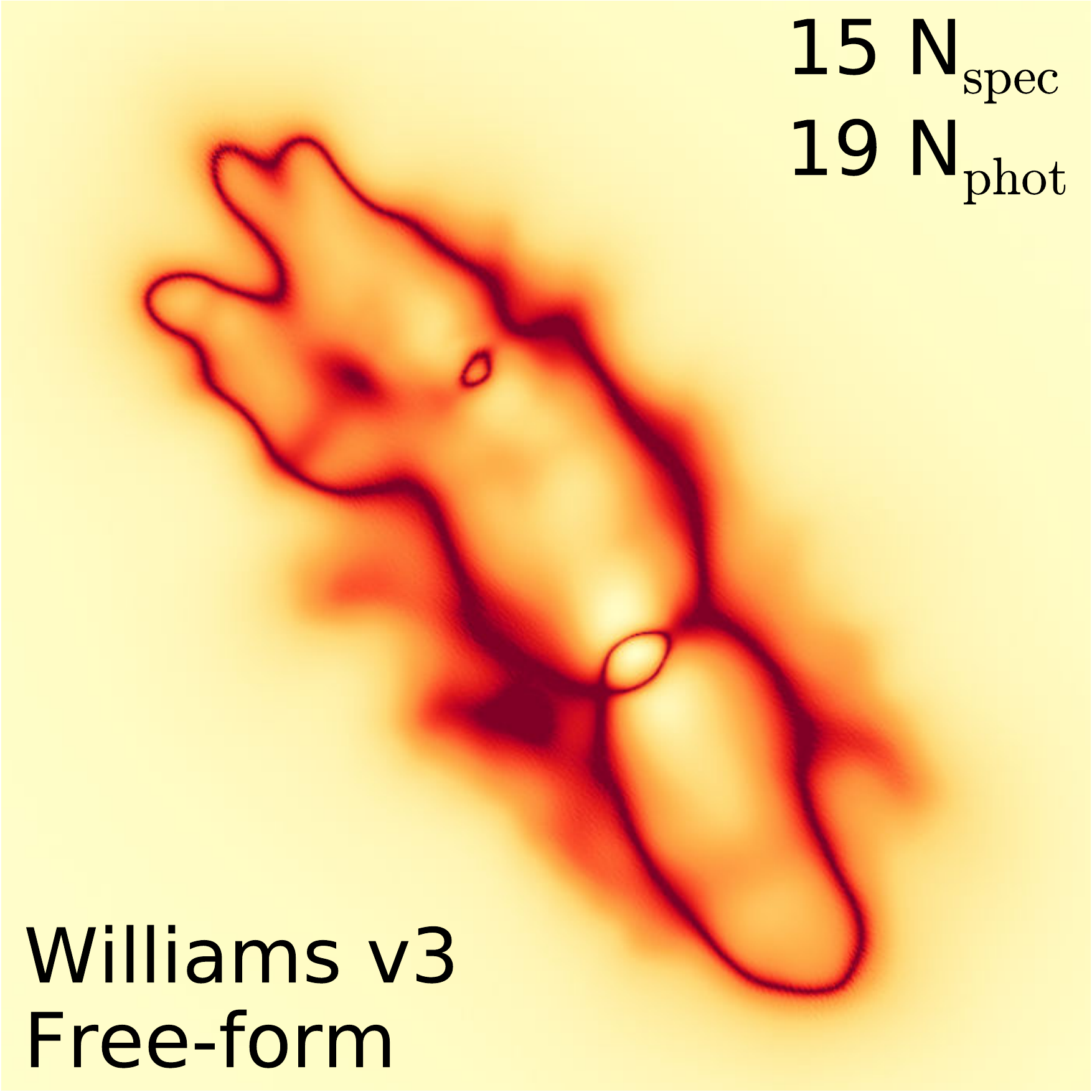}
  \includegraphics[clip=True, width = 0.49\columnwidth]{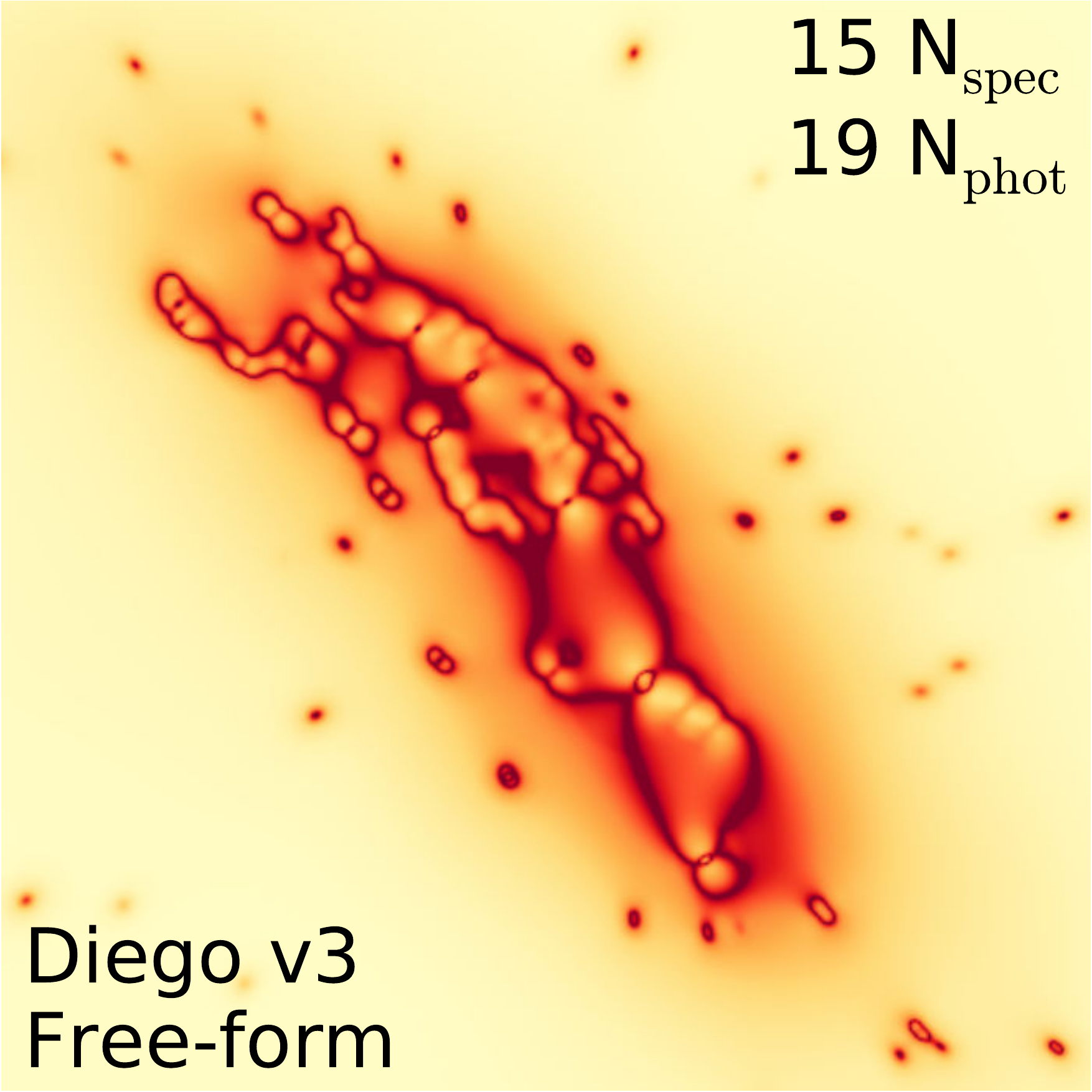}
  
  \includegraphics[clip=True, width = 0.49\columnwidth]{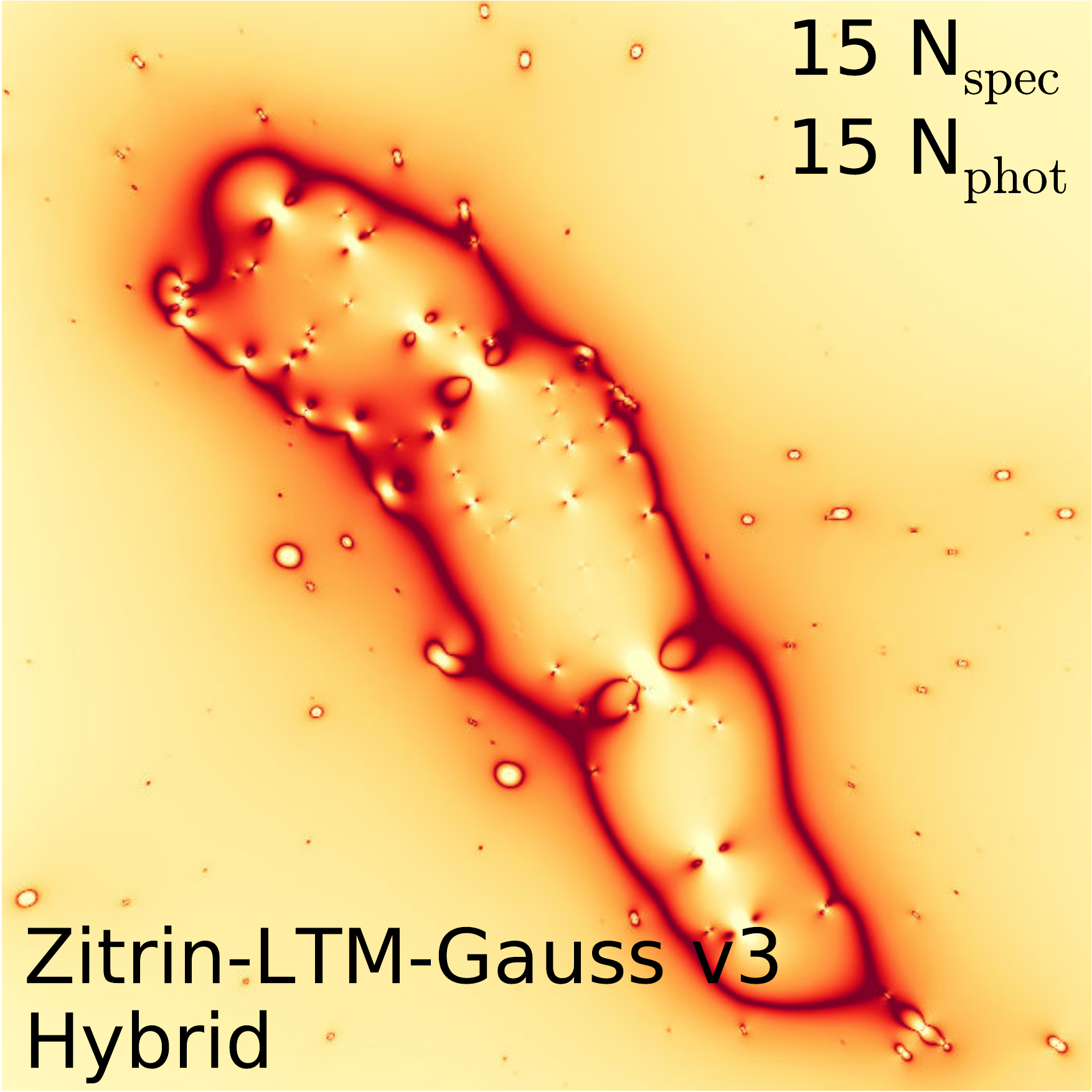}
  \includegraphics[clip=True, width = 0.49\columnwidth]{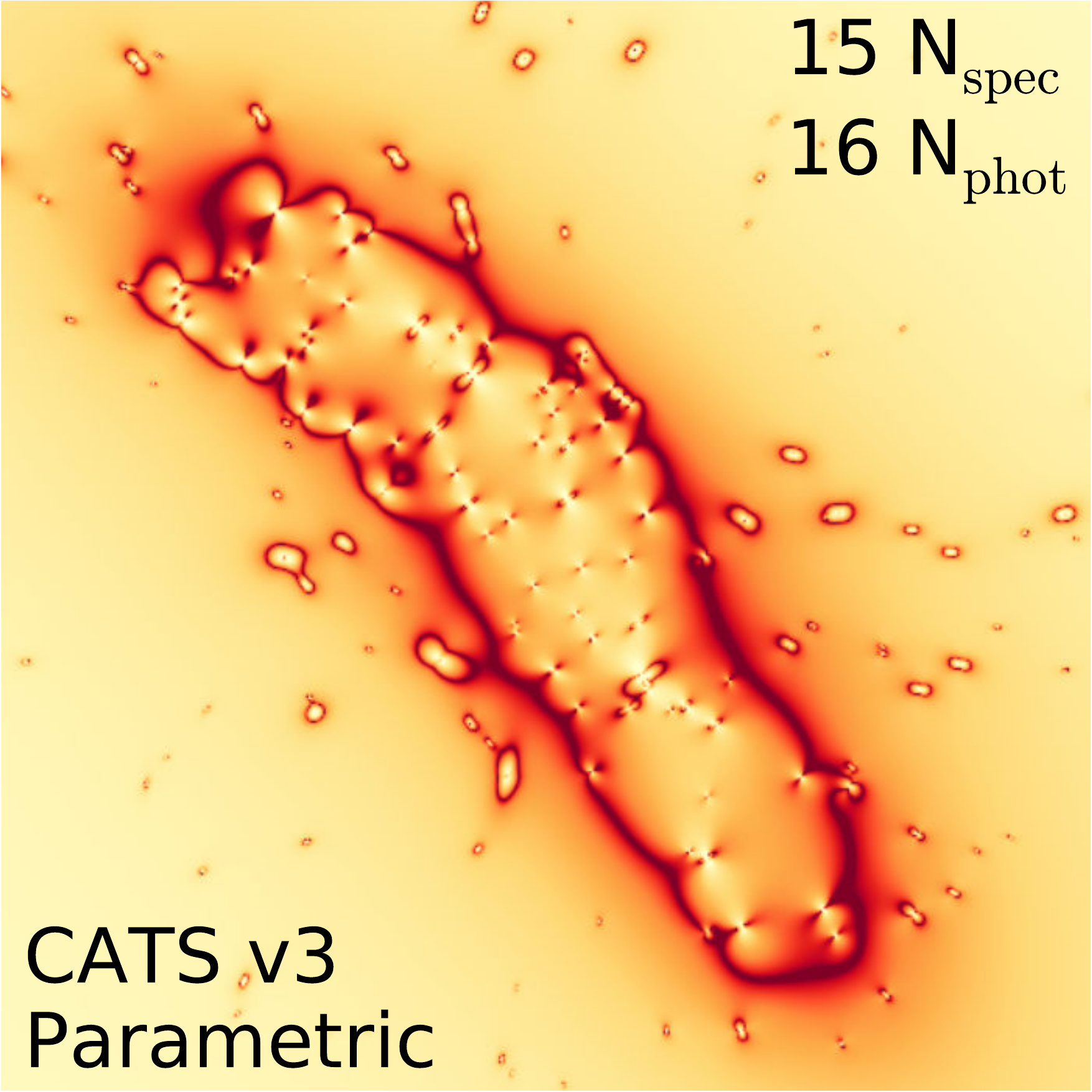}

  \includegraphics[clip=True, width = 0.49\columnwidth]{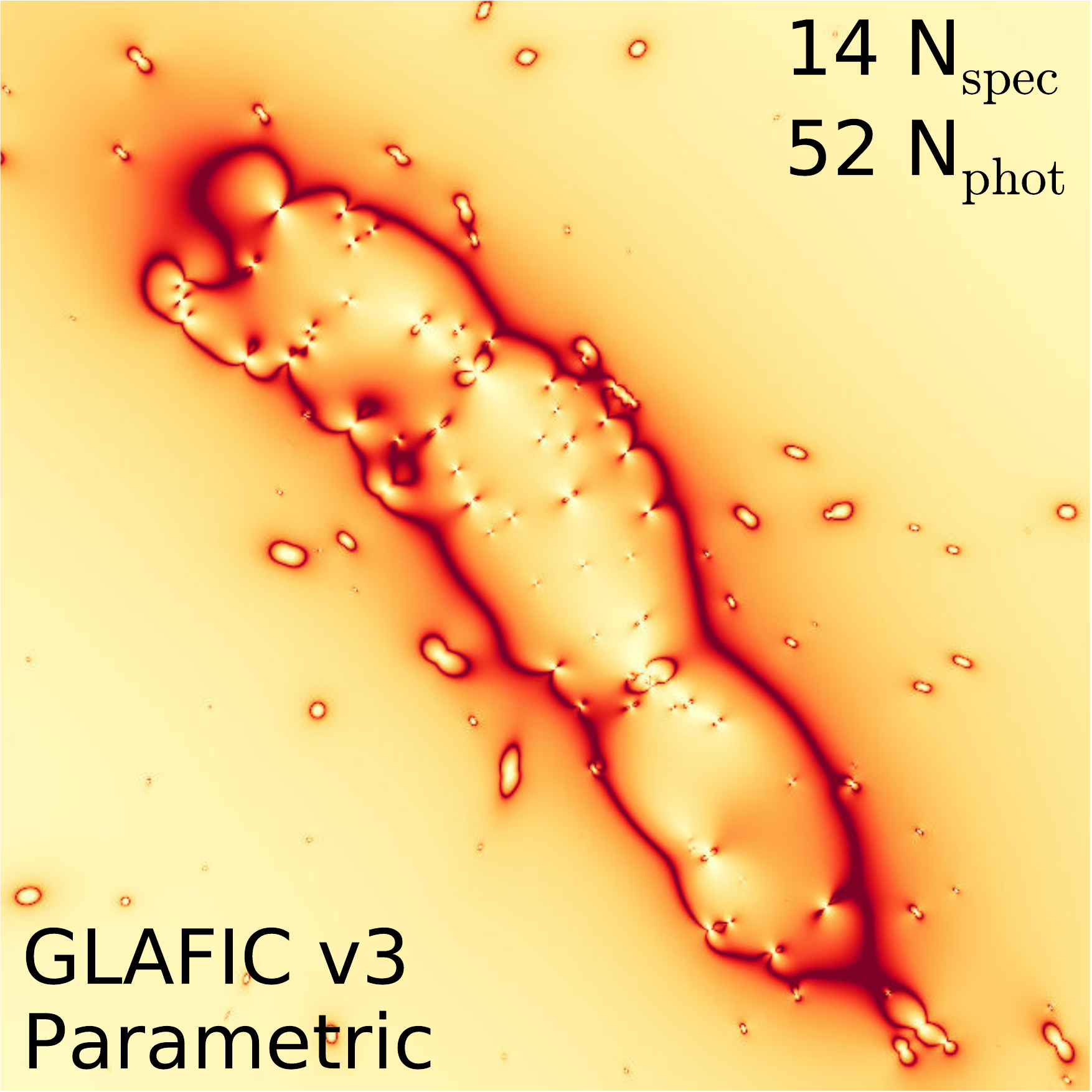}
  \includegraphics[clip=True, width = 0.49\columnwidth]{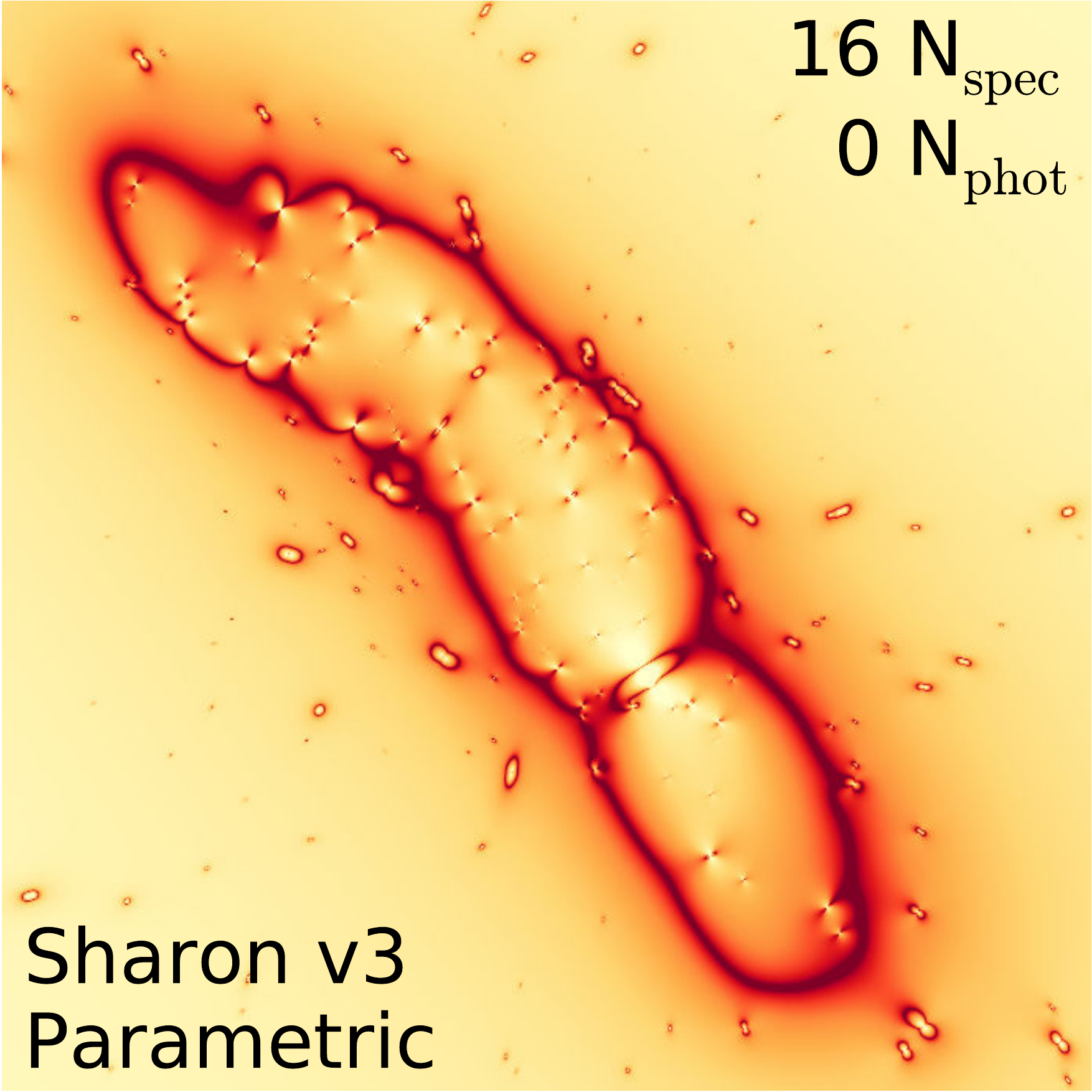}

  \includegraphics[clip=True, width = 0.99 \columnwidth]{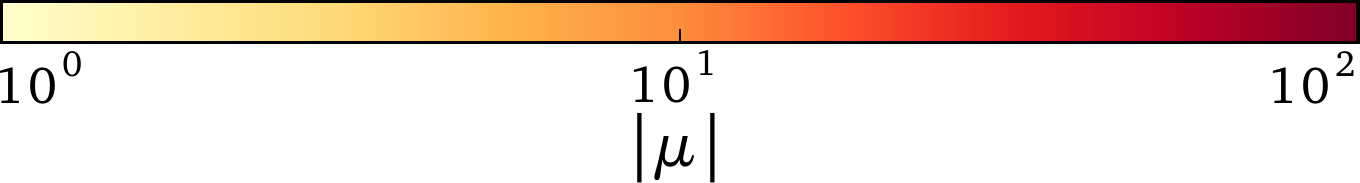}

  \caption{Magnification maps of MACS~0416 for a source at redshift 4. We show the HFF strong lensing models to compare with the work presented here (first panel). For each model, we indicate the number of multiple image families based on spectroscopic ($\rm N_{spec}$) and photometric ($\rm N_{phot}$) information.
  All panels are centered at the same position and are $2\arcmin.1$ across.}
  \label{fig:mag_map}
\end{figure}

In addition, we compare our magnification map for a source at redshift 4 with the HFF models in Figure \ref{fig:mag_map}.
The models cover different methodologies: 1) free-form, meaning that no parametric form is assumed for the total mass distribution, Bradac~v3 \citep{2016arXiv160300505H}, Willians~v3 \citep[using the GRALE software][]{2006MNRAS.367.1209L, 2010ascl.soft11021L} and Diego~v3 \citep[a pre-HFF modeling is presetend in][]{2015MNRAS.447.3130D}; 2) hybrid, that scales the smooth dark matter component with the light, Zitrin-LTM-Gauss~v3 \citep{2005ApJ...621...53B, 2009MNRAS.396.1985Z}; 3) parametric models using different codes, Sharon~v3 \citep[see][for their pre-HFF modeling]{2014ApJ...797...48J} and CATS~v3 \citep[pre-HFF presented in][]{2014MNRAS.444..268R} with \emph{lenstool} and GLAFIC~v3 \citep{2016ApJ...819..114K}.
Differences in parametric and free-form models are expected on small spatial scales.
In the case of the parametric models, the main difference is related to the north-east region due to the presence of the third halo, however the overall shape of the critical lines is very similar.

We emphasize that a full comparison of these maps becomes meaningful only when the input constraints (number and quality) of all models are the same or at least very similar.
Notice that some modelers consider several knots of the same background source as different multiple image systems, increasing the number of $\rm N_{spec}$.
Moreover, as underscored in this and previous works \citep[Gr15, Ca16, ][]{2016ApJ...817...60T, 2016ApJ...820...50R, 2016arXiv160808713J} a large number of spectroscopic families is critical for the robustness of the lens model, removing misidentification of multiple images and alleviating model degeneracies.
Also note that the use of different selection of member galaxies leads to significant deviations of the magnification on small scales.
In our case, a highly pure and complete sample of members is provided by MUSE spectroscopy.
For a complete comparison of the HFF lensing models see \citet{2016arXiv160507621P} and \citet{2016arXiv160604548M} in the context of strong lensing simulations of galaxy clusters.

\section{Conclusions}

In this article, we have significantly extended the panoramic VIMOS
spectroscopic campaign of MACS 0416, presented in Gr15 and \citet{2015arXiv151102522B}, with
data from the MUSE integral-field spectrograph on the VLT, which has
yielded 208 new secure redshift measurements in the central 2
arcmin$^2$ region of the cluster. Notably, a new large set of multiply
lensed sources was identified using two MUSE archival pointings,
extending the work of Gr15 and \citet{2016arXiv160300505H} and
bringing the number of spectroscopically identified multiple-image
systems from 15 to 37. This was possible by measuring 59 new redshifts
to very faint magnitude, thanks to the sensitivity of MUSE to line
fluxes as faint as $10^{-19}\, {\rm erg}\, {\rm s}^{-1}\, {\rm
cm}^{-2}\, \AA^{-1}$ \citep[see][for the study of a similar set of
low-luminosity Lyman-$\alpha$ emitters with MUSE observations of the
HFF cluster AS1063]{{2016arXiv160601471K}}. This new sample also
extends the redshift range of known multiple images, with five additional
systems at $z>5$, one of which is at $z=6.145$ (13 images with measured
redshift at $z>5$).  The MUSE observations also allowed us to secure
redshifts of 144 member galaxies over an area
of $\sim\! 0.2\, {\rm Mpc}^2$. Three-quarters of the
cluster galaxies selected down to $mag_{F160W}=24$ (corresponding to
$M_*\approx 3\times 10^8 \,\rm M_\odot$) are now spectroscopically
confirmed.

With such a large set of 102 spectroscopic multiple images and a much
improved sample of galaxy members in the cluster core, we
have built a new strong lensing model and obtained an accurate
determination of the projected total mass distribution of MACS 0416.
The main results of this study can be summarized as follows:
\begin{enumerate}

\item We can reproduce the observed multiple-image positions with an
  accuracy of $\Delta_{\rm rms}=0\arcsec.59$, which is somewhat larger
  than the one obtained by Gr15 ($0\arcsec.36$), who however used less
  than one-third of the multiple images.
\item The large-scale component of the total mass distribution was initially
  modeled with two cored elliptical pseudo-isothermal profiles around the two
  BCGs, as in Gr15, however larger positional offsets $\Delta$ in the
  NE portion of the cluster led us to introduce a third floating cored
  halo in the model. We find interesting that, besides significantly
  reducing the $\Delta_{\rm rms}$, the best-fit position of this third
  halo is very close to a peak of the convergence map obtained by
  \citet{2016arXiv160300505H} with an independent free-form lensing
  model, which also exploits the weak-lensing shear.  Although this
  third halo is centered on a relatively small overdensity of cluster
  galaxies, it could not be identified in the phase-space analysis of 
  \citet{2015arXiv151102522B}, most probably because of the combination of projection effects
  and the absence of clear separation in the projected velocity space.

\item The new best-fitting centers of the two main halos are now found
  within $\sim\! 2\arcsec$ from the respective BCGs,
  further reducing the halo-BCG offset when compared with the Gr15
  model. As described in \citet{2015arXiv151102522B}, such a concentric distribution of light and dark-matter mass, when compared with the distribution of the X-ray emitting gas whose main peak is at the position of the northern BCG, is consistent with a pre-merging scenario.

\item The cumulative projected total mass profile is found in excellent
  agreement with the one of Gr15, and in good agreement with the dynamical and X-ray mass
 which was however obtained with the simple
  approximation of a single spherical halo \citep[see][]{2015arXiv151102522B}. Together with
  the point 2. above, this suggests that owing to a significant
  enhancement of constraints in the strong lensing model we are now able
  to better resolve the mass distribution of the smooth cluster halo.

\item The overall scaling of the total mass-to-light ratio for the
  sub-halo population, traced by the new highly complete and pure
  sample of cluster galaxies, is found consistent with the one of Gr15.
  Our new model therefore corroborates the evidence found in Gr15
  that a sub-halo mass function is significantly suppressed when
  compared to simulations, particularly at the high-mass end. A similar
  result has recently been obtained in an independent study \citep{2016arXiv160701023M} of the
  Abell 2142 galaxy cluster with SDSS data.  A detailed analysis of the sub-halo
  population and different mass components in the core of MACS 0416,
  which takes advantage of the internal velocity dispersions of
  cluster galaxies \citep[see e.g.,][]{2015MNRAS.447.1224M,2016arXiv160208491M} is deferred to a
  future paper, where we also plan a detailed comparison with the
  study of \citet{2016arXiv160300505H}.

\end{enumerate}

Remarkably, the new spectroscopic identifications with MUSE
observations of MACS 0416 match in some cases the continuum magnitude
limit of the HFF data for Lyman-$\alpha$ emitters \citep[see
also][]{2016arXiv160601471K}, and complement the HST NIR GRISM
spectroscopy of the GLASS survey. Not surprisingly, this cluster now
becomes one of the best test bench for strong lensing modeling (see
Figure~\ref{fig:mag_specz}), which we argue need to rely largely, or
entirely, on spectroscopically confirmed multiple-image systems for
high-precision modeling.

  The accuracy we have reached in reproducing the observed multiple-image positions with this new model, on the other hand suggests that
  it will be challenging to further improve on these results by simply
  introducing more mass components in parametric
  models. Interestingly, the large number of constraints for this
  cluster should allow free-form models to become more
  effective, for example in discovering extra mass clumps with unusual total
  mass-to-light ratios. As already noted in Ca16 \cite[see
  also][]{2016ApJ...817...60T}, with the current high-quality set of
  strong lensing constraints we seem to have hit the limit of the
  single-plane lensing approximation, so that the next step in
  precision strong-lensing modeling inevitably will have to properly
  take into account the effects of the structure along the line of
  sight, adequately sampled by spectroscopic data.

  As previously done with CLASH-VLT VIMOS observations of HFF
  clusters, we make public the new extended redshift catalog \footnote{the full redshift catalog including VIMOS and MUSE measurments can be found in the electronic journal and at the link: \url{https://sites.google.com/site/vltclashpublic/data-release}}, which
  includes secure redshift determinations from the MUSE data, in the
  effort to add further value to the entire HFF dataset.

\begin{acknowledgements}
  The authors thank the anonymous referee for the useful comments on the manuscript.
  We acknowledge financial support from PRIN-INAF 2014 1.05.01.94.02.
  C.G. acknowledges support by VILLUM FONDEN Young Investigator
  Program grant 10123.  This work made use of the CHE cluster,
  managed and funded by ICRA/CBPF/MCTI, with financial support from
  FINEP (grant 01.07.0515.00 from CT-INFRA - 01/2006) and FAPERJ
  (grants E-26/171.206/2006 and E-26/110.516/2012).  P.R. acknowledges
  the hospitality and support of the visitor program of the DFG
  cluster of excellence “Origin and Structure of the Universe”. This
  work made use of data taken under the ESO programs ID
  094.A-0115(B) and ID 094.A-0525(A), and also 186.A-0798.
\end{acknowledgements}

\bibliographystyle{aa}
\bibliography{references}

%\begin{appendix}

%\section{Multiple image properties and spectra}

\vspace{0.5cm}
\noindent {\bf \large Appendix A: Multiple image properties and spectra}

In this appendix, we present in table \ref{tab:multiple_images} the information about the multiple images used in the strong lensing modeling.
The full redshift catalog, combining the MUSE and VIMOS measurements, is available in the eletronic version of the paper, as well as at the link \url{https://sites.google.com/site/vltclashpublic/data-release}.
In Figure \ref{fig:specs}, we show the MUSE spectra around relevant spectral features and image cutouts for multiple images.

\renewcommand{\thetable}{A.\arabic{table}}
\renewcommand{\thefigure}{A.1}

\longtab{
\setcounter{table}{0}
\begin{longtable}{ccccccc}
\caption{\label{tab:multiple_images} Information on spectroscopically identified multiple images in MACS~0416.}\\
\hline\hline
 ID & RA & DEC & $z_{\rm MUSE}$ & $z_{\rm previous}$ & ID$_{\rm ref}$ & $mag_{F814W}$\\
\hline
\endfirsthead
\caption{continued.}\\
\hline\hline
 ID & RA & DEC & $z_{\rm MUSE}$ & $z_{\rm previous}$ & ID$_{\rm ref}$ & $mag_{F814W}$\\
\hline
\endhead
\hline
\endfoot
1a  & 64.049084 & $-$24.062862  & 3.2355 & --- & 26.3 & $27.87 \pm 0.13$   \\
1b  & 64.046959 & $-$24.060797  & 3.2355 & --- & 26.2 & $27.43 \pm 0.09$ \\
1c  & 64.046449 & $-$24.060397  & 3.2355 & 2.185$^a$ & 26.1 & $26.53 \pm 0.06$ \\
\hline
2a  & 64.050865 & $-$24.066538  & 6.1452 & --- & 207.2 & $28.20 \pm 1.09^*$\\
2b  & 64.048179 & $-$24.062406  & 6.1452 & --- & ---   & $29.06 \pm 1.09^*$\\
2c  & 64.043572 & $-$24.059004  & 6.1452 & --- & 206.1 & $28.66 \pm 1.09^*$\\
\hline
3a  & 64.049232 & $-$24.068174 & 3.2885 & --- & 44.3 & $27.50 \pm 0.11$\\
3b  & 64.045269 & $-$24.062763 & 3.2885 & --- & 44.1 & $25.69 \pm 0.05$ \\
3c  & 64.041556 & $-$24.059997 & 3.2885 & --- & 44.2 & $26.13 \pm 0.06$ \\
\hline
4a  & 64.048126 & $-$24.066957  & --- & ---        & 27.1 & $25.73 \pm 0.03$\\
4b  & 64.047468 & $-$24.066039  & --- & 2.107$^{a}$ & 27.2 & $24.13 \pm 0.03$\\
4c  & 64.042209 & $-$24.060541  & --- & ---        & 27.3 & $25.81 \pm 0.04$\\
\hline
5a & 64.047463 & $-$24.068822  & 1.8950 & 1.893$^{d,e,f}$ & 2.3 & --- \\
5b & 64.043071 & $-$24.063080  & 1.8950 & 1.893$^{d,e,f}$ & 2.2 & $23.88 \pm 0.01^*$\\
5c & 64.041089 & $-$24.061806  & 1.8950 & 1.893$^{d,e,f}$ & 2.1 & $23.67 \pm 0.01^*$\\
\hline
6a & 64.047808 & $-$24.070164  & 3.6065 & --- & --- & $28.53 \pm 0.23$\\
6b & 64.043657 & $-$24.064401  & 3.6065 & --- & --- & $26.73 \pm 0.20$\\
6c & 64.037676 & $-$24.060756  & 3.6065 & --- & --- & $28.00 \pm 0.17$\\
\hline
7a & 64.047098 & $-$24.071105  & ---  & 2.085$^{d,e,f}$ & 7.3 & $28.05 \pm 0.14$\\
7b & 64.040664 & $-$24.063586  & 2.0881  & 2.085$^{d,e,f}$ & 7.2 & $25.29 \pm 0.03$\\
7c & 64.039795 & $-$24.063081  & 2.0881  & 2.085$^{d,e,f}$ & 7.1 & $25.43 \pm 0.04$\\
\hline
8a  & 64.044624 & $-$24.071488  & --- & 2.282$^{a}$   & 29.3  & $25.81 \pm 0.06$\\
8b  & 64.040485 & $-$24.066330  & --- & 2.267$^{b}$ & 29.2a & ---\\
8c  & 64.034256 & $-$24.062997  & --- & ---        & 29.1  & $25.76 \pm 0.05$\\
\hline
9a  & 64.045112 & $-$24.072341  & 3.2882 & --- & 32.1 & $26.11 \pm 0.05 ^*$ \\
9b  & 64.040079 & $-$24.066738  & 3.2882 & --- & 32.2 & $25.33 \pm 0.04 ^*$ \\
(9c)&(64.033157)&($-$24.062893) & --- & --- & --- & ---\\
\hline
10a & 64.044564 & $-$24.072092  & ---  & 2.094$^{a}$ & 23.1 & $24.92 \pm 0.02$ \\
10b & 64.039576 & $-$24.066623  & ---  & ---      & 23.2 & $25.20 \pm 0.03$ \\
10c & 64.034336 & $-$24.063734  & ---  & 2.091$^{a}$ & 23.3 & $25.23 \pm 0.03$ \\
\hline
(11a)&(64.046676)&($-$24.075312) & --- & --- & --- & ---\\
11b & 64.038515 & $-$24.065965  & 3.2922 & --- & 55.3 & --- \\
11c & 64.035223 & $-$24.064731  & 3.2922 & --- & 55.1 & $27.93 \pm 0.11$\\
\hline
(12a)&(64.041365)&($-$24.070852)& 0.9397 & ---      & 28.3 & --- \\
12b & 64.036843 & $-$24.067457  & 0.9397 & 0.937$^a$ & 28.2 & --- \\
12c & 64.036507 & $-$24.067028  & 0.9397 & 0.937$^a$ & 28.1 & --- \\
\hline
13a & 64.039245 & $-$24.070383  & 1.0054 & 1.005$^c$ & 11.1 & --- \\
13b & 64.038301 & $-$24.069728  & 1.0054 & 1.005$^c$ & 11.2 & --- \\
13c & 64.034234 & $-$24.066016  & 1.0054 & ---    & 11.3 & $26.79 \pm 0.12$\\
\hline
14a & 64.034483 & $-$24.066956  & 3.2215 & --- & D22.1 & ---\\
14b & 64.034190 & $-$24.066488  & 3.2215 & --- & D22.2 & ---\\
14c & 64.034001 & $-$24.066445  & 3.2215 & --- & D22.3 & ---\\
(14d)&(64.045886)&($-$24.076722)& --- & --- & --- & ---\\
(14e)&(64.035165)&($-$24.067968)& --- & --- & --- & ---\\
\hline
15a & 64.041802 & $-$24.075731  & 1.9894 & 1.989$^{d,e,f}$ & 3.3 & $23.99 \pm 0.01$\\
15b & 64.035249 & $-$24.070989  & 1.9894 & 1.989$^{d,e,f}$ & 3.2 & $23.94 \pm 0.02$\\
15c & 64.030769 & $-$24.067129  & 1.9894 & 1.989$^{d,e,f}$ & 3.1 & $24.17 \pm 0.01$\\
\hline
16a & 64.033523 & $-$24.069448  & 2.0948 & ---       & 5.3 & $24.11 \pm 0.02$\\
16b & 64.032656 & $-$24.068663  & 2.0948 & 2.092$^b$ & 5.2 & $24.75 \pm 0.02$\\
16c & 64.032410 & $-$24.068414  & 2.0948 & ---       & 5.1 & --- \\
(16e)&(64.043470)&($-$24.076860)& --- & --- & --- & ---\\
\hline
17a & 64.040489 & $-$24.078380  & 3.9663 & --- & --- & $27.97 \pm 0.27$ \\
17b & 64.035107 & $-$24.073864  & ---   & --- & --- & $25.29 \pm 0.17^*$ \\ 
17c & 64.027171 & $-$24.068224  & ---   & --- & --- & $27.47 \pm 0.17^*$ \\
\hline
18a & 64.040177 & $-$24.079872  & 3.8710 & --- & 49.2 & $27.12 \pm 0.09$\\
18b & 64.033937 & $-$24.074565  & 3.8710 & --- & 49.1 & $27.08 \pm 0.15$\\
(18c)&(64.026991)&($-$24.069620)& --- & --- & --- & ---\\
\hline
19a & 64.040140 & $-$24.080305  & 4.1032 & --- & 51.1 & $26.70 \pm 0.08^*$\\
19b & 64.033667 & $-$24.074762  & 4.1032 & --- & 51.2 & $25.38 \pm 0.11^*$\\
19c & 64.026633 & $-$24.070476  & 4.1032 & --- & 51.3 & --- \\
\hline
20a & 64.040351 & $-$24.081482  &  ---  & 3.223$^{d,e,f}$ & 13.3 & $25.75 \pm 0.03$\\
20b & 64.032157 & $-$24.075108  & 3.2175 & 3.223$^{d,e,f}$ & 13.2 & --- \\
20c & 64.027572 & $-$24.072673  & 3.2175 & 3.223$^{d,e,f}$ & 13.1 & $24.64 \pm 0.02$\\
\hline
(21a)&(64.042028)&($-$24.081835)& --- & --- & --- & ---\\
\;\;21b$^\dagger$ & 64.030906 & $-$24.074341  & 5.1060 & --- & 34.2 & --- \\ 
\;\;21c$^\dagger$ & 64.029176 & $-$24.073382  & 5.1060 & --- & 34.1 & --- \\ 
\hline
(22a)&(64.040114)&($-$24.082217)& --- & --- & --- & ---\\
22b & 64.030997 & $-$24.077173  & 3.9230 & --- & 60.3 & ---\\
22c & 64.027127 & $-$24.073572  & 3.9230 & --- & ---  & $29.54 \pm 0.43$\\
\hline
23a & 64.035668 & $-$24.079920  & 2.5425 & 2.545$^{b}$ & 45.1 & $26.27 \pm 0.05$ \\
23b & 64.032638 & $-$24.078508  & 2.5425 & ---         & ---  & --- \\
(23c)&(64.024668)&($-$24.071076)& --- & --- & --- & ---\\
\hline
24a & 64.035833 & $-$24.081321  & 1.6333 & 1.637$^{d}$ & 14.3 & $23.19 \pm 0.01$\\
24b & 64.031039 & $-$24.078953  & 1.6333  & 1.637$^{d}$ & 14.2 & $23.34 \pm 0.01$\\
24c & 64.026239 & $-$24.074337  & 1.6333  & 1.637$^{d}$ & 14.1 & $23.41 \pm 0.01$\\
\hline
25a & 64.038073 & $-$24.082404  & 3.1103 & --- & 67.1 & $27.84 \pm 0.13^*$ \\
25b & 64.030366 & $-$24.079015  & 3.1103 & --- & 67.3 & --- \\
25c & 64.025446 & $-$24.073648  & 3.1103 & --- & 67.2 & $28.06 \pm 0.13$ \\
\hline
26a & 64.037722 & $-$24.082388  & 3.0773 & --- & 58.2 & $27.58 \pm 0.09$\\
26b & 64.030484 & $-$24.079222  & 3.0773 & --- & 58.3 & $26.49 \pm 0.08$\\
26c & 64.025186 & $-$24.073575  & 3.0773 & --- & 58.1 & $27.46 \pm 0.08$\\
\hline
27a & 64.037469 & $-$24.083657  & 3.4909 & --- & 35.1 & --- \\
27b & 64.029409 & $-$24.079889  & 3.4909 & --- & 35.2 & $26.37 \pm 0.05$\\
27c & 64.024946 & $-$24.075021  & 3.4909 & --- & 35.3 & $26.43 \pm 0.04$ \\
\hline
28a & 64.038350 & $-$24.084126  & ---    & --- & ---  & $28.76 \pm 0.26$ \\
28b & 64.028322 & $-$24.079004  & 3.2526 & --- & 47.2 & ---\\
28c & 64.026330 & $-$24.076705  & 3.2526 & --- & 47.1 & $25.93 \pm 0.05$\\
\hline
29a & 64.036702 & $-$24.083855  & --- & 2.298$^{d,e,f}$ & 10.3 & $25.60 \pm 0.03$\\
29b & 64.028504 & $-$24.079755  & --- & 2.298$^{d,e,f}$ & 10.2 & $24.99 \pm 0.02$\\
29c & 64.025993 & $-$24.077080  & --- & 2.298$^{d,e,f}$ & 10.1 & $24.57 \pm 0.02$\\
\hline
30a & 64.033628 & $-$24.083185  & 3.4406 & --- & 38.1 & $27.20 \pm 0.08$ \\
30b & 64.031251 & $-$24.081904  & 3.4406 & --- & 38.2 & $26.98 \pm 0.10 ^*$\\
30c & 64.022699 & $-$24.074595  & 3.4406 & --- & 38.3 & $27.94 \pm 0.14 ^*$\\
\hline
31a & 64.035486 & $-$24.084679  & 4.1218 & --- & 48.1 & $26.21 \pm 0.08$\\
31b & 64.029234 & $-$24.081813  & 4.1218 & --- & 48.2 & $24.76 \pm 0.06^*$\\
31c & 64.023412 & $-$24.076125  & 4.1218 & --- & 48.3 & $25.39 \pm 0.06$\\
\hline
32a & 64.035054 & $-$24.085504  & 5.3650 & --- & 33.2 & $27.95 \pm 1.09$ \\
32b & 64.028403 & $-$24.082993  & 5.3650 & --- & 33.1 & $26.60 \pm 0.70 ^*$\\
32c & 64.022988 & $-$24.077265  & 5.3650 & --- & 33.2 & $26.03 \pm 0.20 ^*$ \\
\hline
33a & 64.032017 & $-$24.084230  & 5.9729 & --- & --- & $27.03 \pm 0.82 ^*$\\
33b & 64.030821 & $-$24.083697  & 5.9729 & --- & --- & --- \\
(33c)&(64.021697)&($-$24.075230)& --- & --- & --- & ---\\
\hline
(34a)&(64.034067)&($-$24.085284)& --- & --- & --- & ---\\
34b & 64.027632 & $-$24.082609  & 3.9228 & --- & --- & $27.57 \pm 0.28$ \\
34c & 64.023731 & $-$24.078477  & 3.9228 & --- & --- & $28.19 \pm 0.45$ \\ 
\hline
35a$^\dagger$ & 64.033681 & $-$24.085855  & 5.6390 & --- & --- & --- \\ 
35b$^\dagger$ & 64.028654 & $-$24.084240  & 5.6390 & --- & --- & --- \\ 
35c$^\dagger$ & 64.022187 & $-$24.077559  & 5.6390 & --- & --- & --- \\ 
\hline
36a & 64.031614 & $-$24.085762  &   ---  & 1.964$^{d,e}$ & 16.3 & $24.35 \pm 0.02$\\
36b & 64.028339 & $-$24.084553  &   ---  & 1.964$^{d,e}$ & 16.2 & $23.48 \pm 0.02$\\
36c & 64.024074 & $-$24.080895  & 1.9614 & 1.964$^{d,e}$ & 16.1 & --- \\
\hline
37a & 64.029809 & $-$24.086363  &  ---  & 2.218$^{d,e,f}$ & 17.1 & $24.32 \pm 0.01$\\
37b & 64.028610 & $-$24.085973  &  ---  & 2.218$^{d,e,f}$ & 17.2 & --- \\
37c & 64.023345 & $-$24.081580  & 2.2182 & 2.218$^{d,e,f}$ & 17.3 & --- \\
\hline\hline

\end{longtable}
\tablefoot{IDs$_{\rm ref}$ corresponds to the ID column in Table 2 of \citet{2016arXiv160300505H}. The F814W magnitudes are based on the ASTRODEEP catalog \citep{2016A&A...590A..31C}. Model-predicted positions are indicated in brackets. MUSE redshifts are generally based on the Ly-$\alpha$ line when present, with exception of IDs 1, 3, 9, 20, 26, and 27 for which CIV and CIII] lines were used to better estimate the systemic redshift. This table includes 56 new redshifts belonging to 22 multiply lensed sources. \\
\tablefoottext{$\dagger$}{ Positions measured in MUSE narrow band images }
\tablefoottext{a}{GLASS redshifts \citet{2016arXiv160300505H}}
\tablefoottext{b}{\citet{2015arXiv151102522B}}
\tablefoottext{c}{Rodney et al., in prep.}
\tablefoottext{d}{ \citet{2015ApJ...800...38G}}
\tablefoottext{e}{ \citet{2014MNRAS.443.1549J} }
\tablefoottext{f}{ \citet{2014MNRAS.444..268R} }
\tablefoottext{g}{ \citet{2013ApJ...762L..30Z} }
\tablefoottext{*}{Objects whose magnitude might be affected by source confusion.}
}

}

\begin{figure*}
   \includegraphics[width = 0.65\columnwidth]{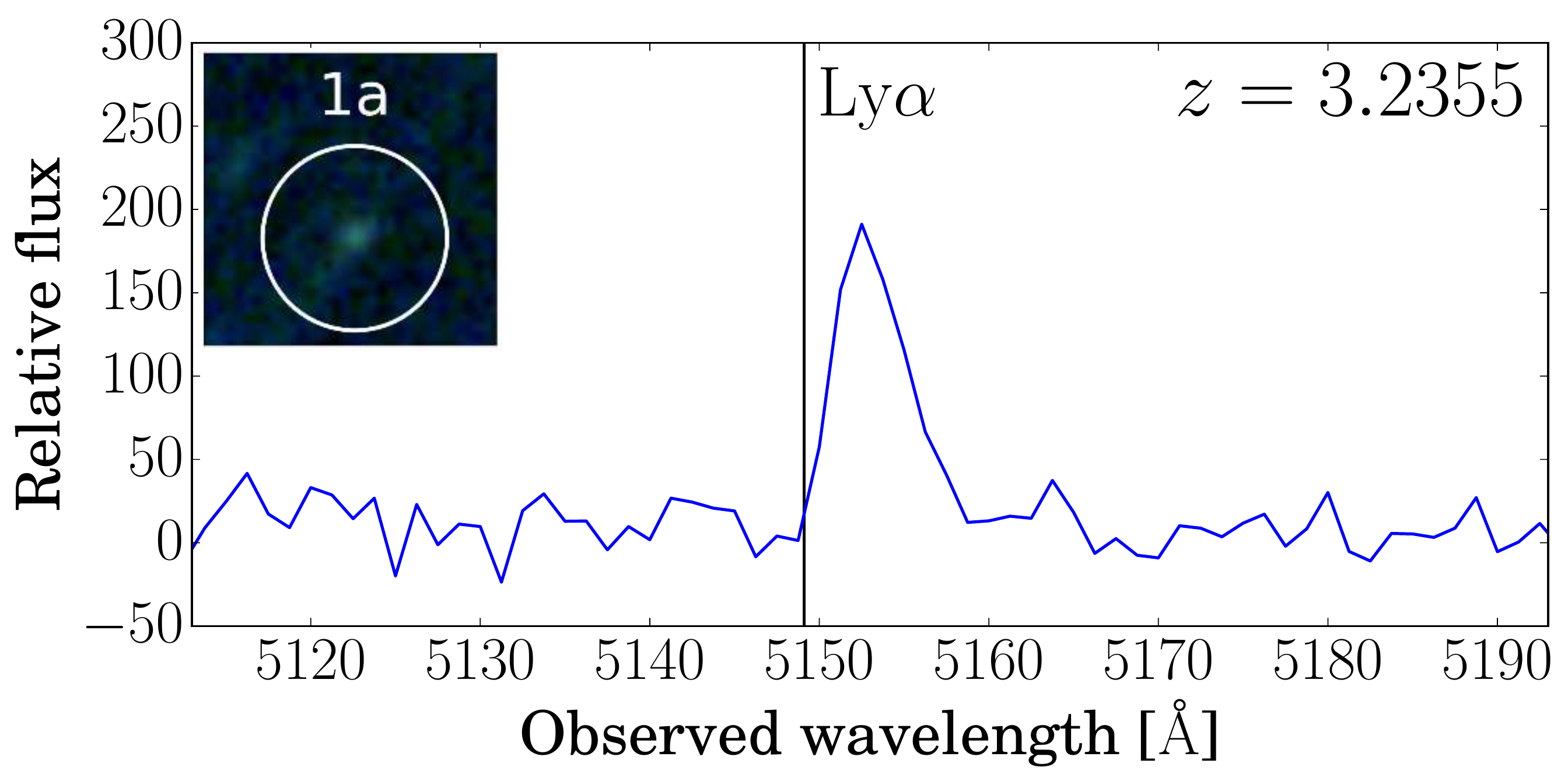}
   \includegraphics[width = 0.65\columnwidth]{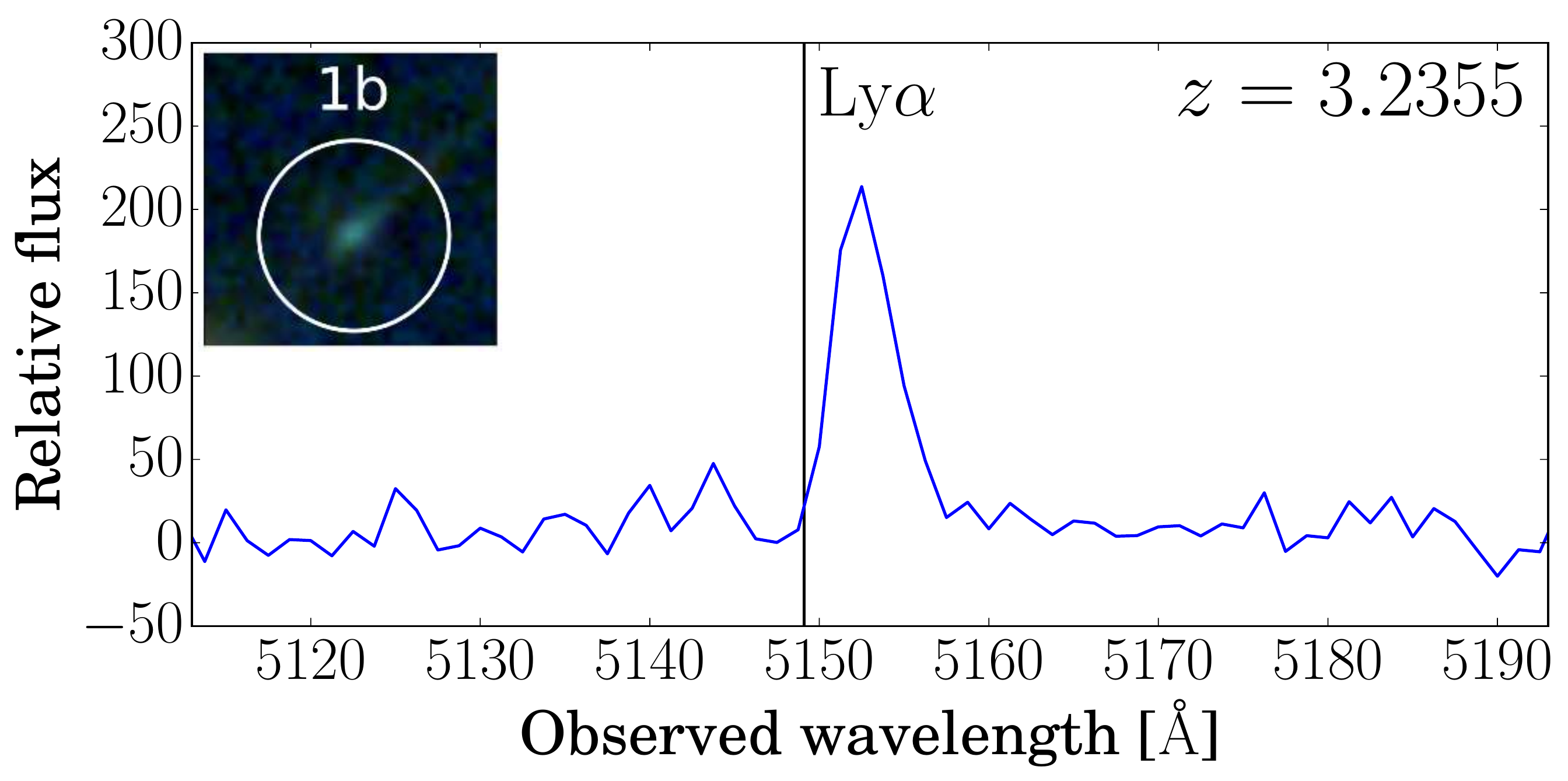}
   \includegraphics[width = 0.65\columnwidth]{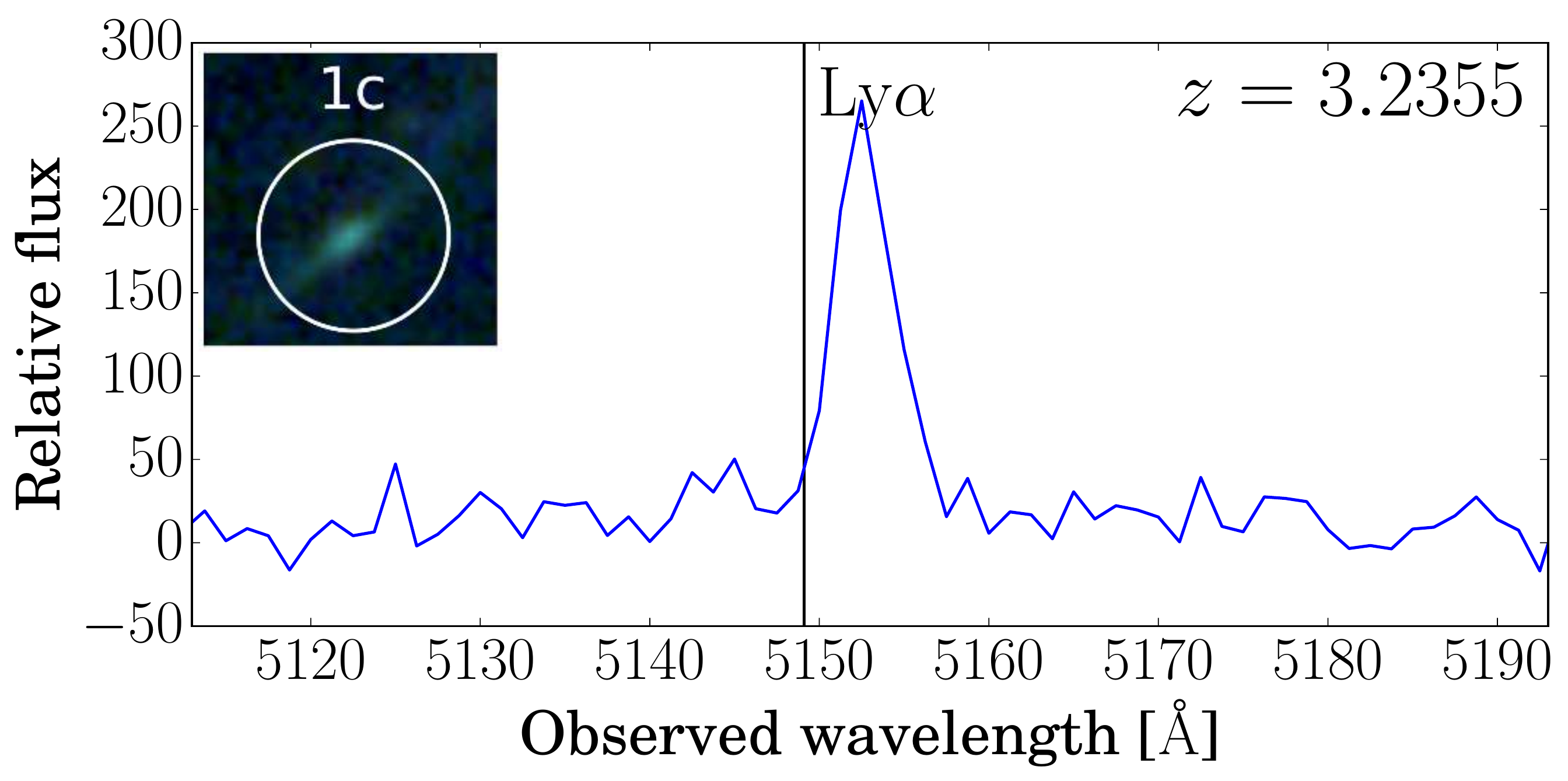}

   \includegraphics[width = 0.65\columnwidth]{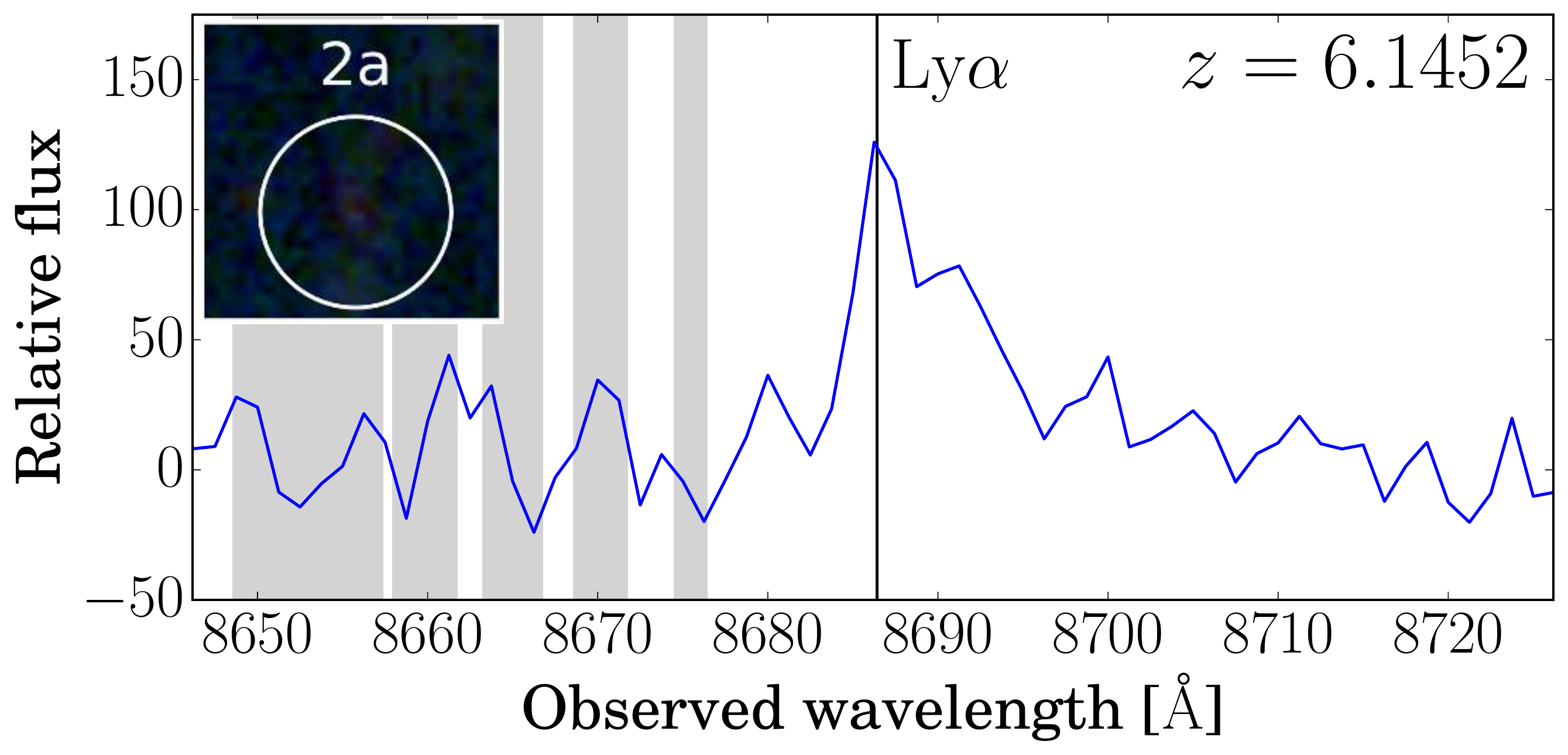}
   \includegraphics[width = 0.65\columnwidth]{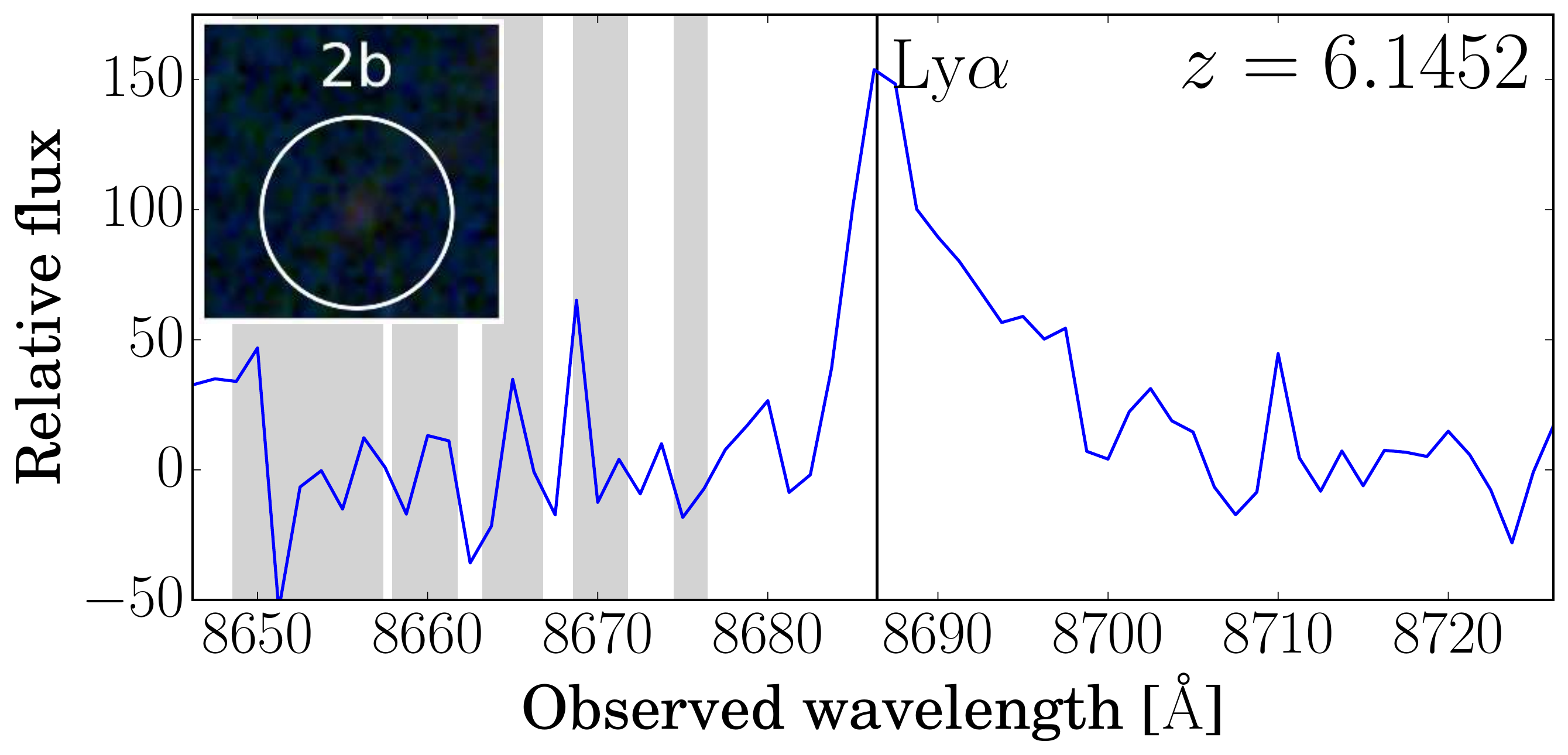}
   \includegraphics[width = 0.65\columnwidth]{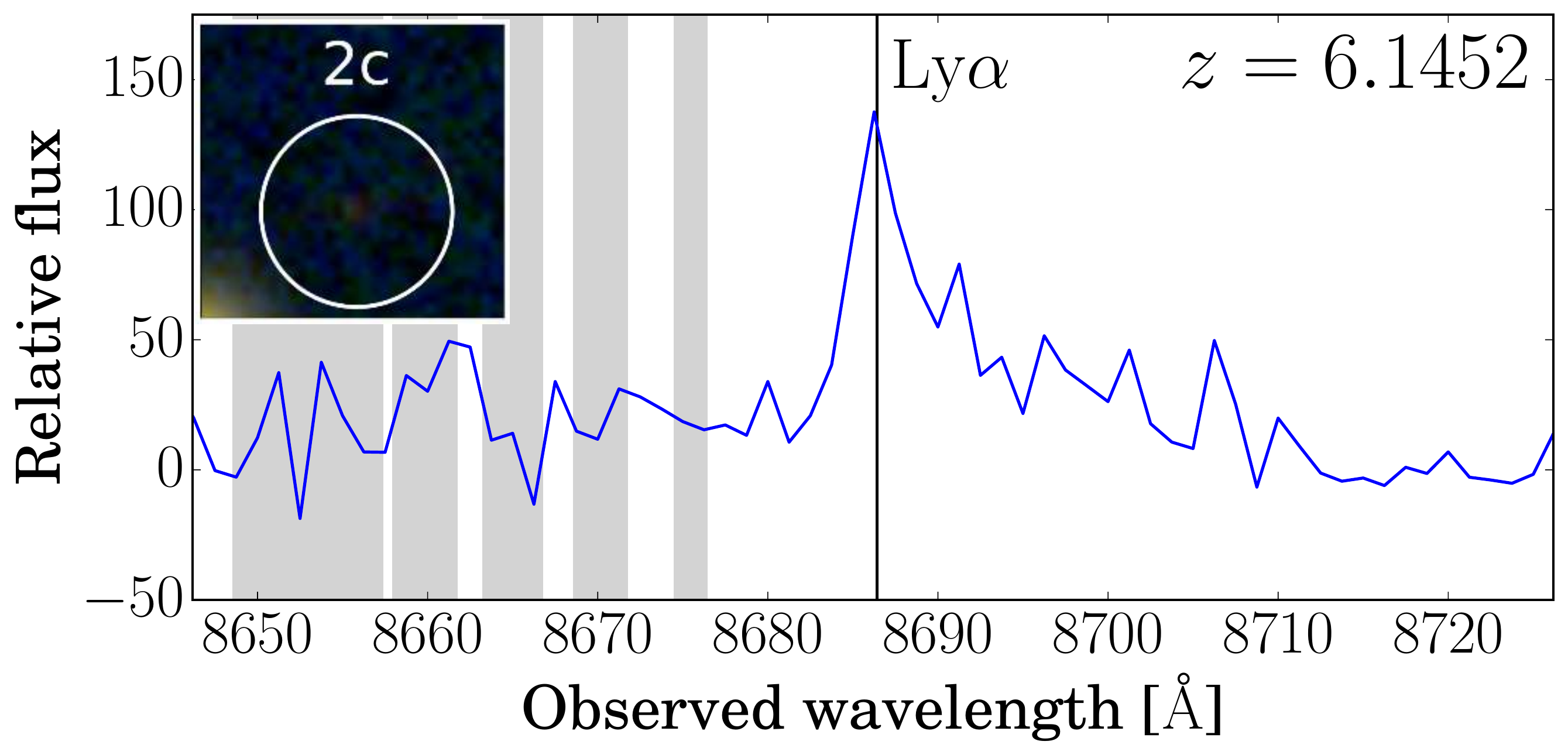}

   \includegraphics[width = 0.65\columnwidth]{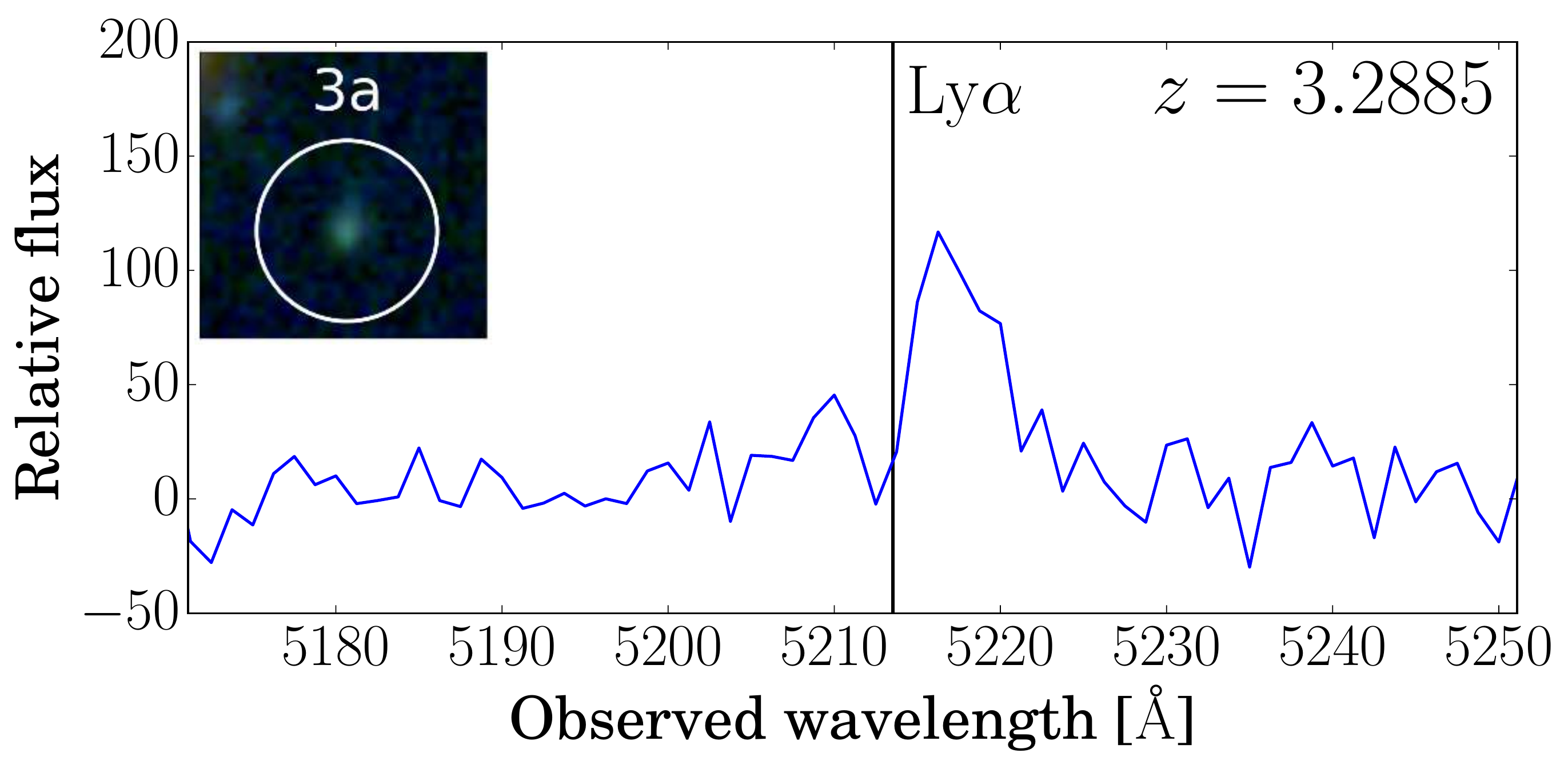}
   \includegraphics[width = 0.65\columnwidth]{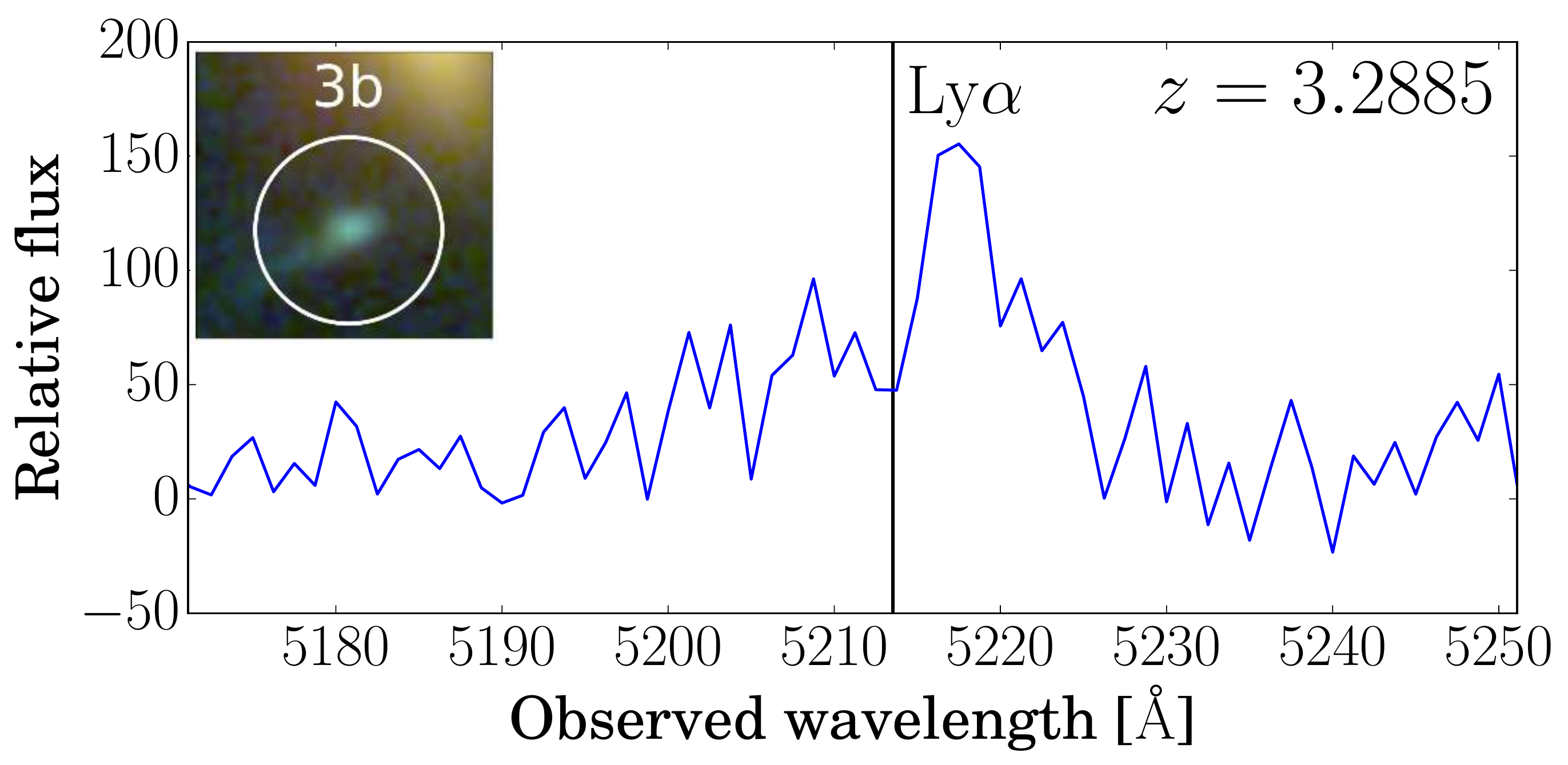}
   \includegraphics[width = 0.65\columnwidth]{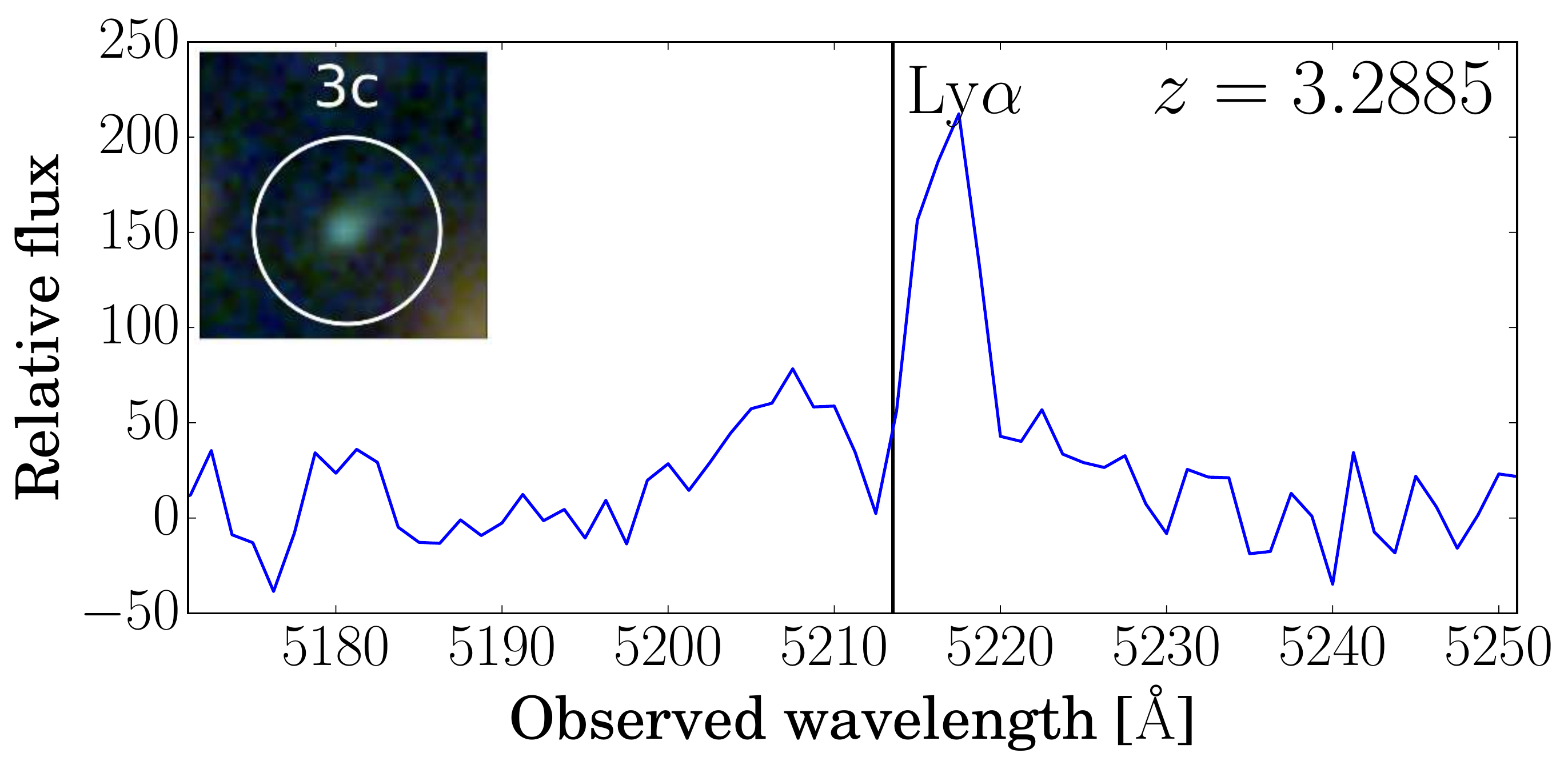}

   \includegraphics[width = 0.65\columnwidth]{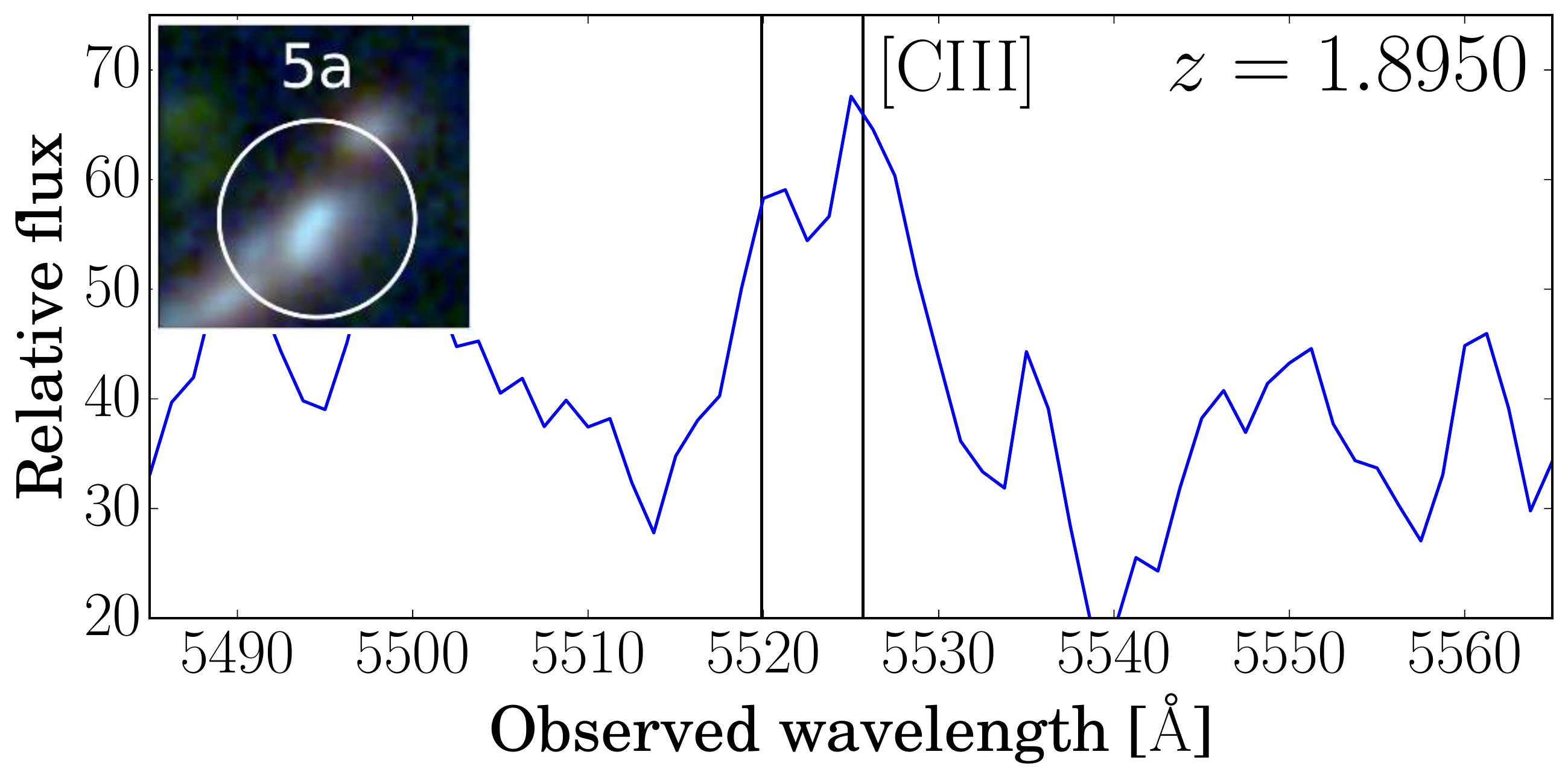}
   \includegraphics[width = 0.65\columnwidth]{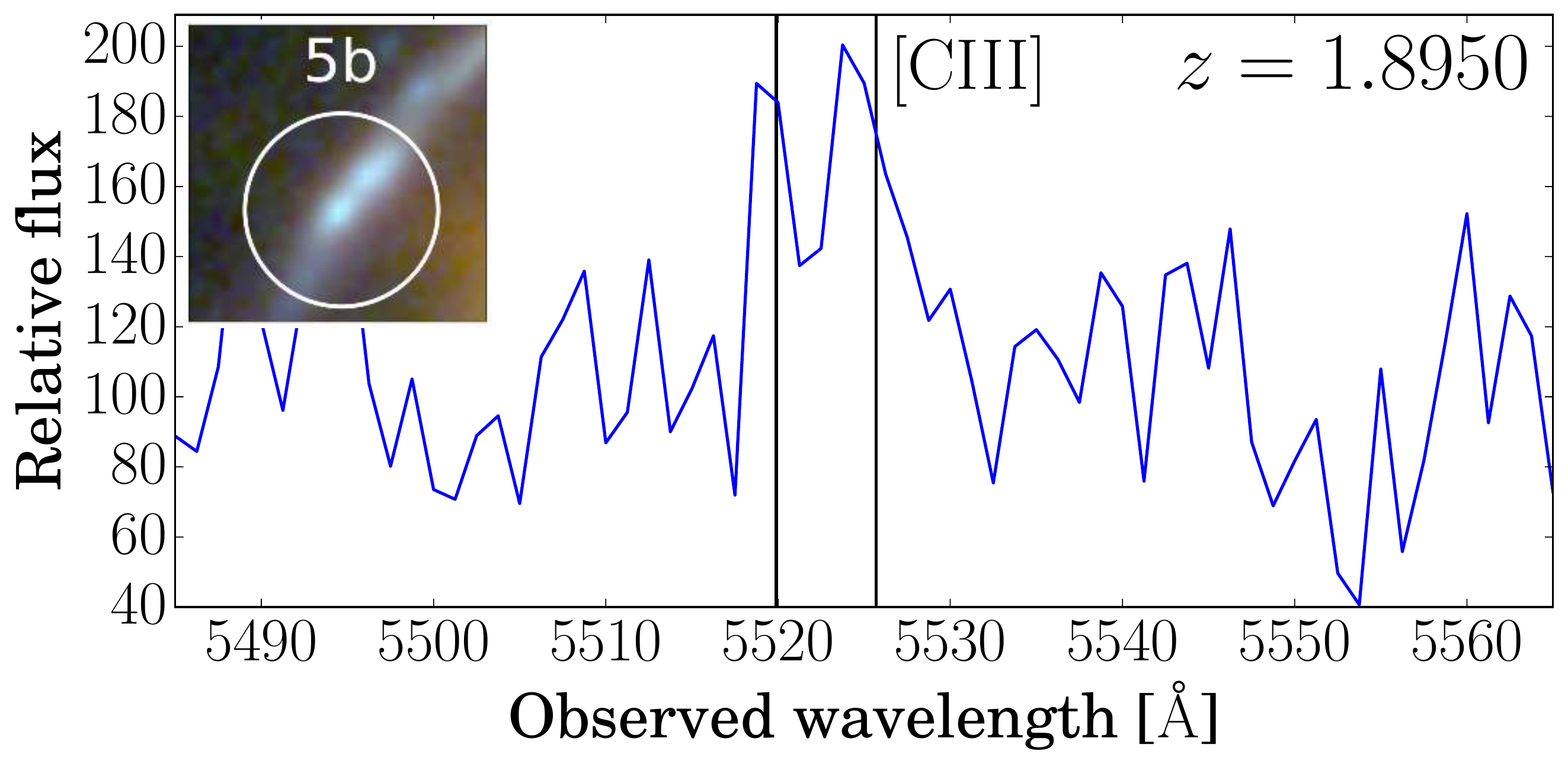}
   \includegraphics[width = 0.65\columnwidth]{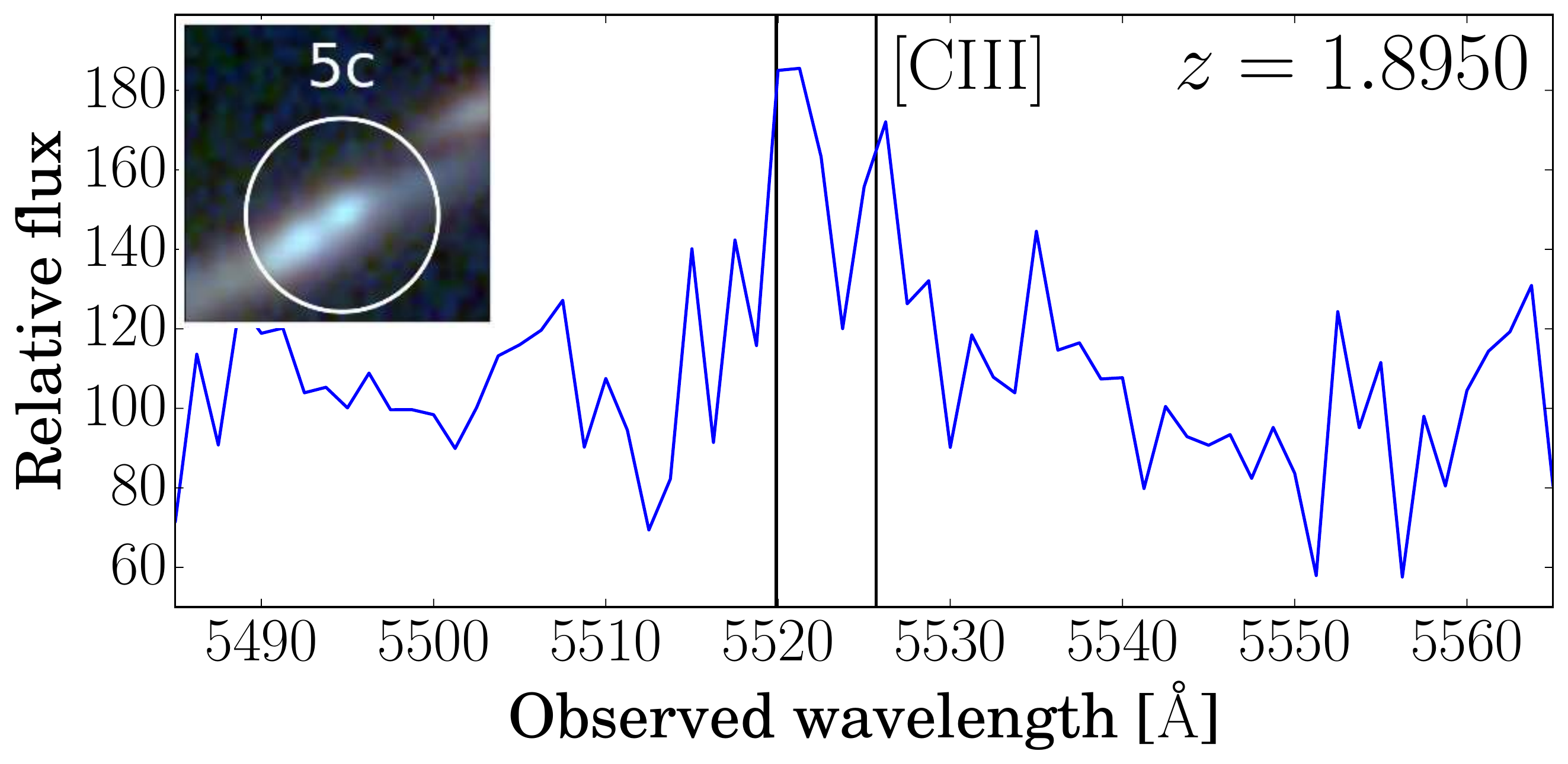}

   \includegraphics[width = 0.65\columnwidth]{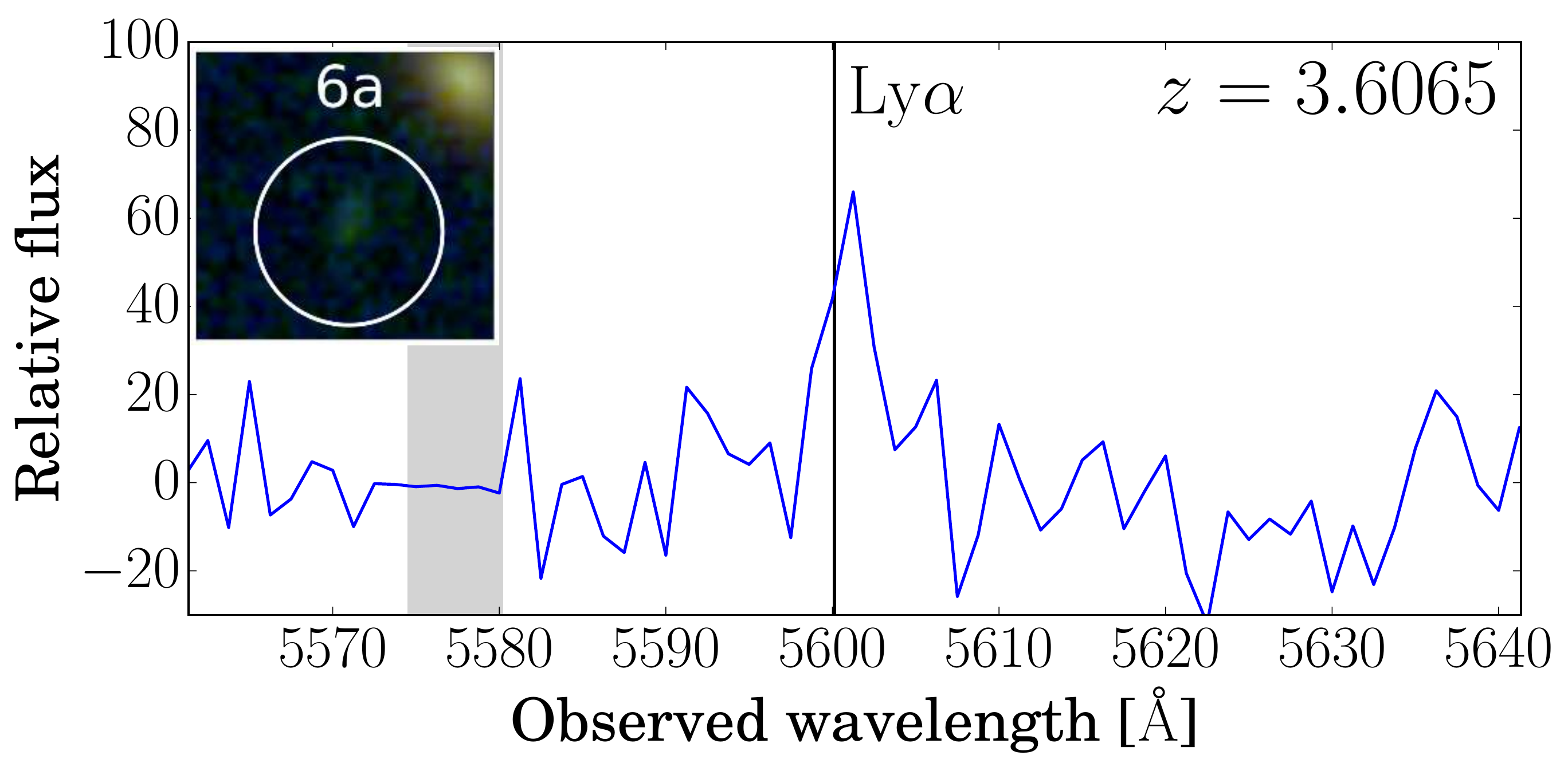}
   \includegraphics[width = 0.65\columnwidth]{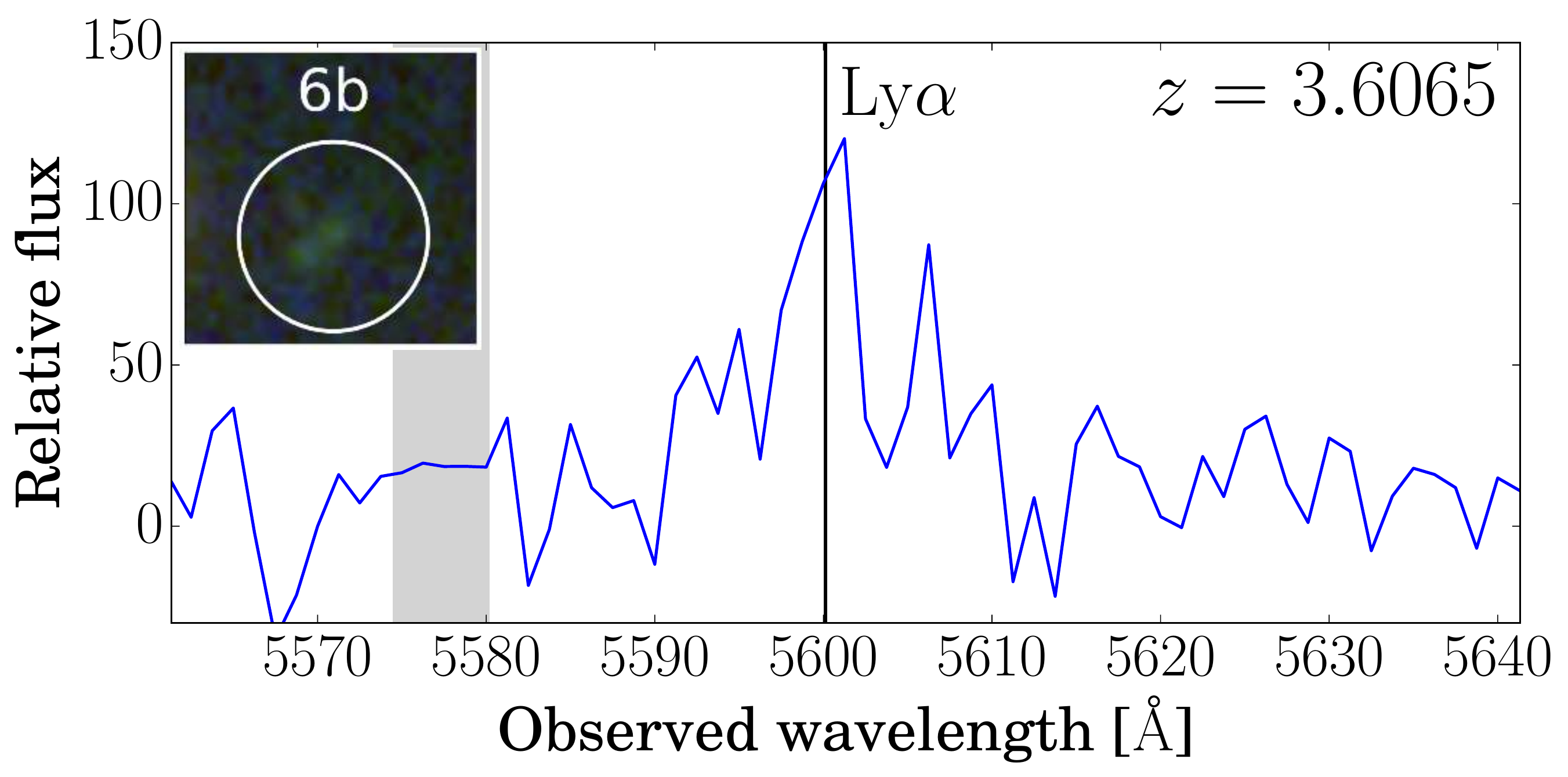}
   \includegraphics[width = 0.65\columnwidth]{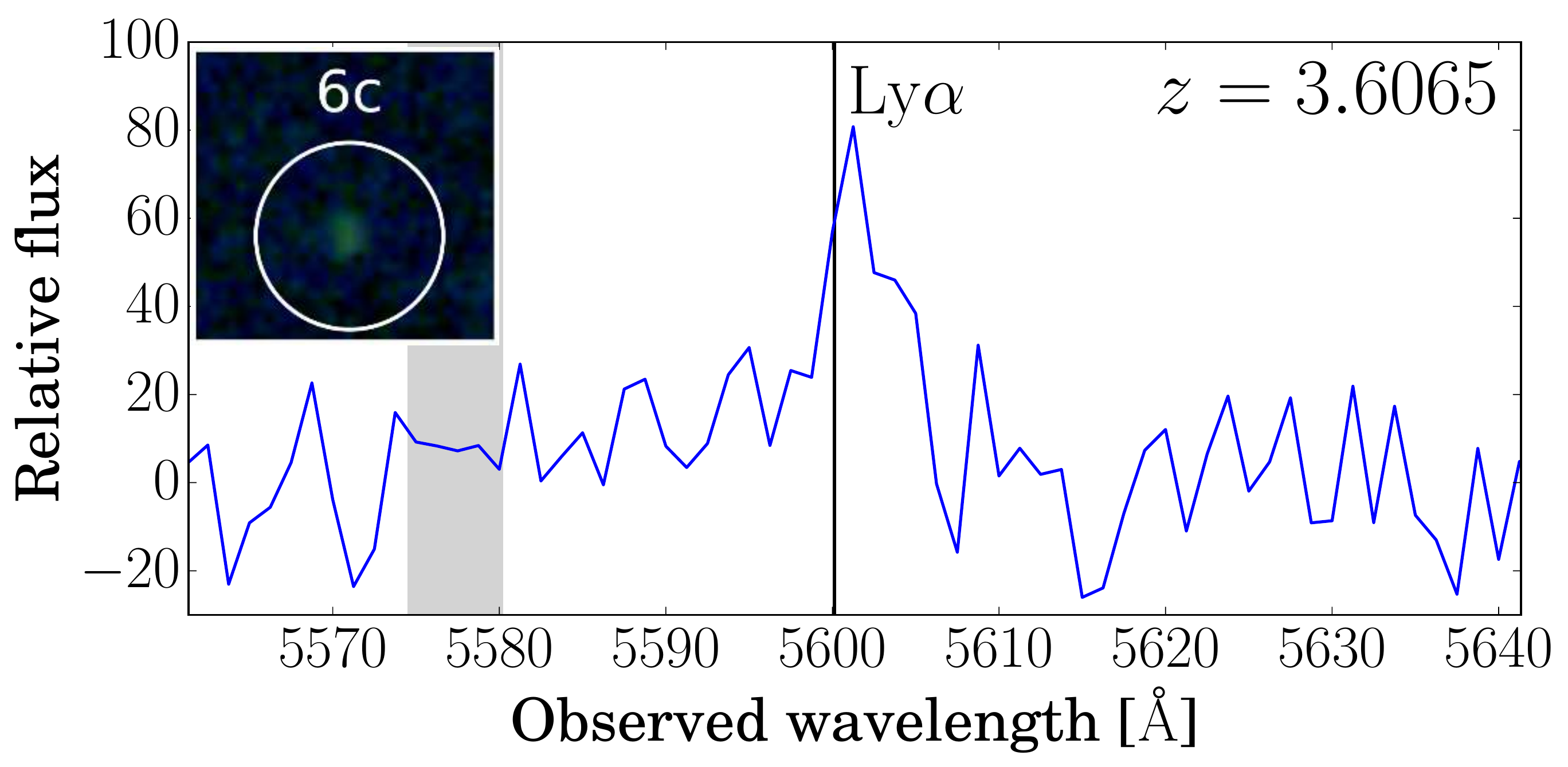}

   \includegraphics[width = 0.65\columnwidth]{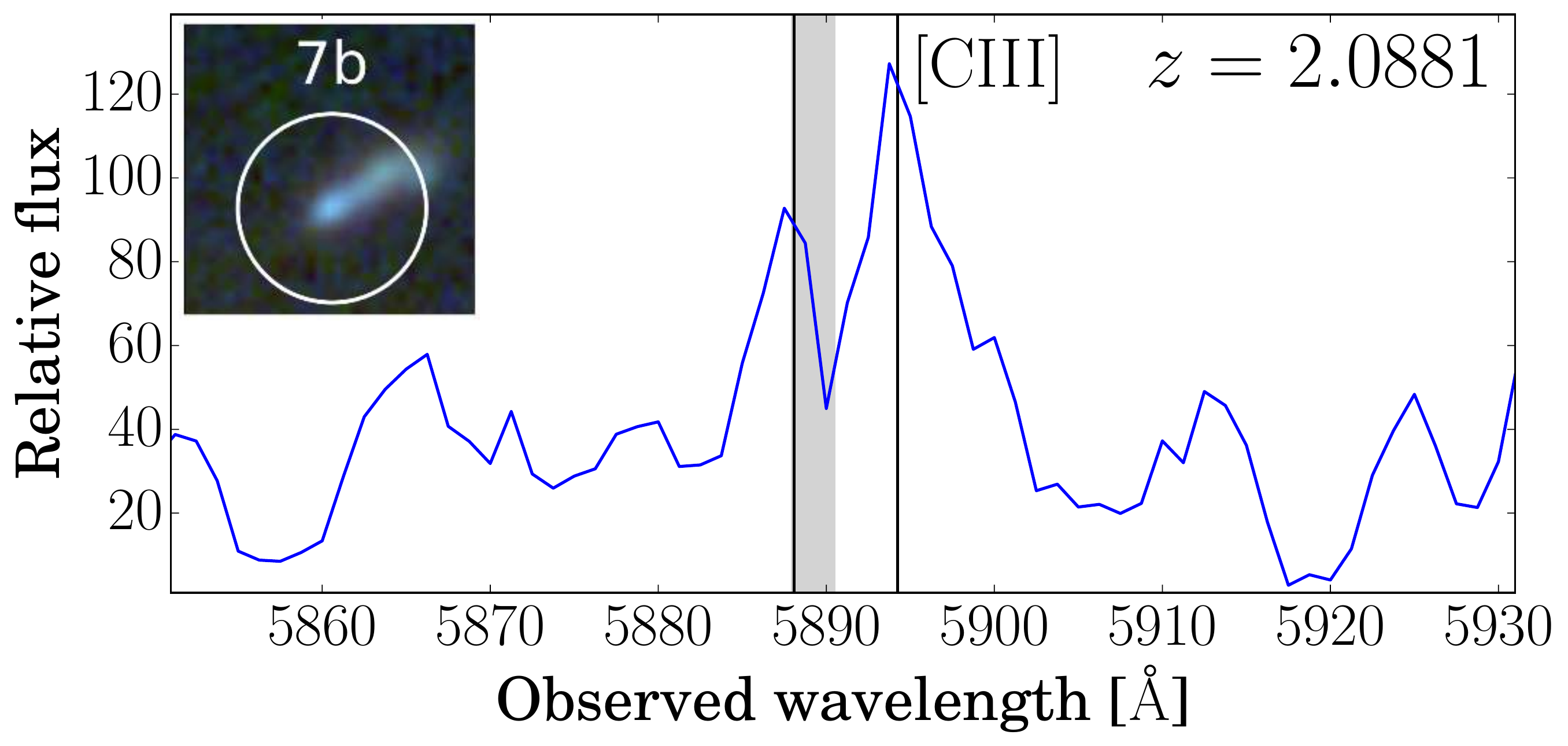}
   \includegraphics[width = 0.65\columnwidth]{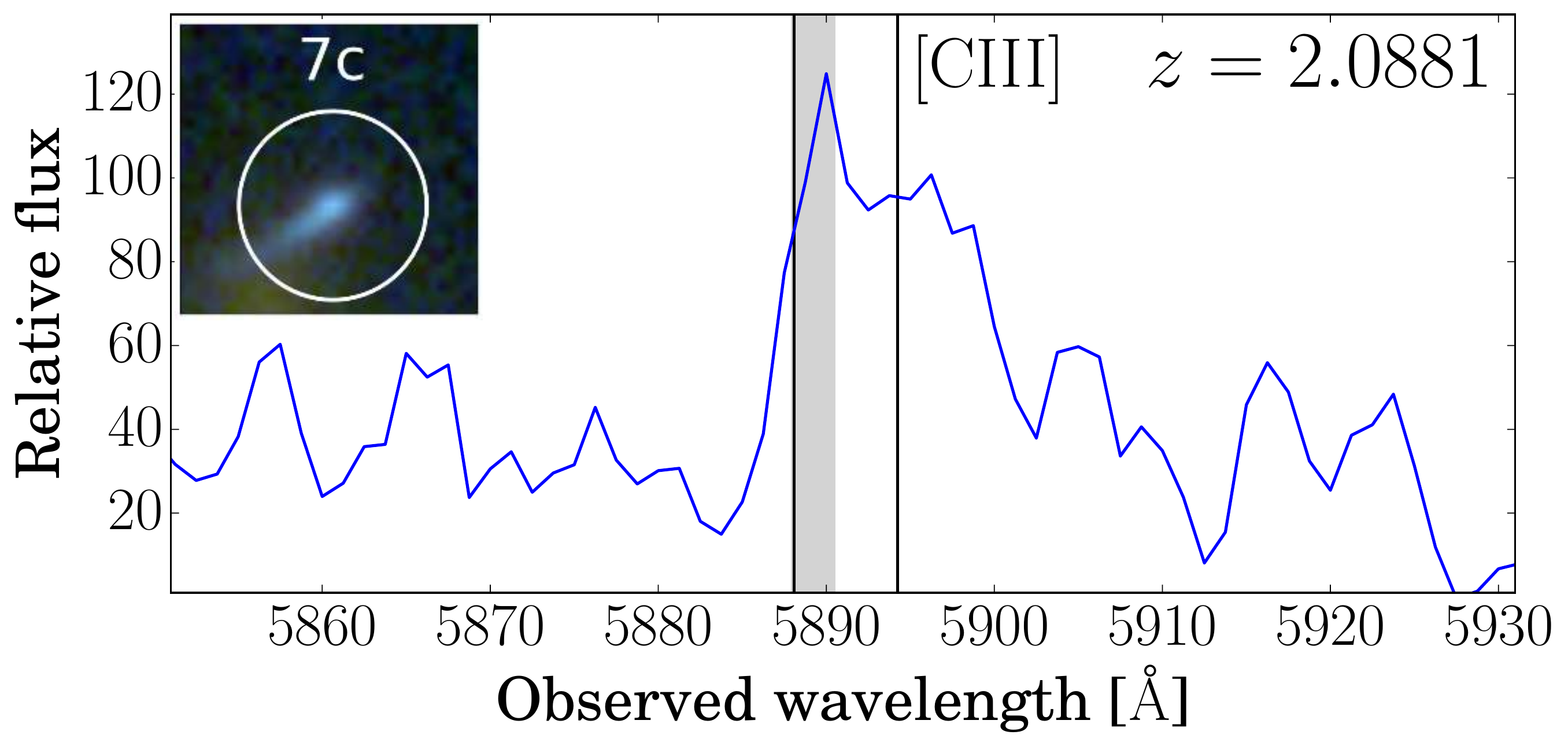}

   \includegraphics[width = 0.65\columnwidth]{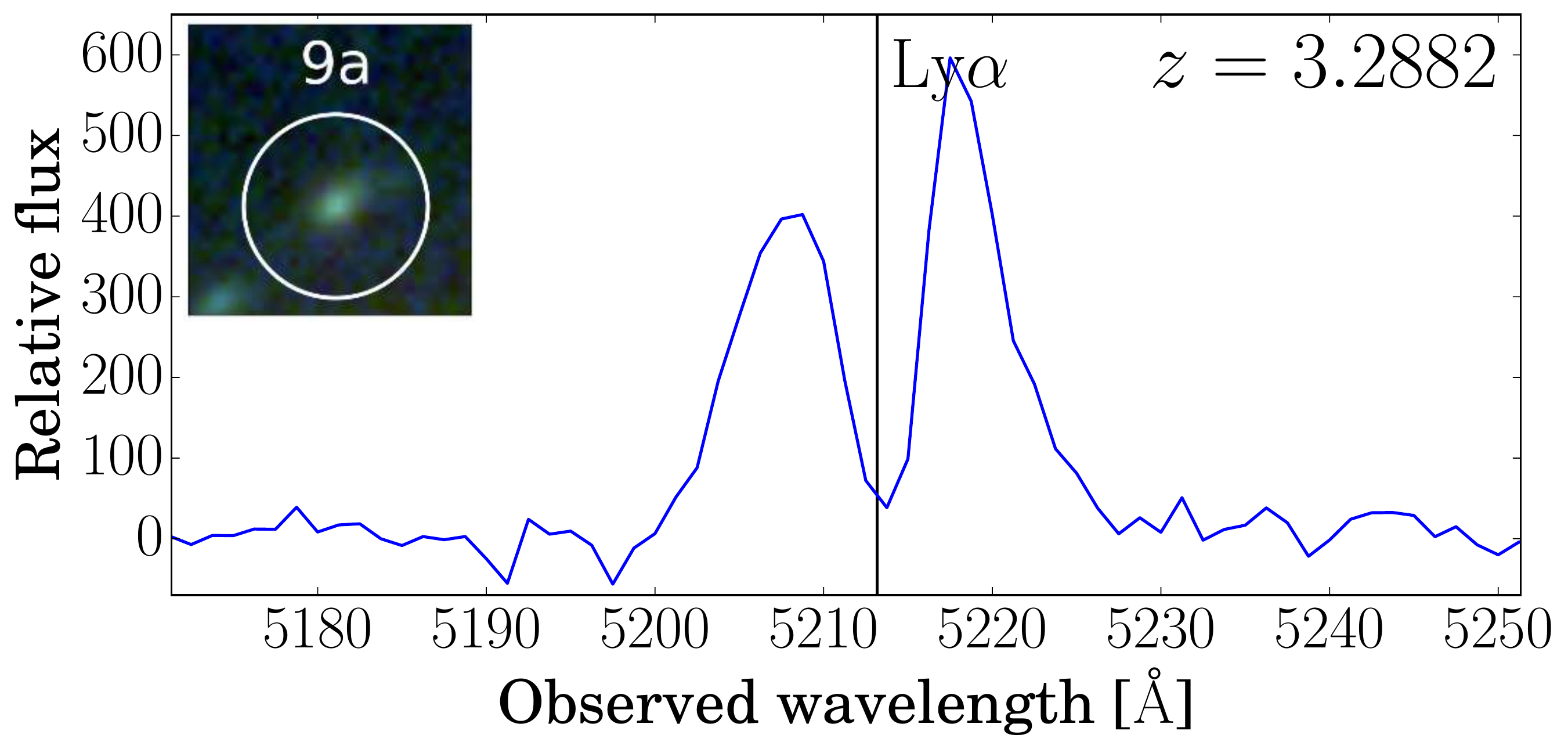}
   \includegraphics[width = 0.65\columnwidth]{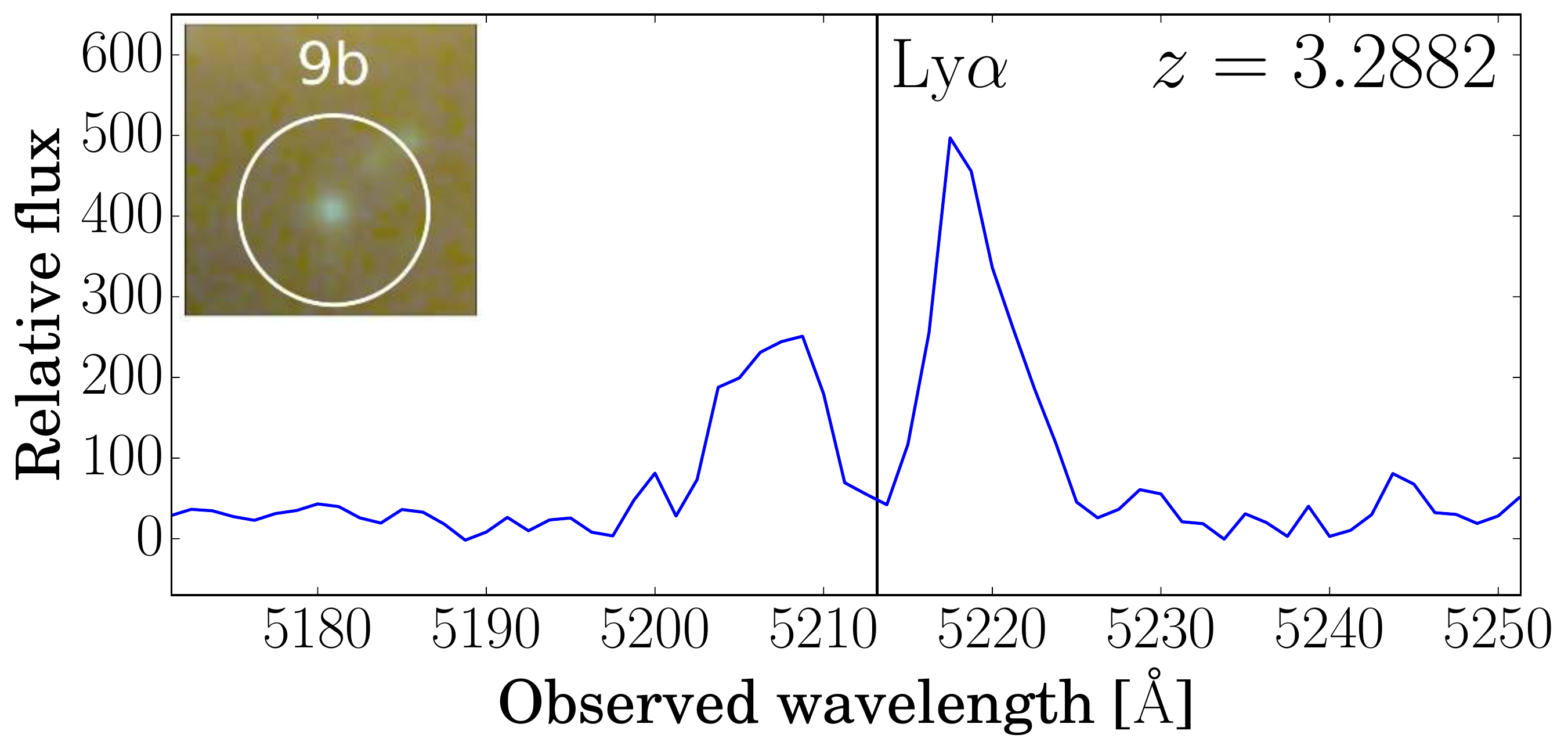}

   \caption{MUSE spectra of multiple images of lensed sources in
     MACS~0416. Panels on the same line show the spectra of multiple
     images belonging to the same family, i.e., associated to the same
     source. The vertical black lines indicate the position of the emission line based on the best estimate of the systemic redshift (see Table \ref{tab:multiple_images} notes).
    Spectral regions with high
     sky contamination are marked in grey; the flux is given in units
     of $\rm 10^{-20} erg\, s^{-1} cm^{-2} \AA^{-1}$. The image cutouts in each
     panel ($2\arcsec$ across) are extracted from the HFF color
     image and show the HST counterparts, or are centered at the
     position of the MUSE emission in the cases of no apparent
     counterparts (see families 21 and 35). }
  \label{fig:specs}
\end{figure*}

\begin{figure*}
  \ContinuedFloat
  %\centering

   \includegraphics[width = 0.65\columnwidth]{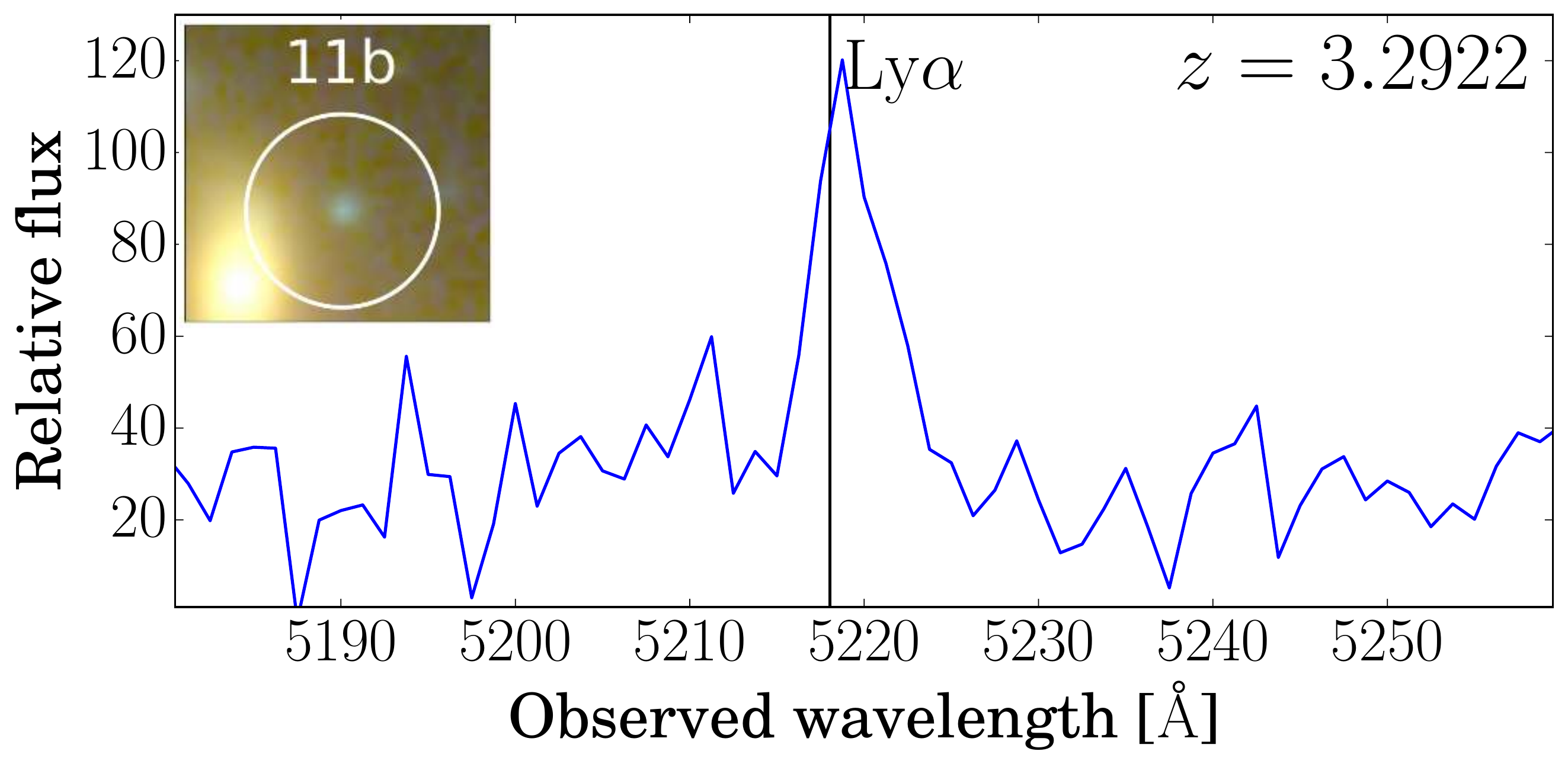}
   \includegraphics[width = 0.65\columnwidth]{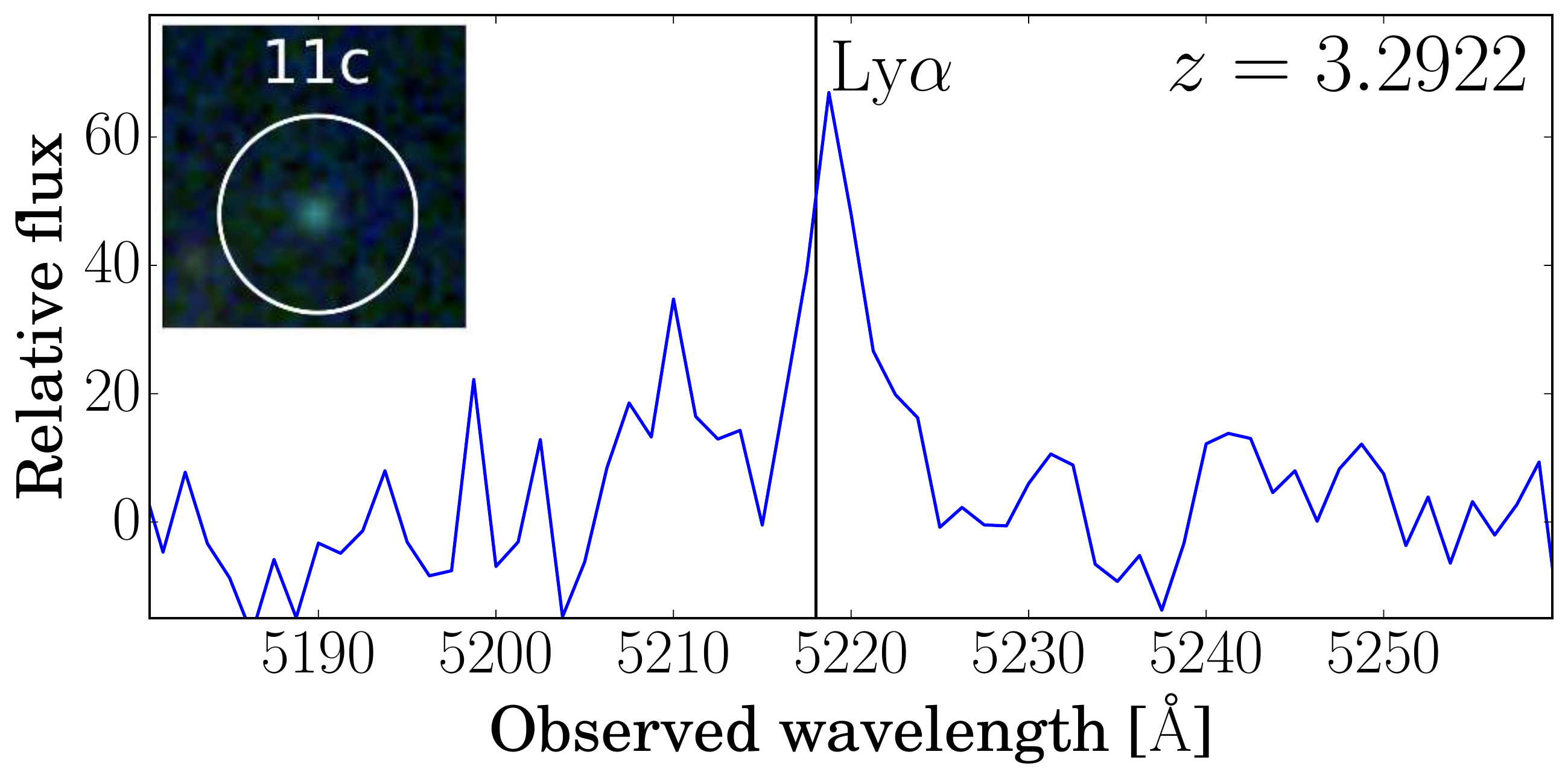}

   \includegraphics[width = 0.65\columnwidth]{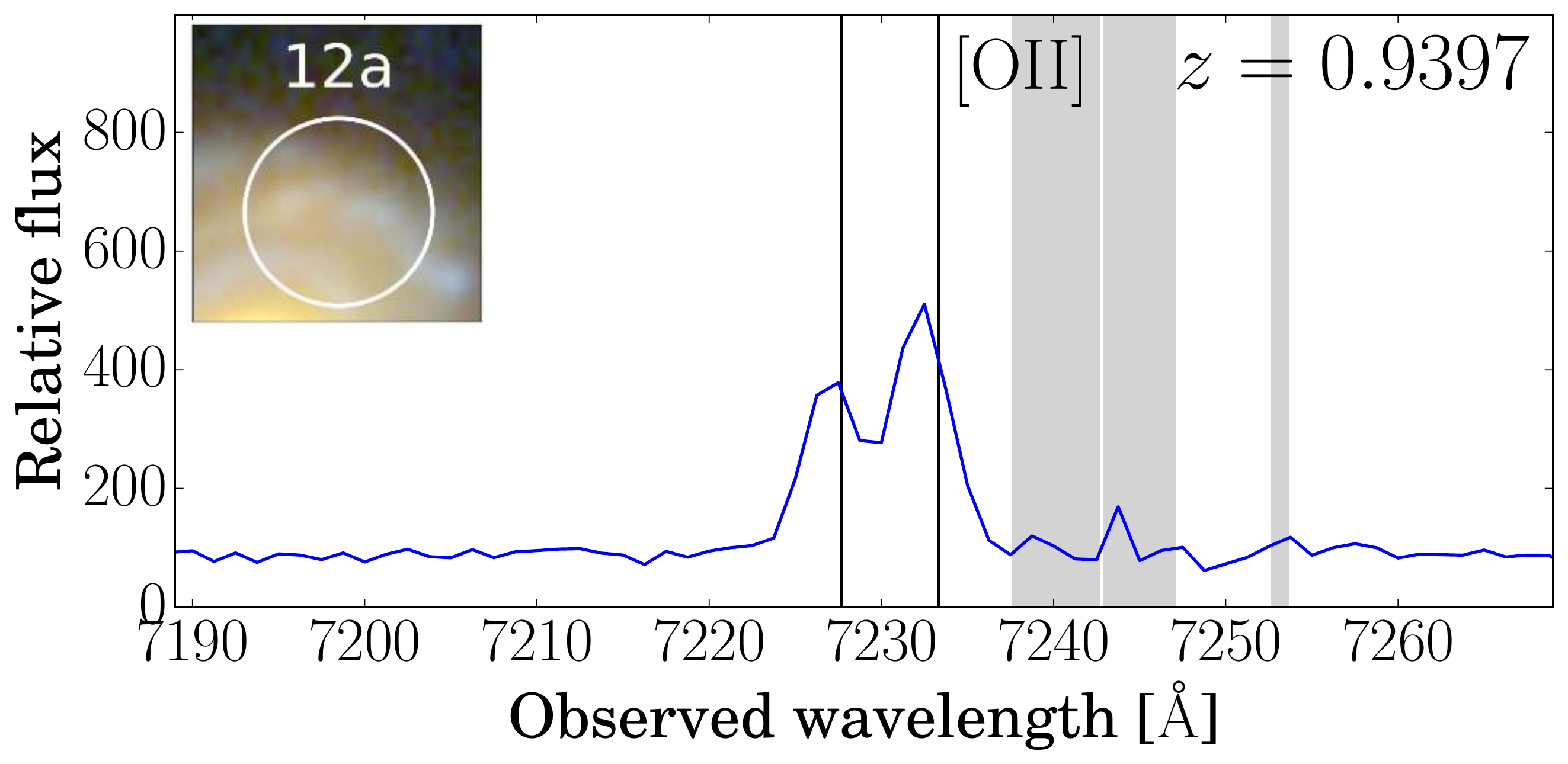}
   \includegraphics[width = 0.65\columnwidth]{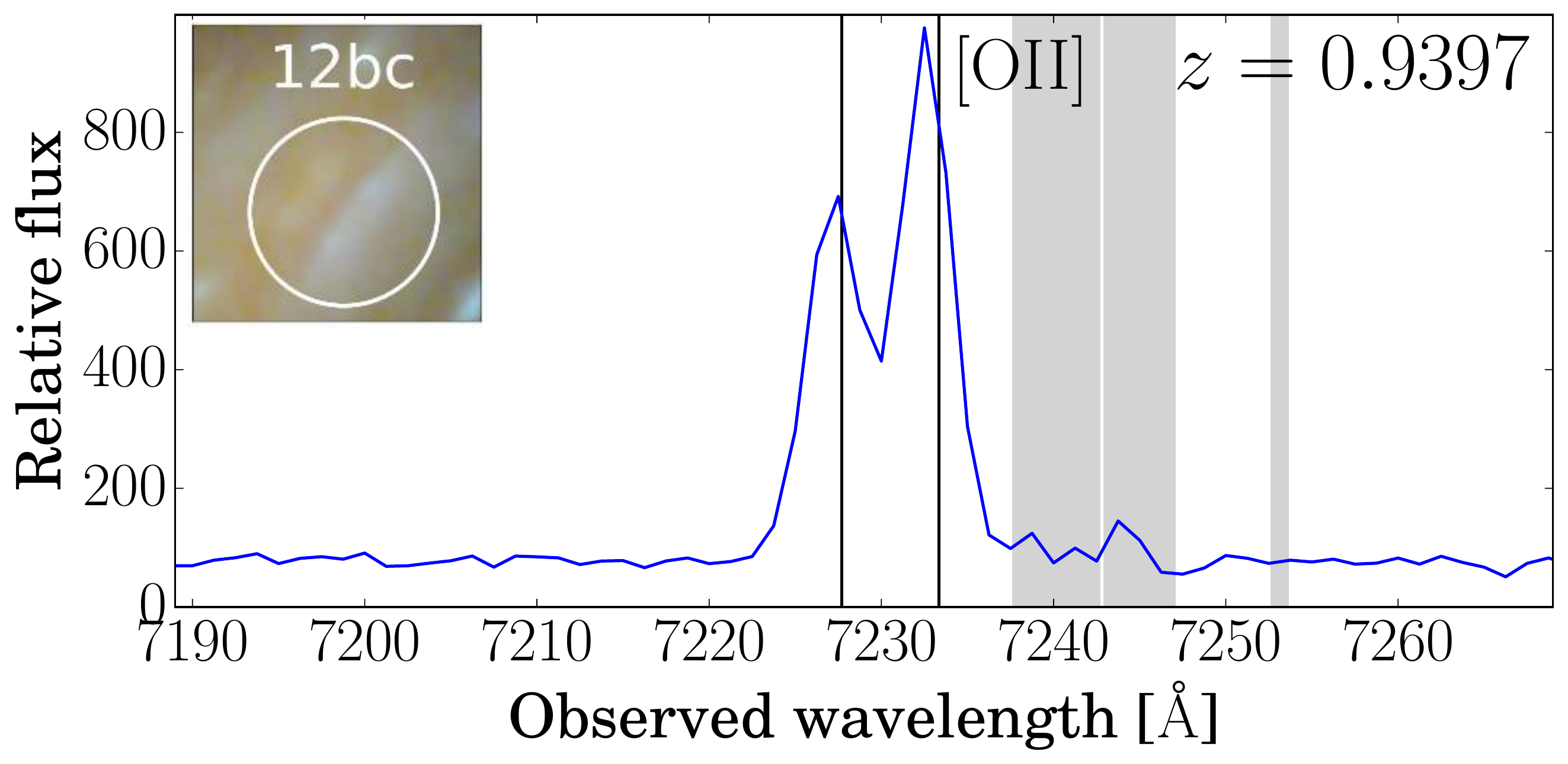}

   \includegraphics[width = 0.65\columnwidth]{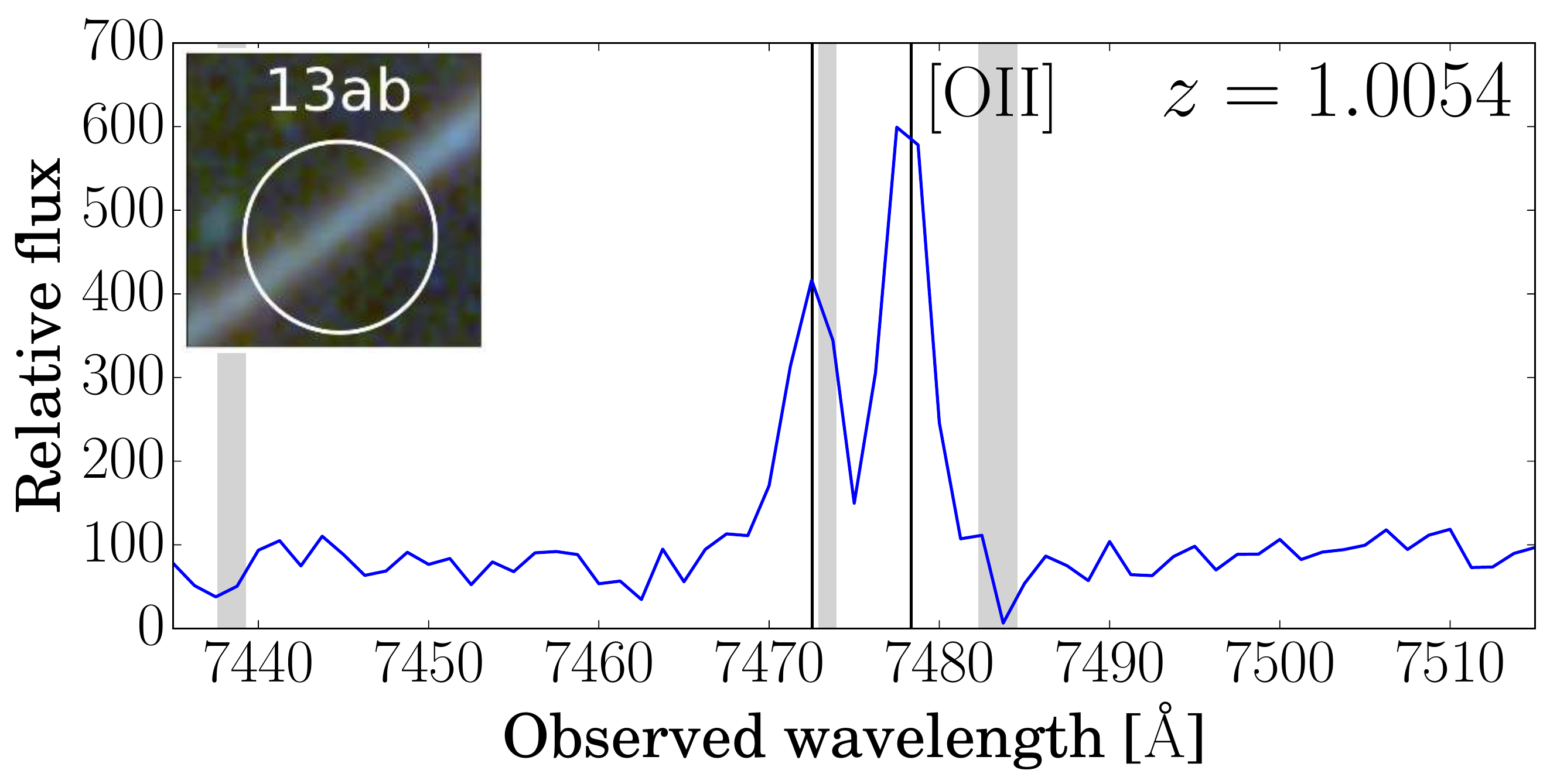}
   \includegraphics[width = 0.65\columnwidth]{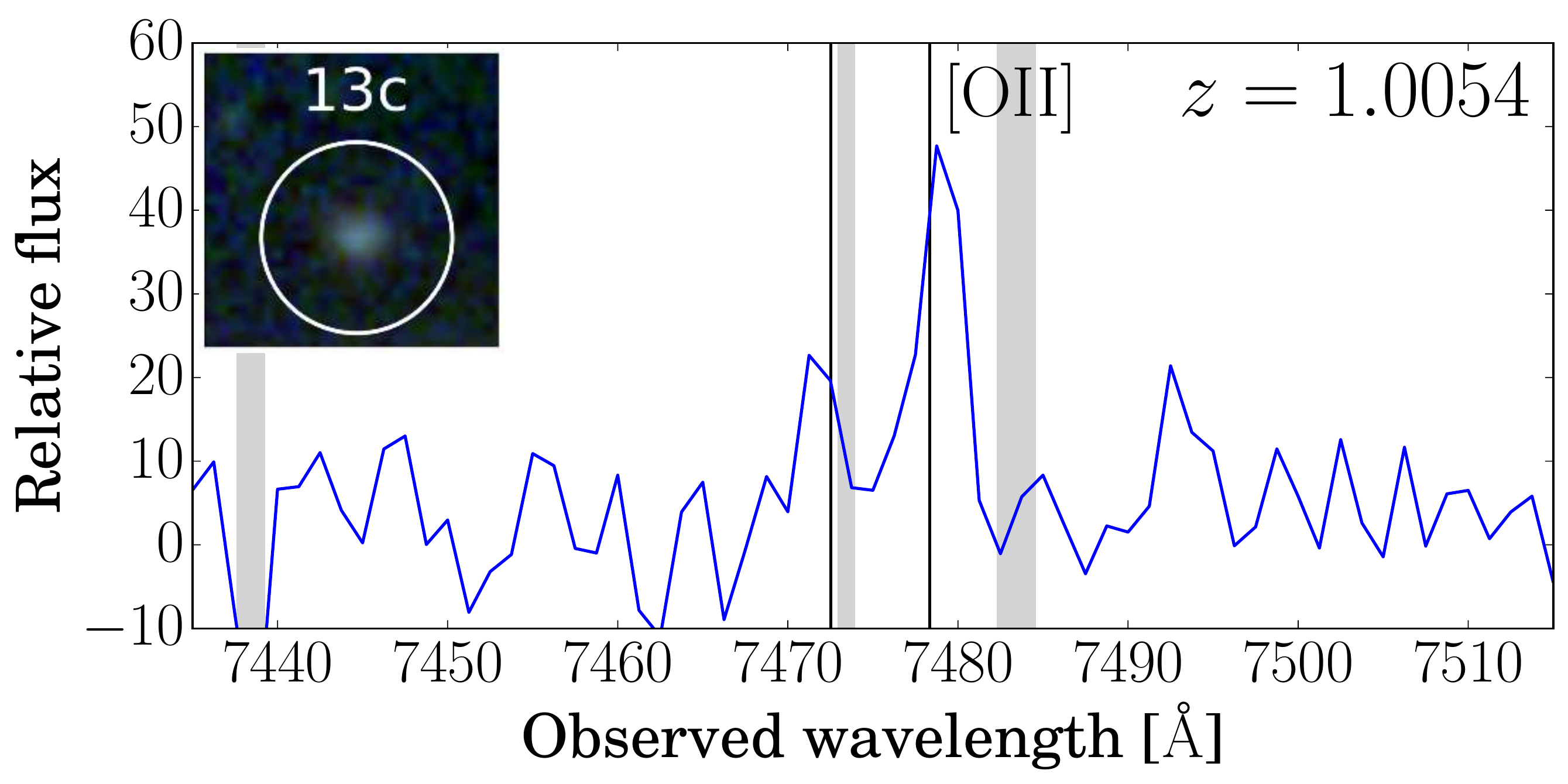}

   \includegraphics[width = 0.65\columnwidth]{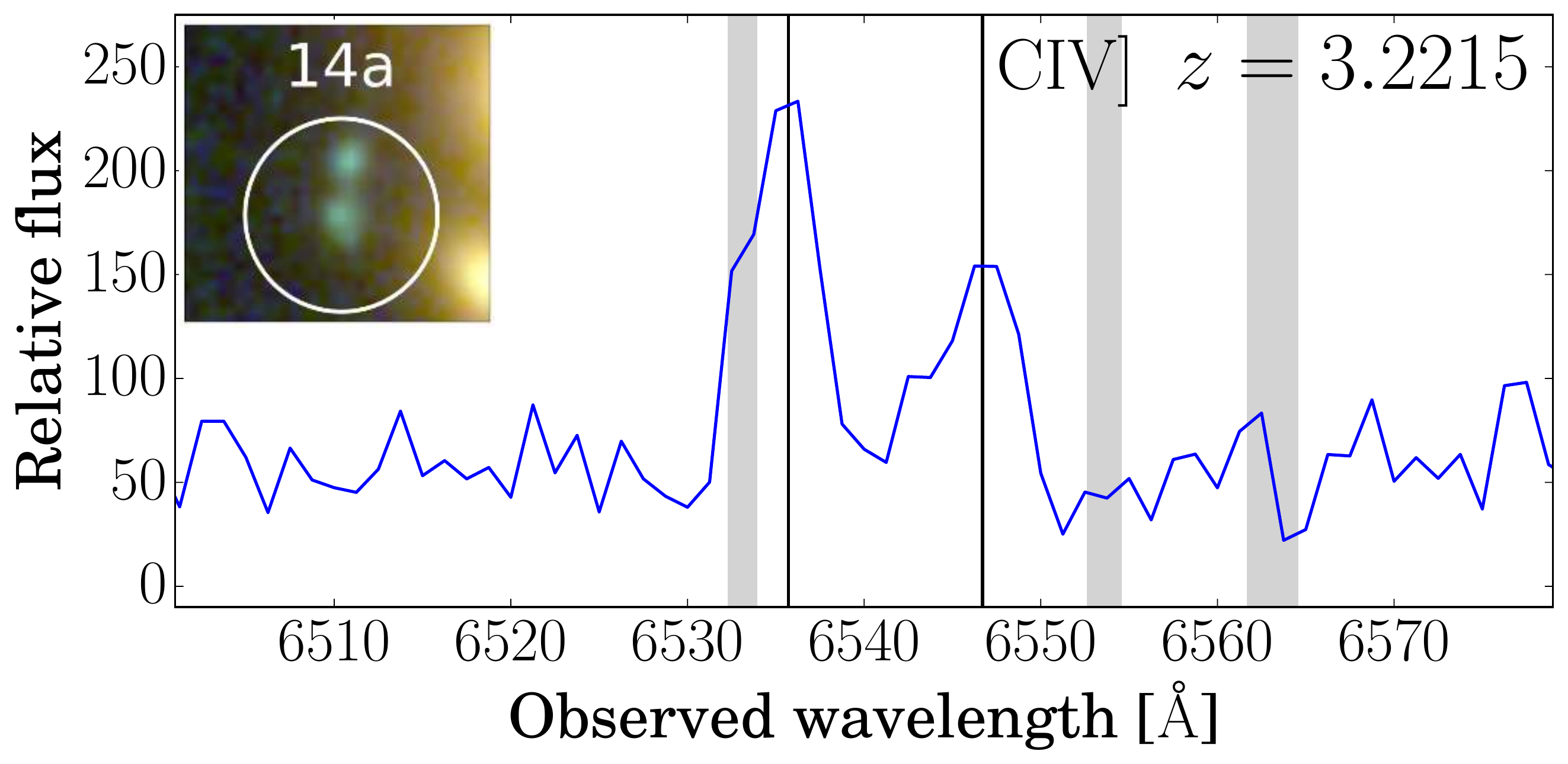}
   \includegraphics[width = 0.65\columnwidth]{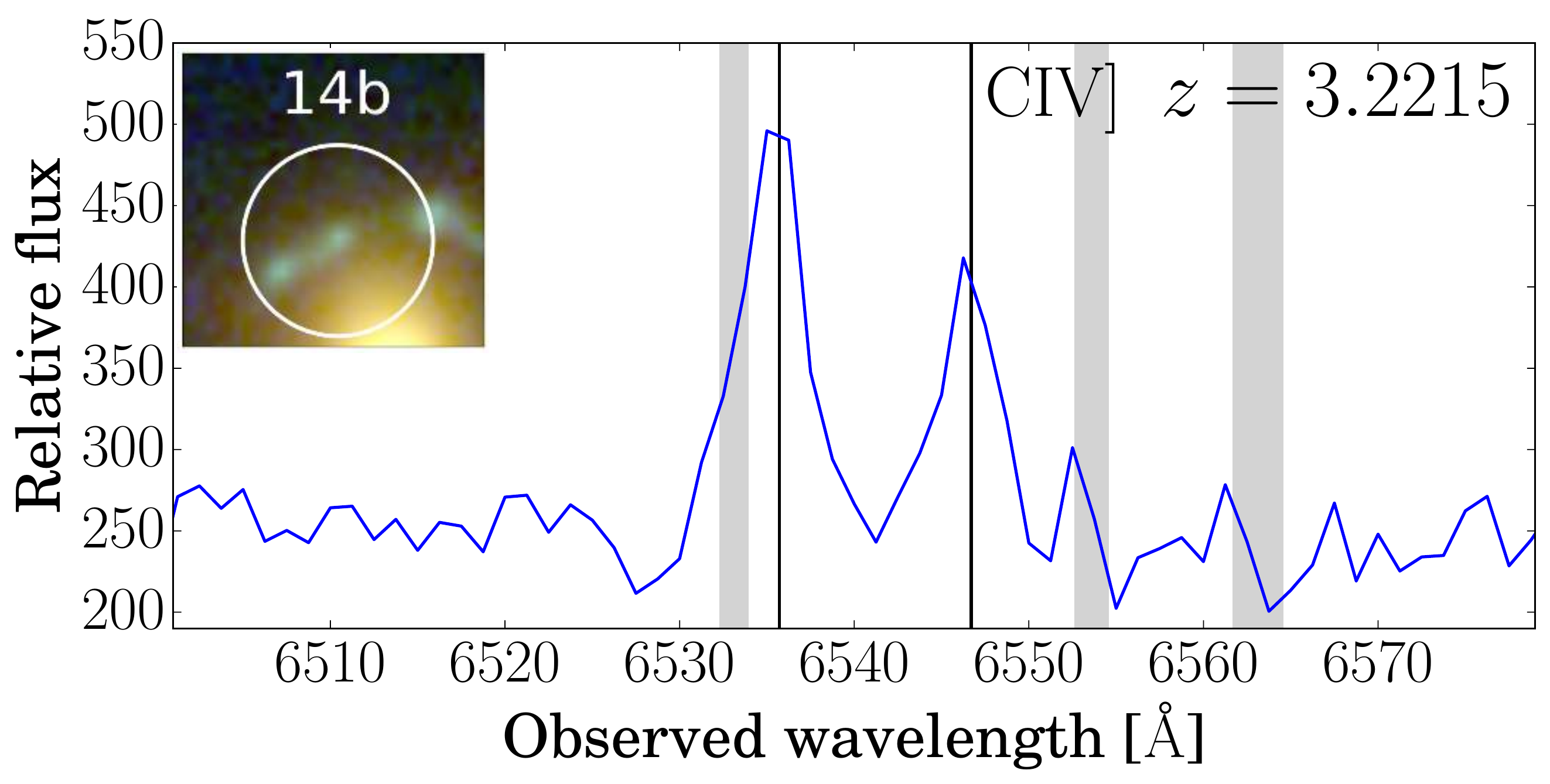}
   \includegraphics[width = 0.65\columnwidth]{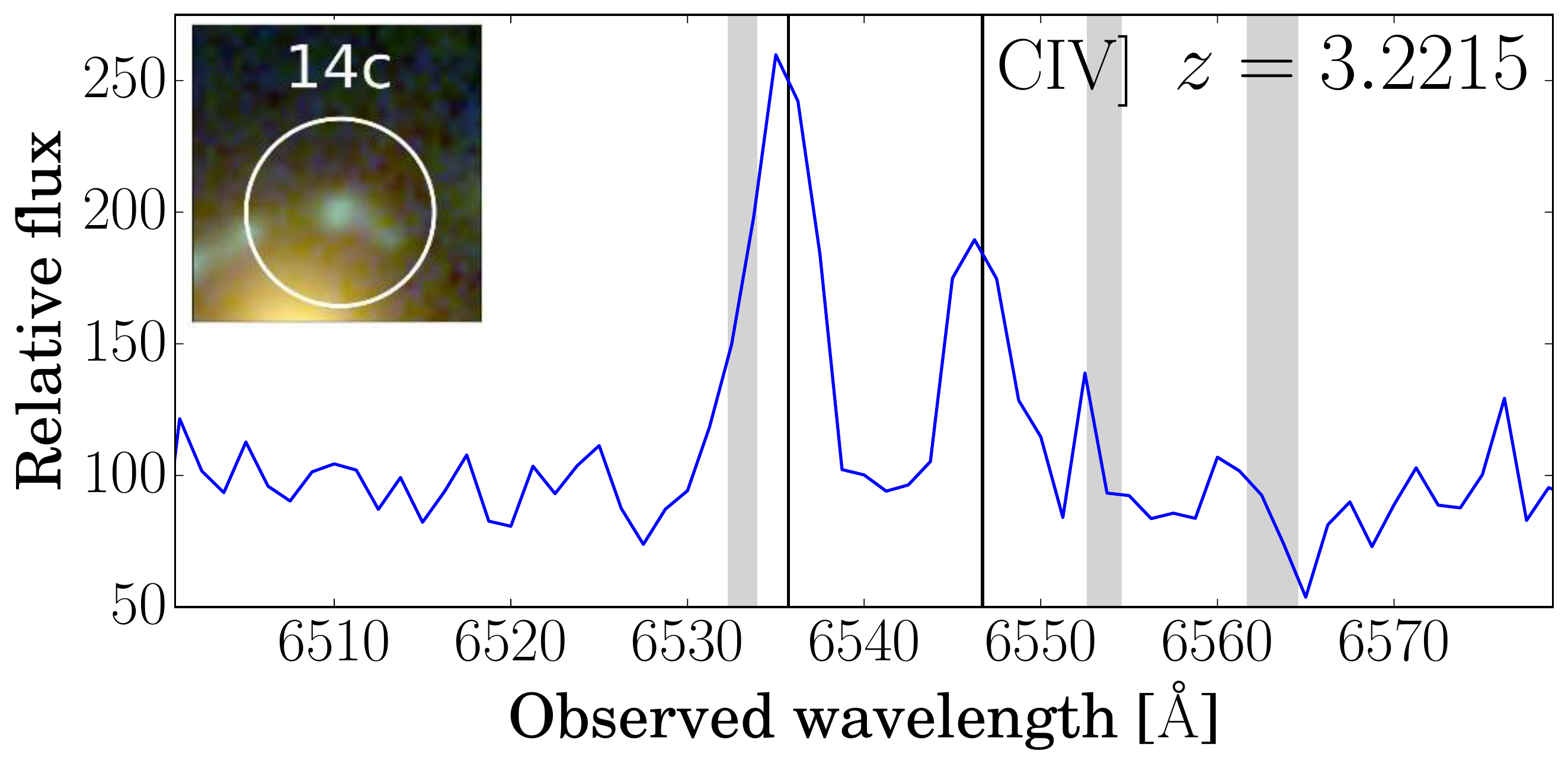}

   \includegraphics[width = 0.65\columnwidth]{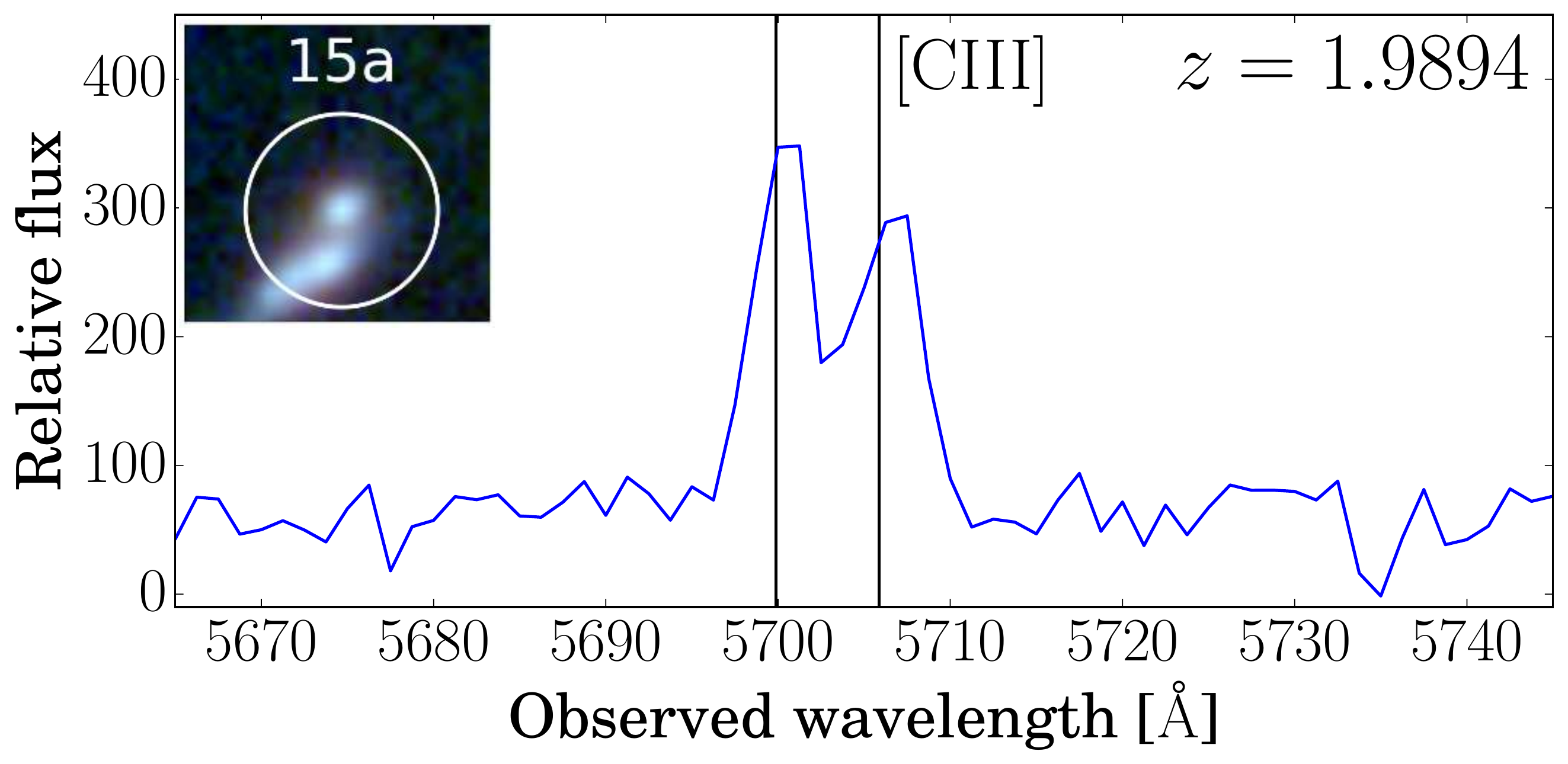}
   \includegraphics[width = 0.65\columnwidth]{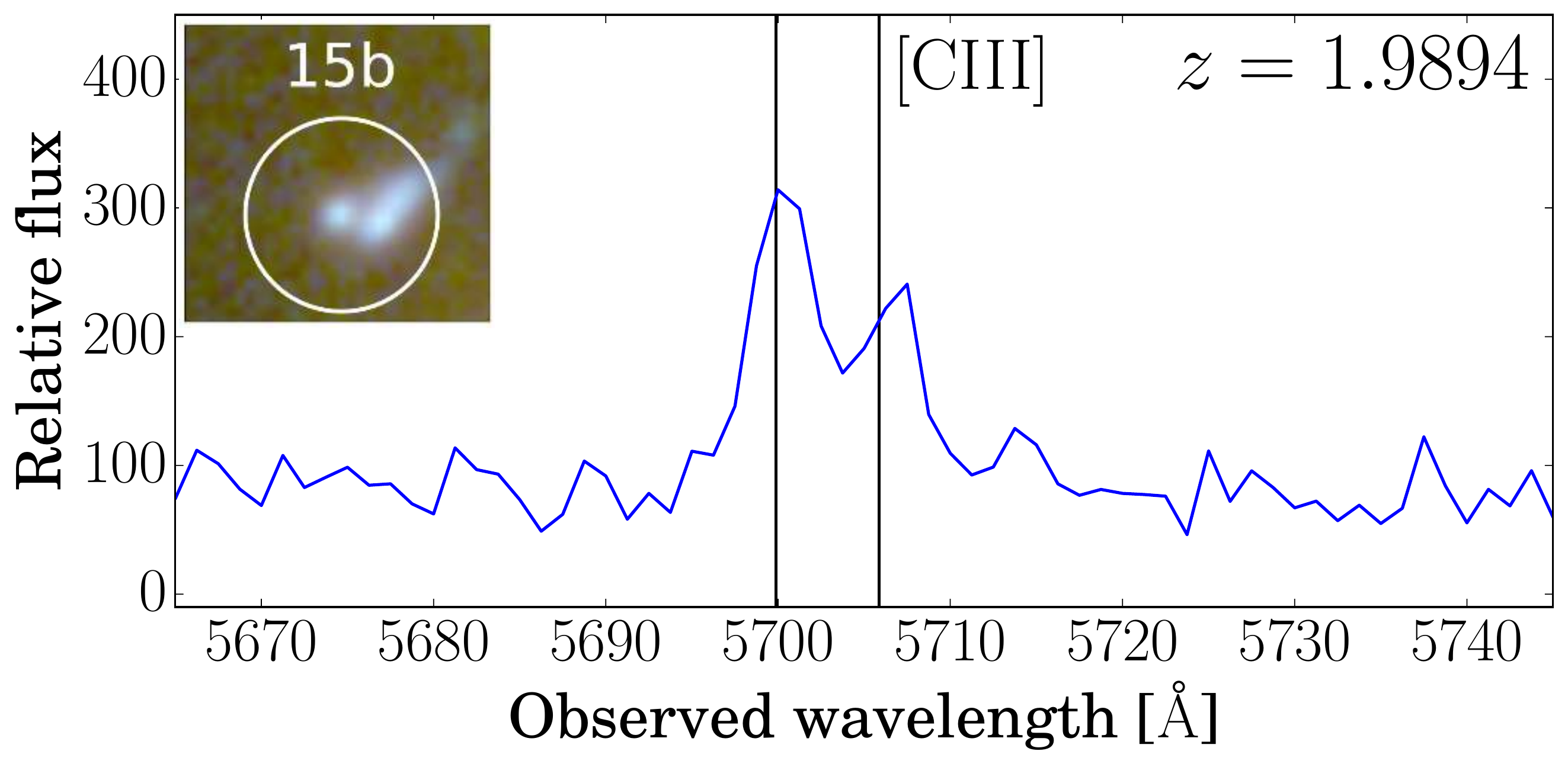}
   \includegraphics[width = 0.65\columnwidth]{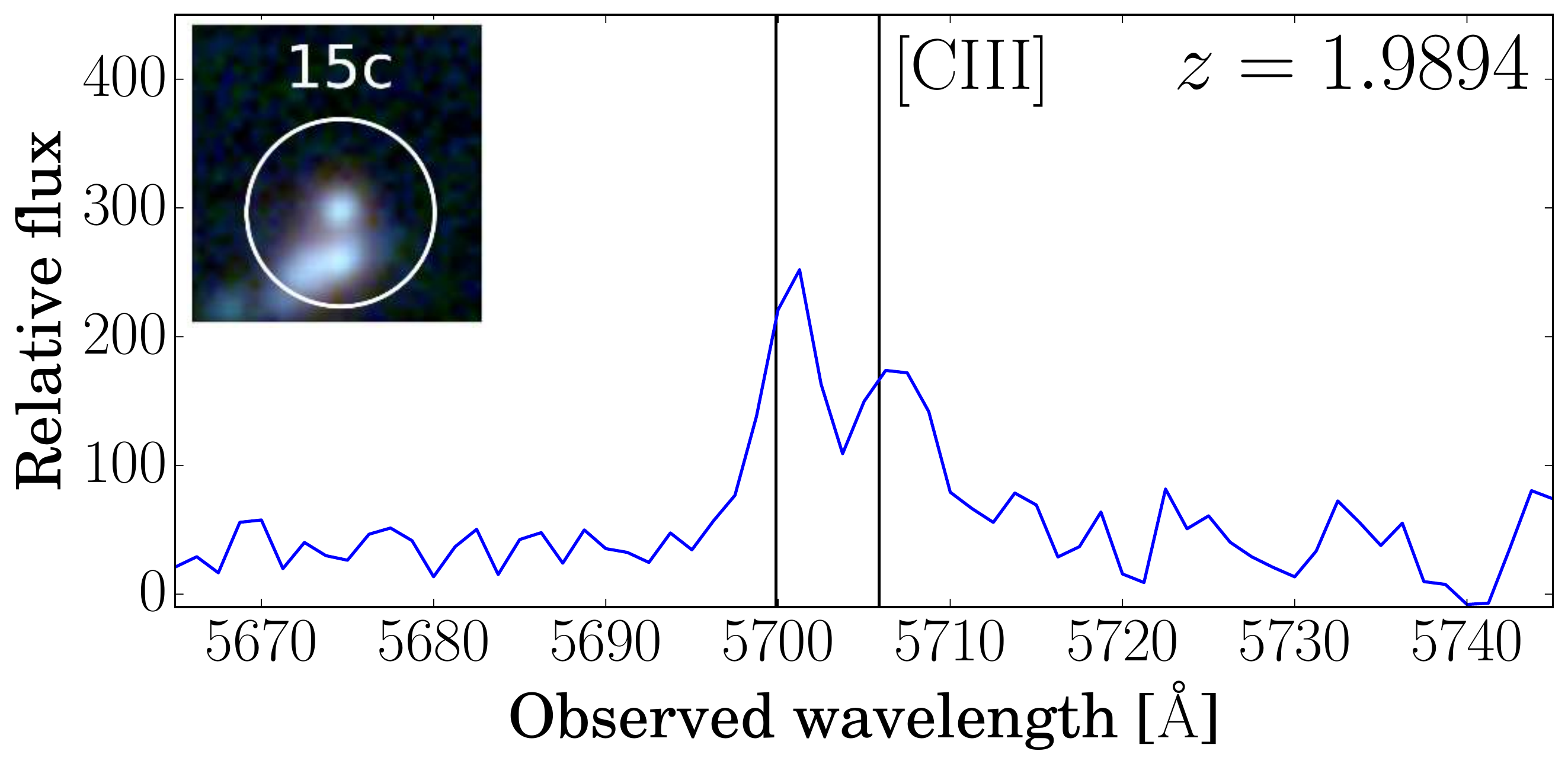}

   \includegraphics[width = 0.65\columnwidth]{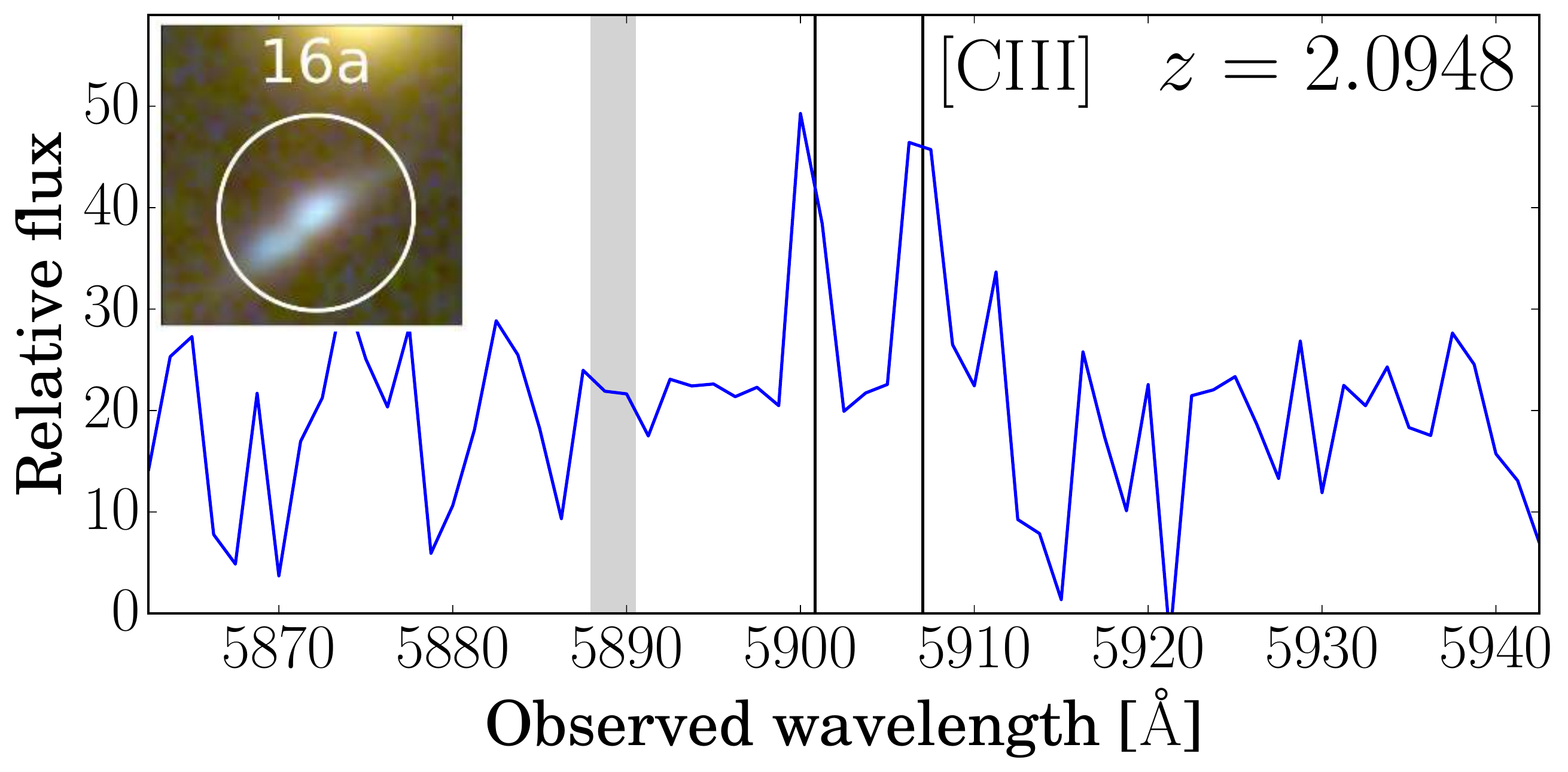}
   \includegraphics[width = 0.65\columnwidth]{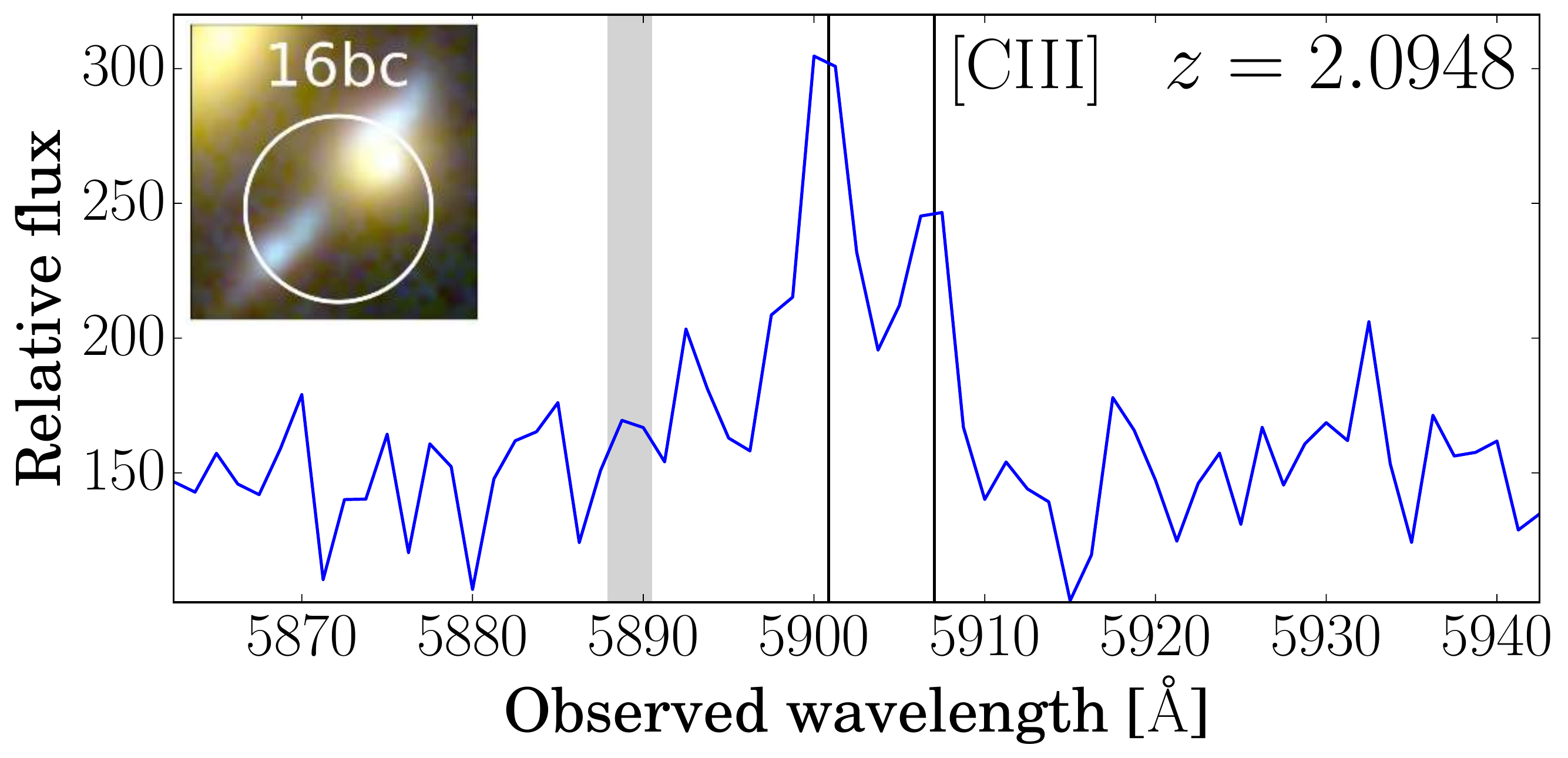}

   \includegraphics[width = 0.65\columnwidth]{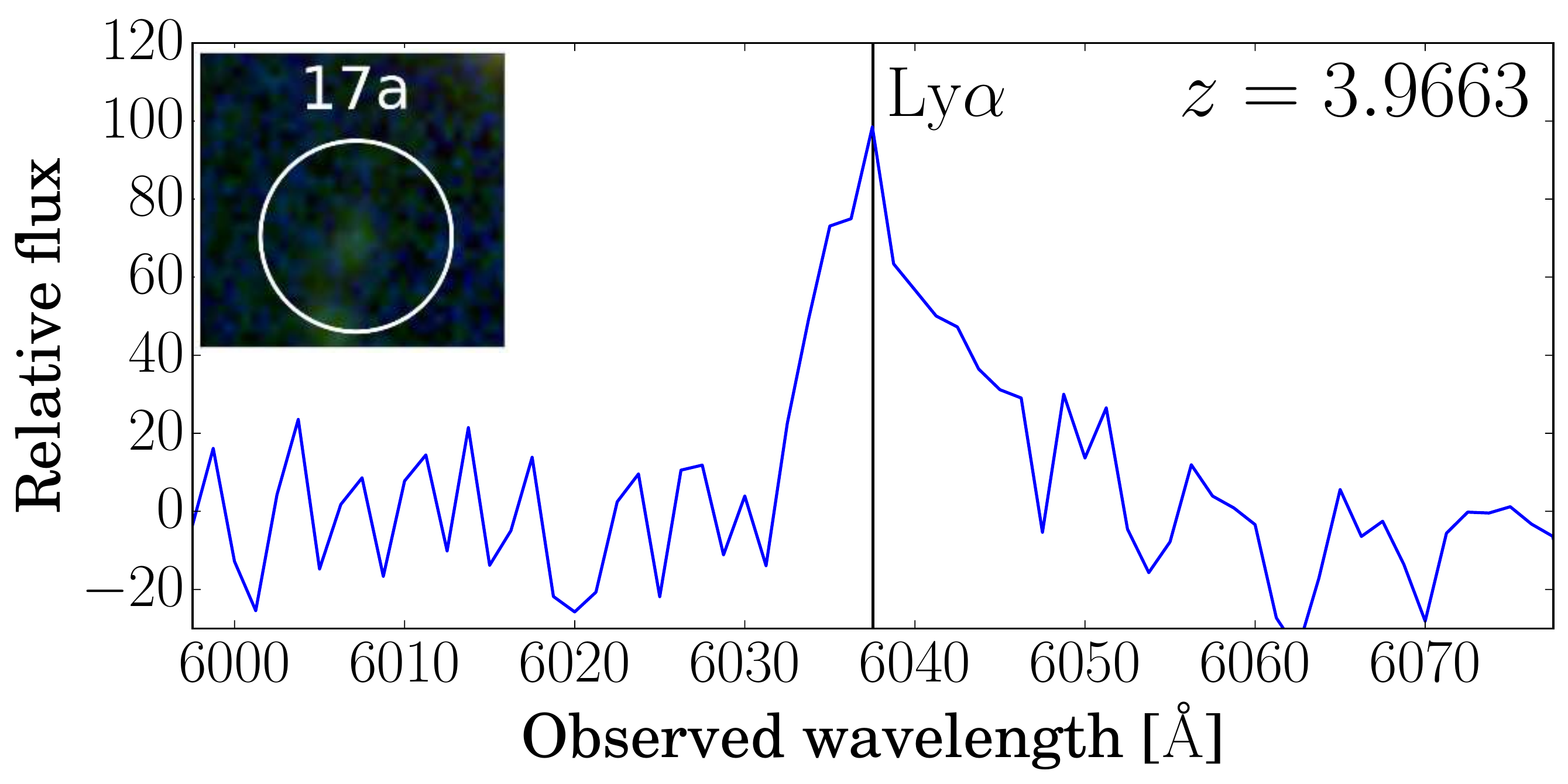}

   \includegraphics[width = 0.65\columnwidth]{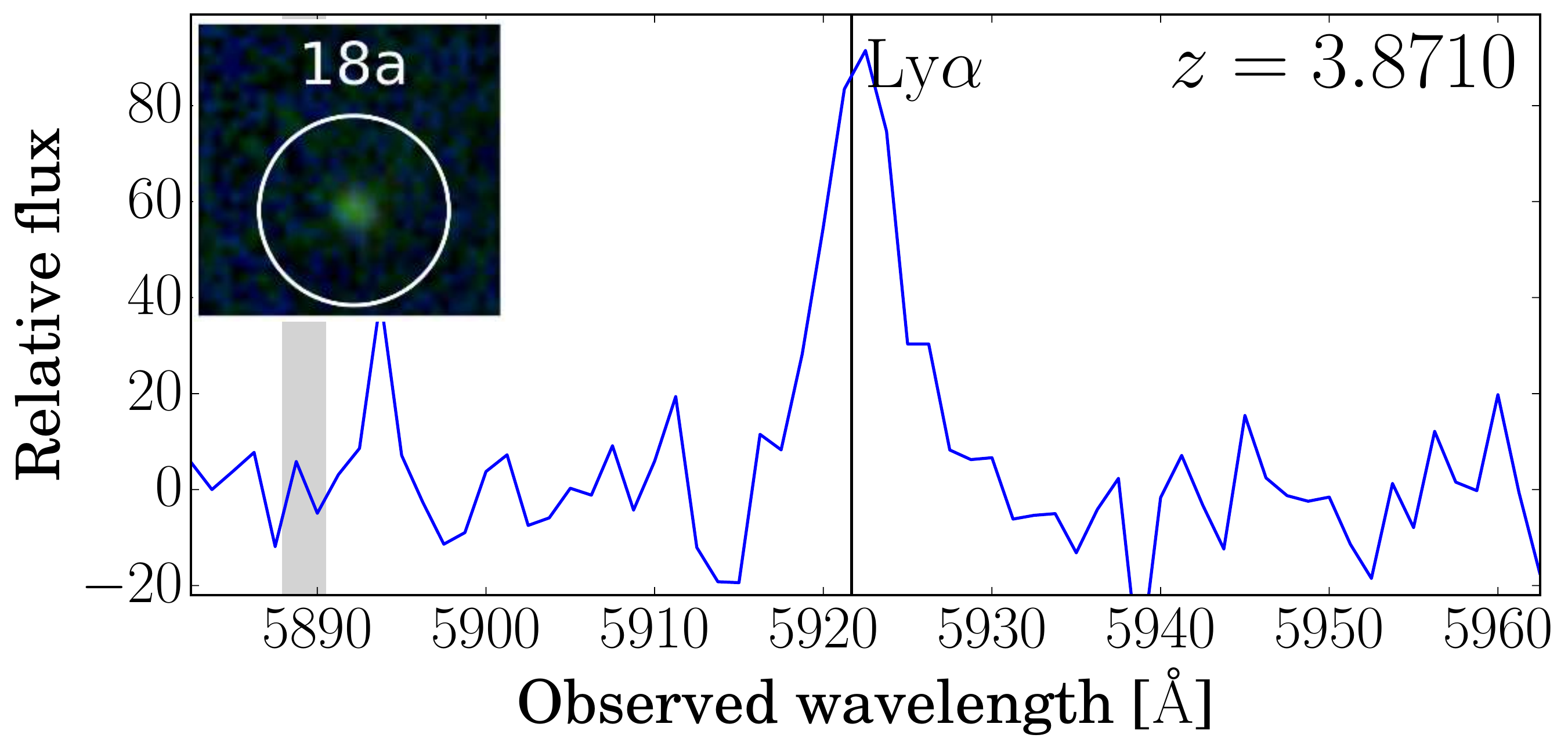}
   \includegraphics[width = 0.65\columnwidth]{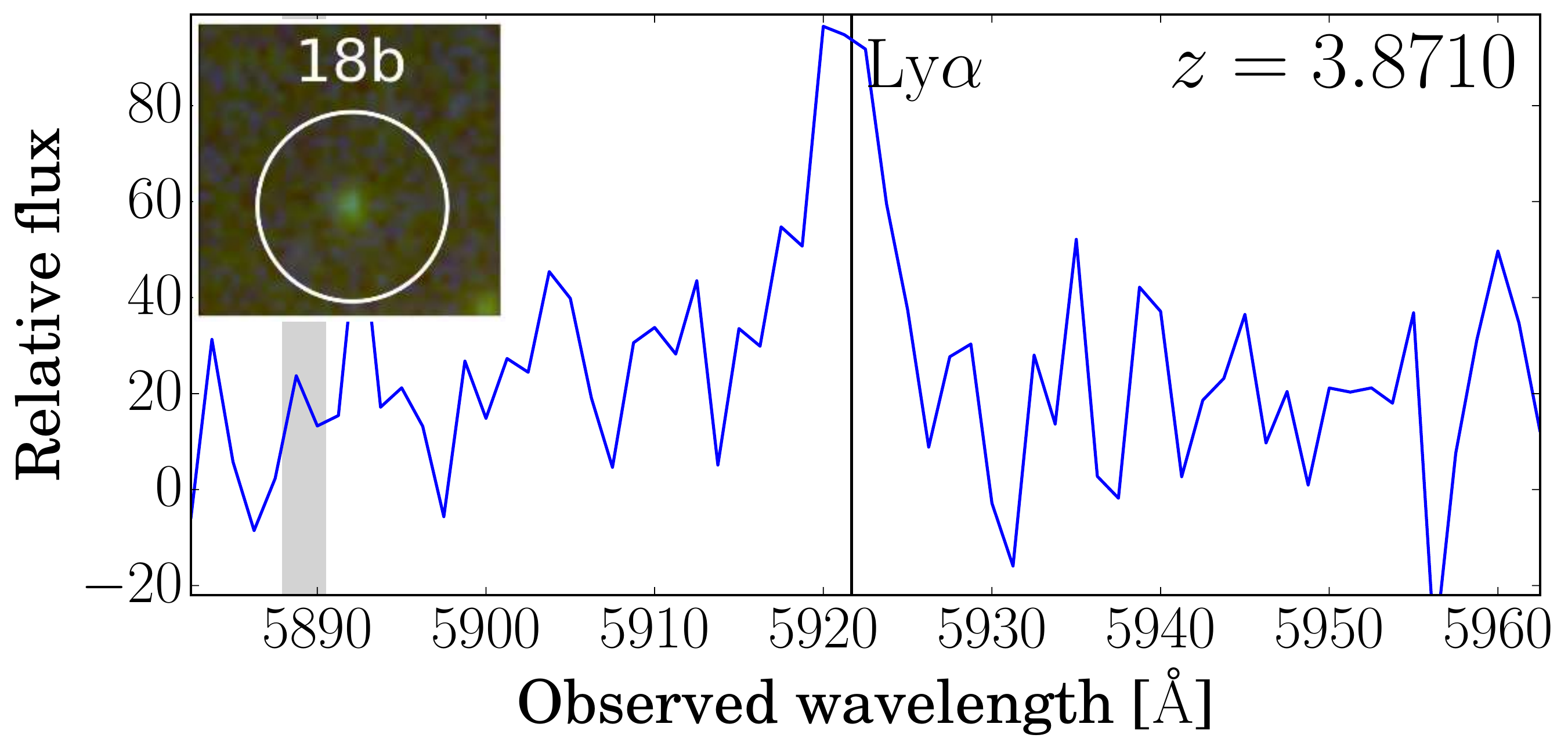}

  \caption{(Continued)}
  \label{fig:specs}
\end{figure*}

\begin{figure*}
  \ContinuedFloat
  %\centering

   \includegraphics[width = 0.65\columnwidth]{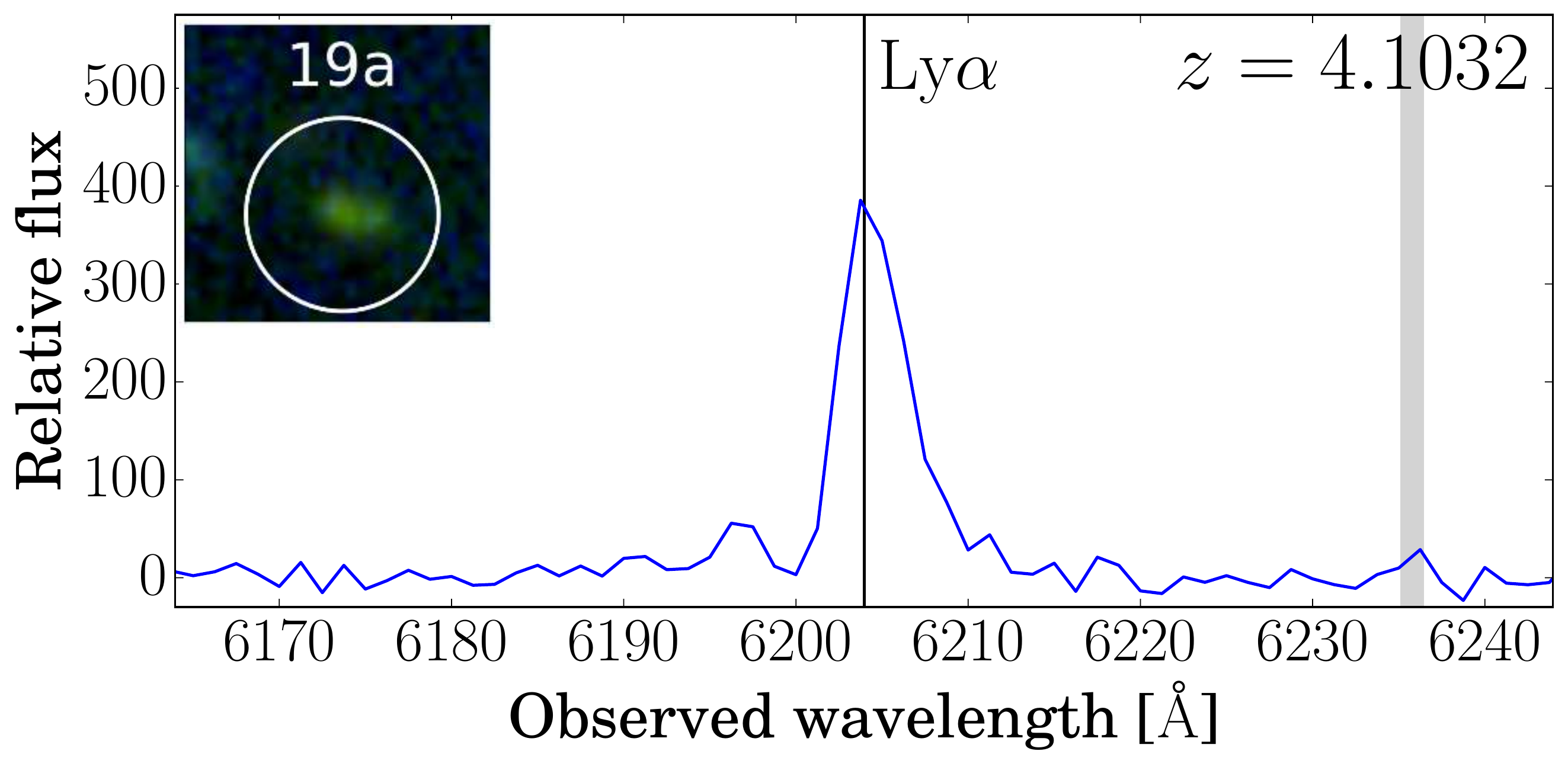}
   \includegraphics[width = 0.65\columnwidth]{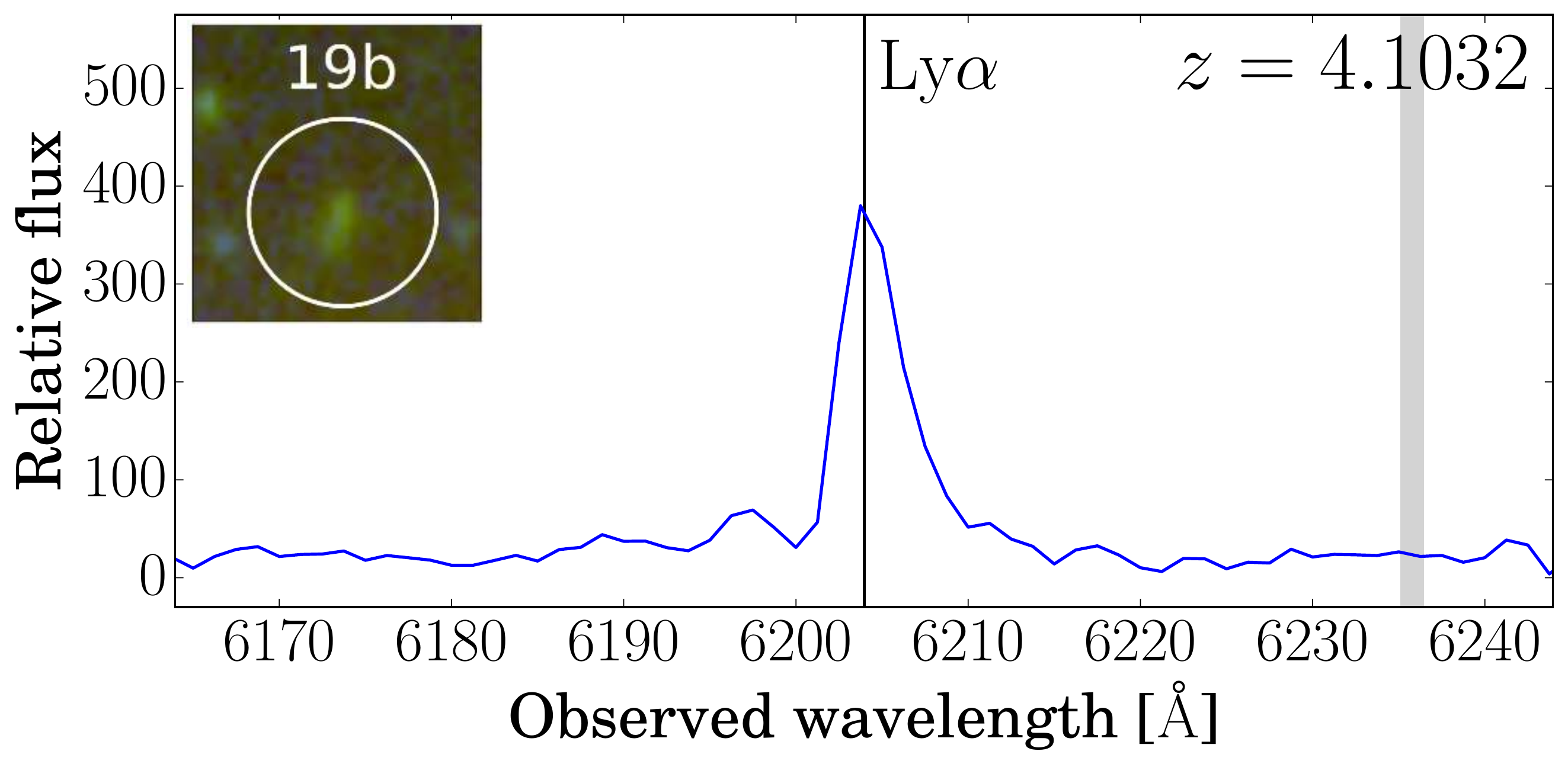}
   \includegraphics[width = 0.65\columnwidth]{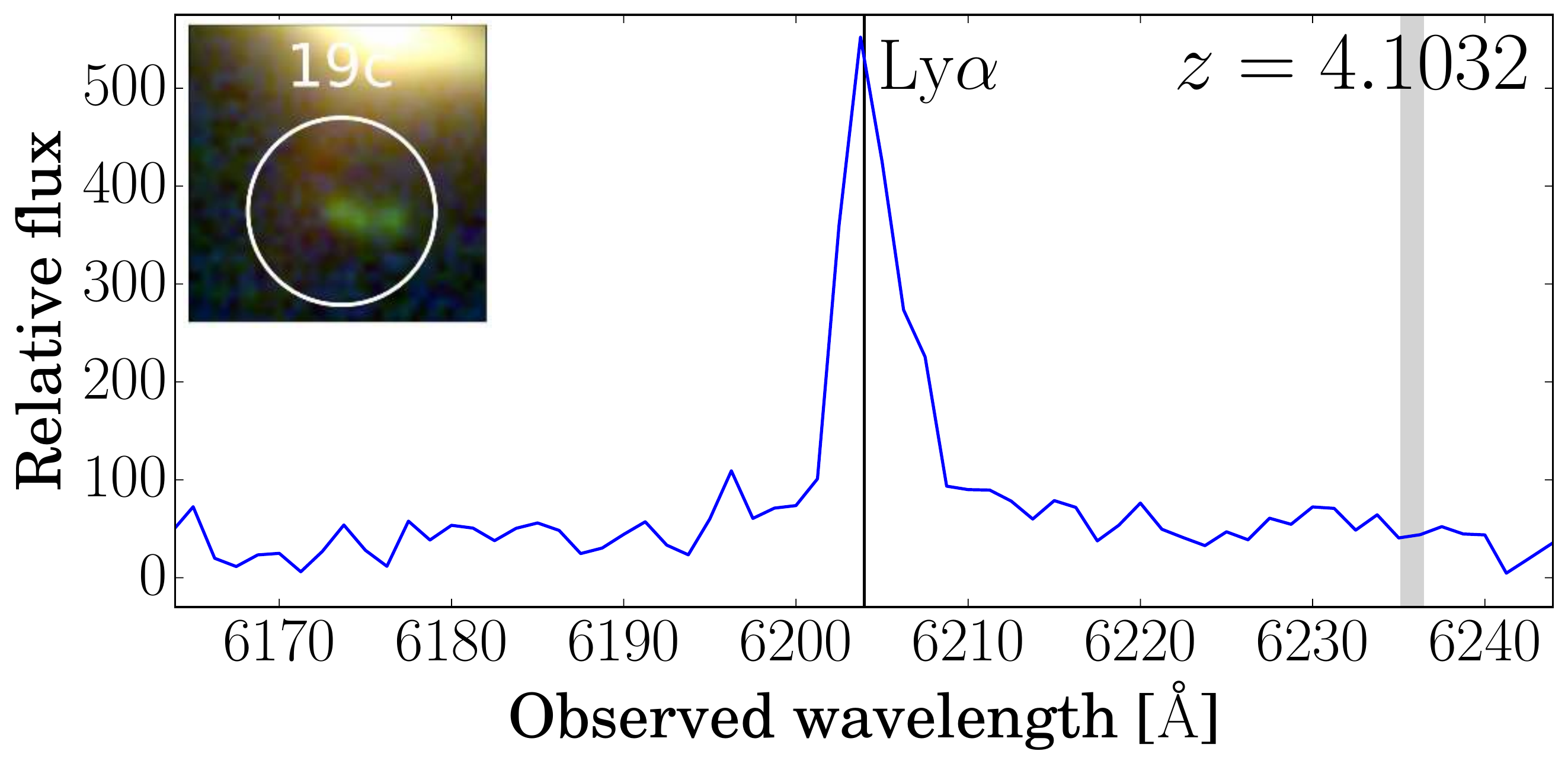}

   \includegraphics[width = 0.65\columnwidth]{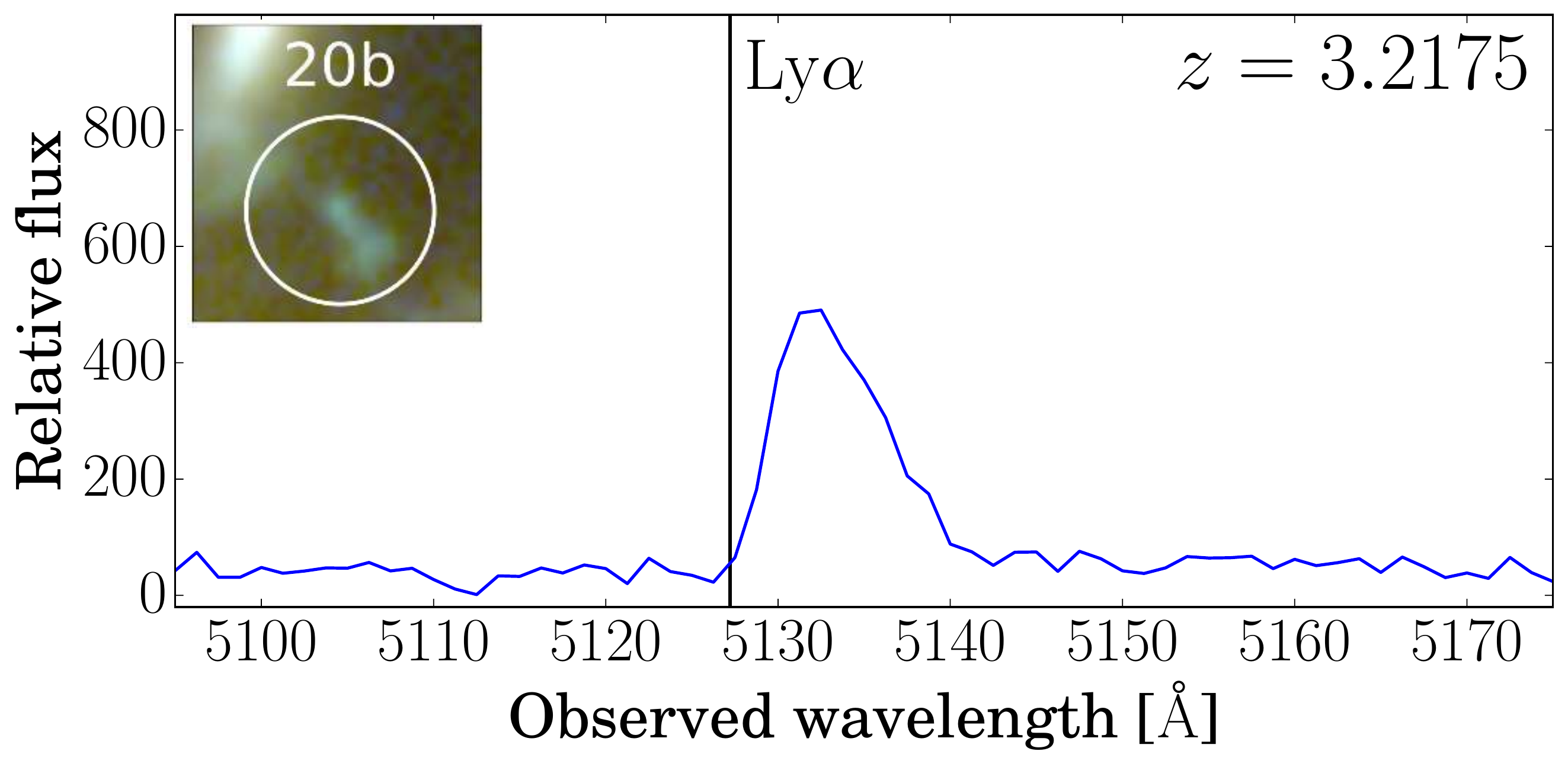}
   \includegraphics[width = 0.65\columnwidth]{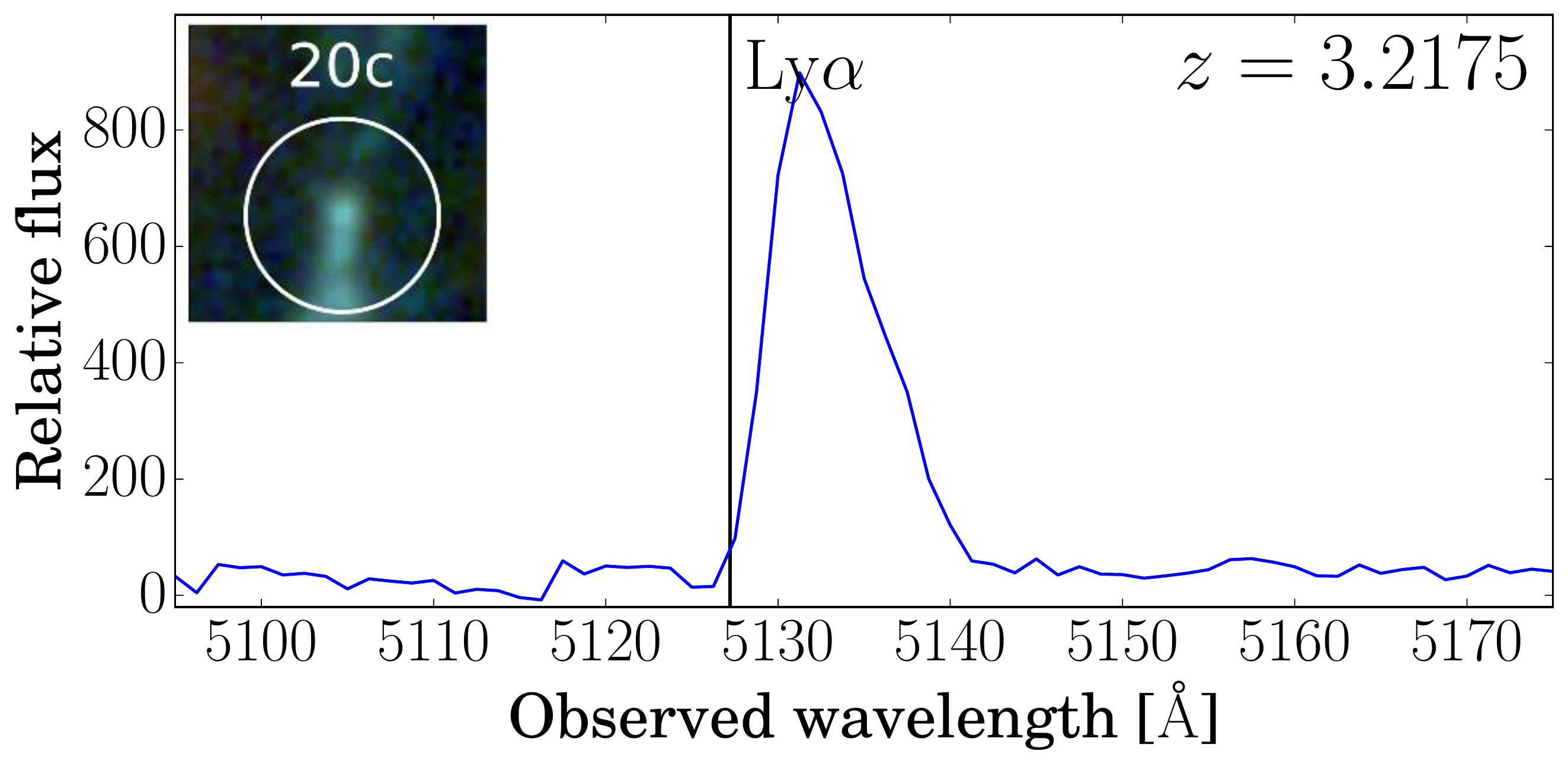}

   \includegraphics[width = 0.65\columnwidth]{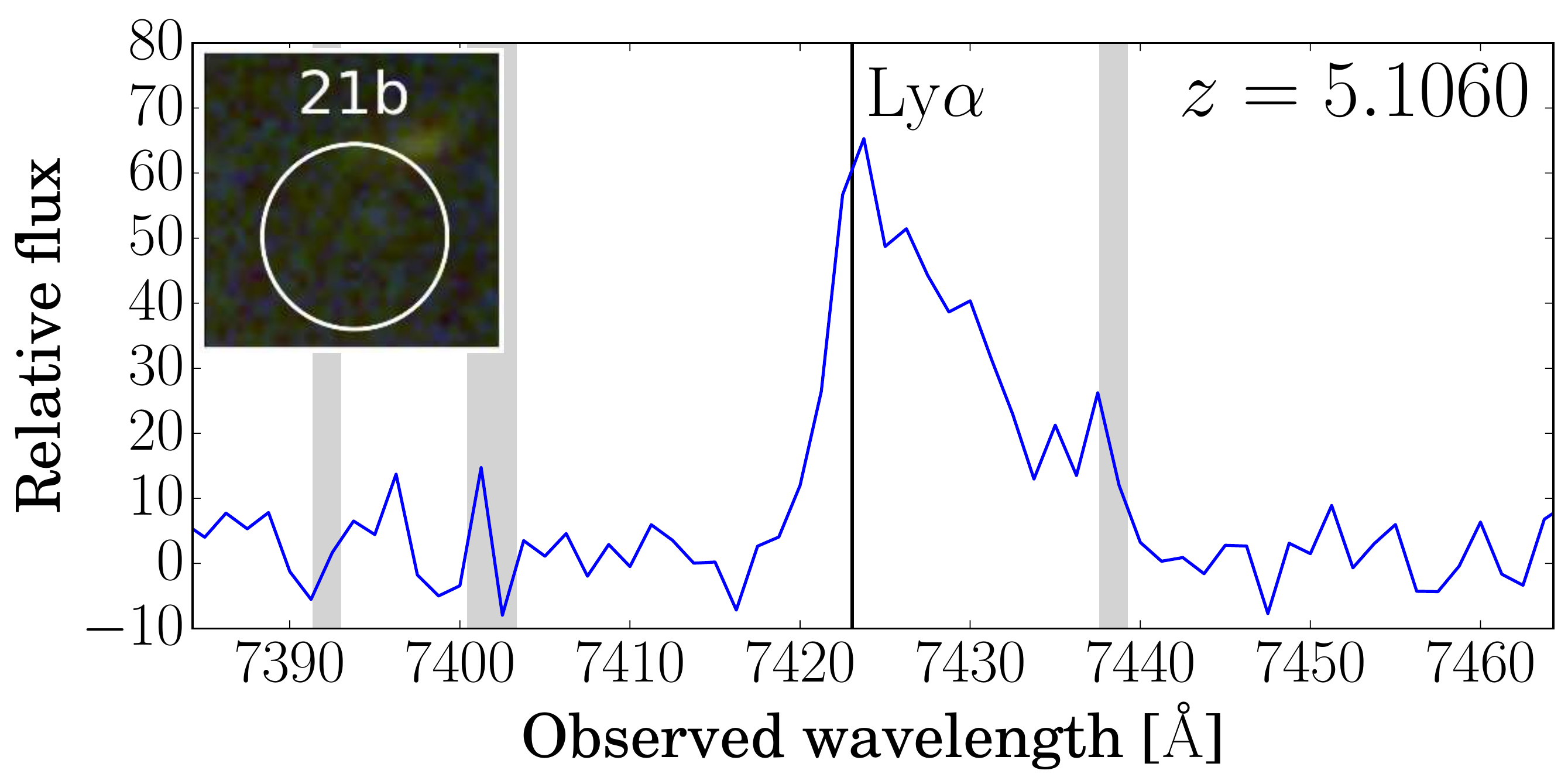}
   \includegraphics[width = 0.65\columnwidth]{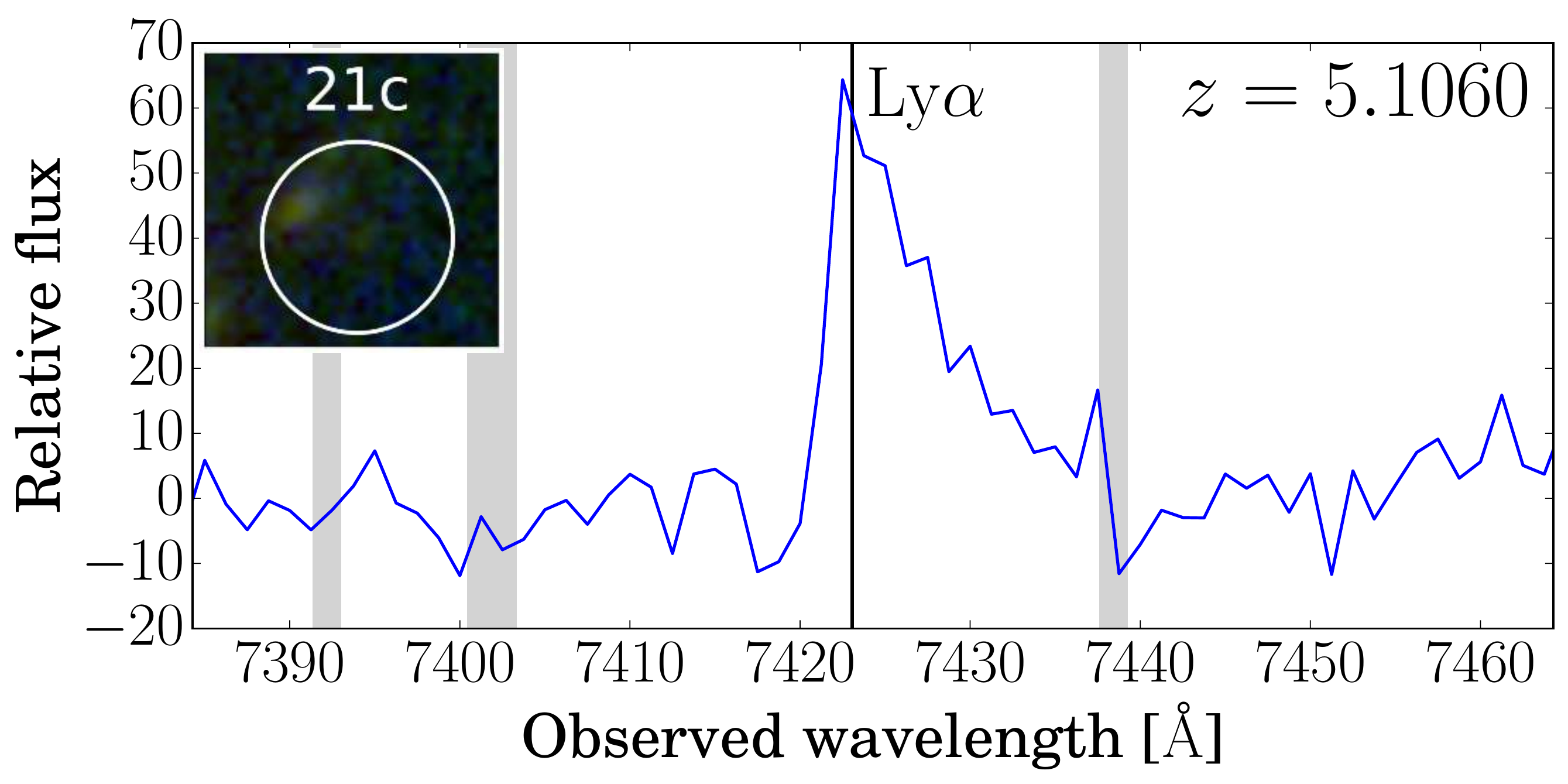}

   \includegraphics[width = 0.65\columnwidth]{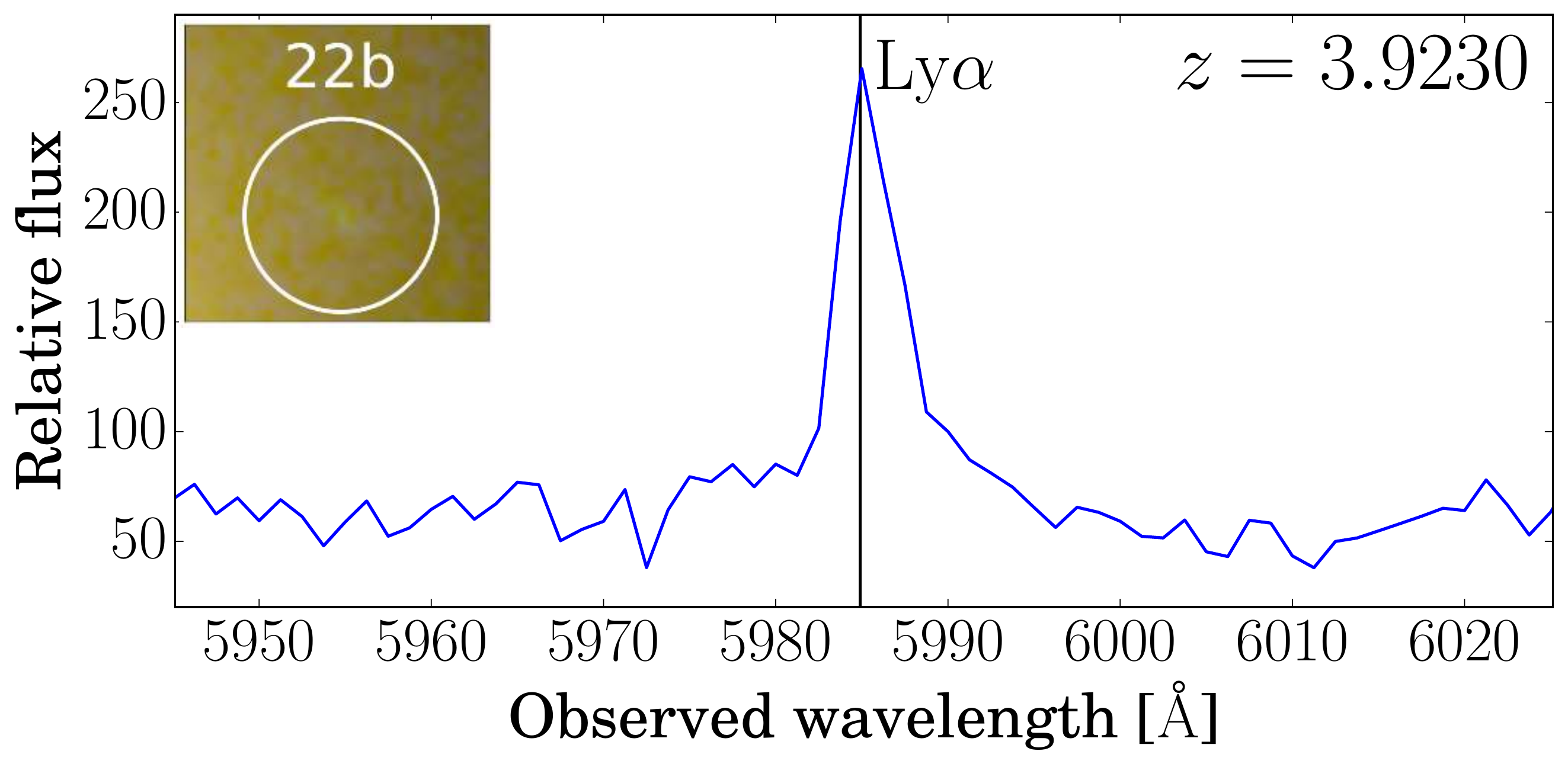}
   \includegraphics[width = 0.65\columnwidth]{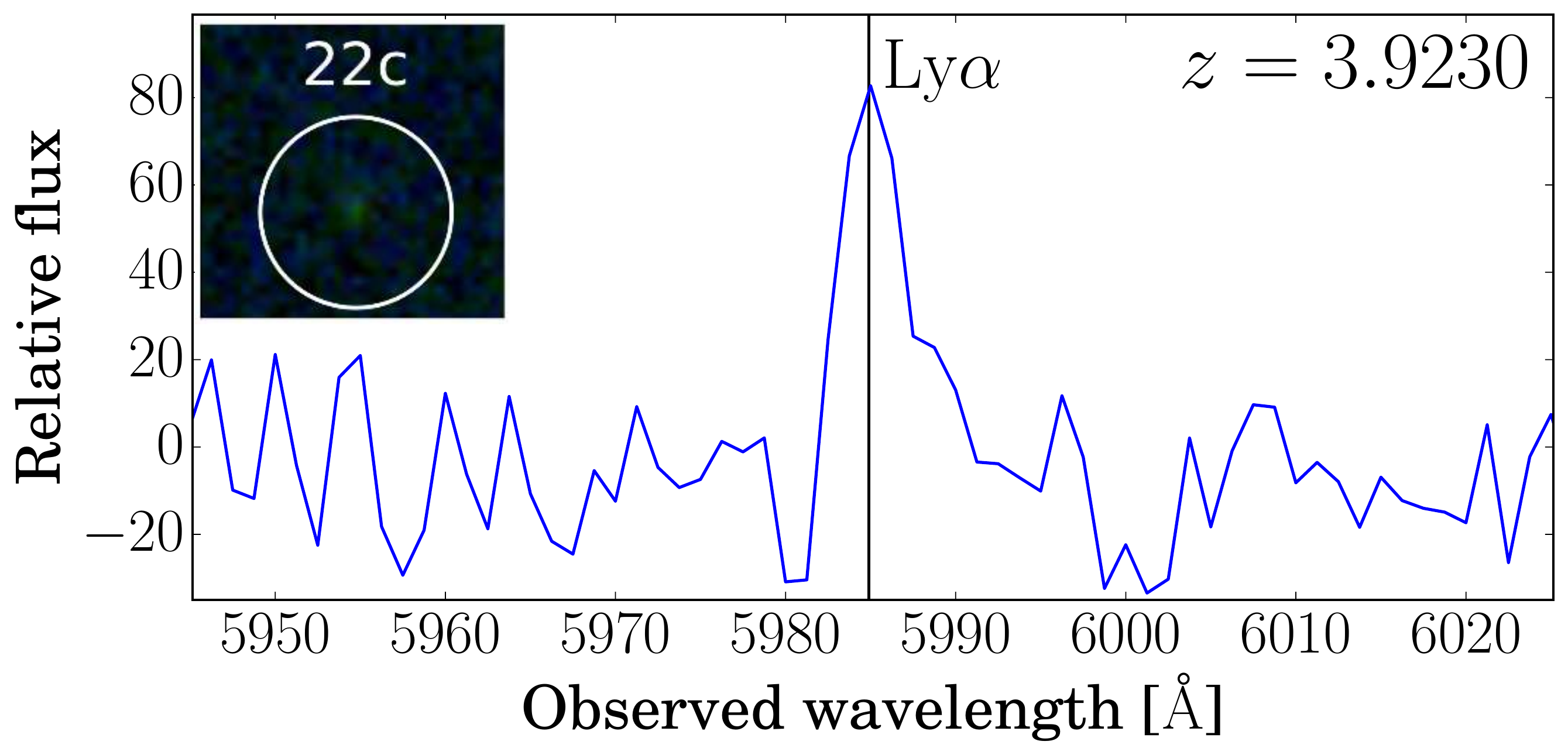}

   \includegraphics[width = 0.65\columnwidth]{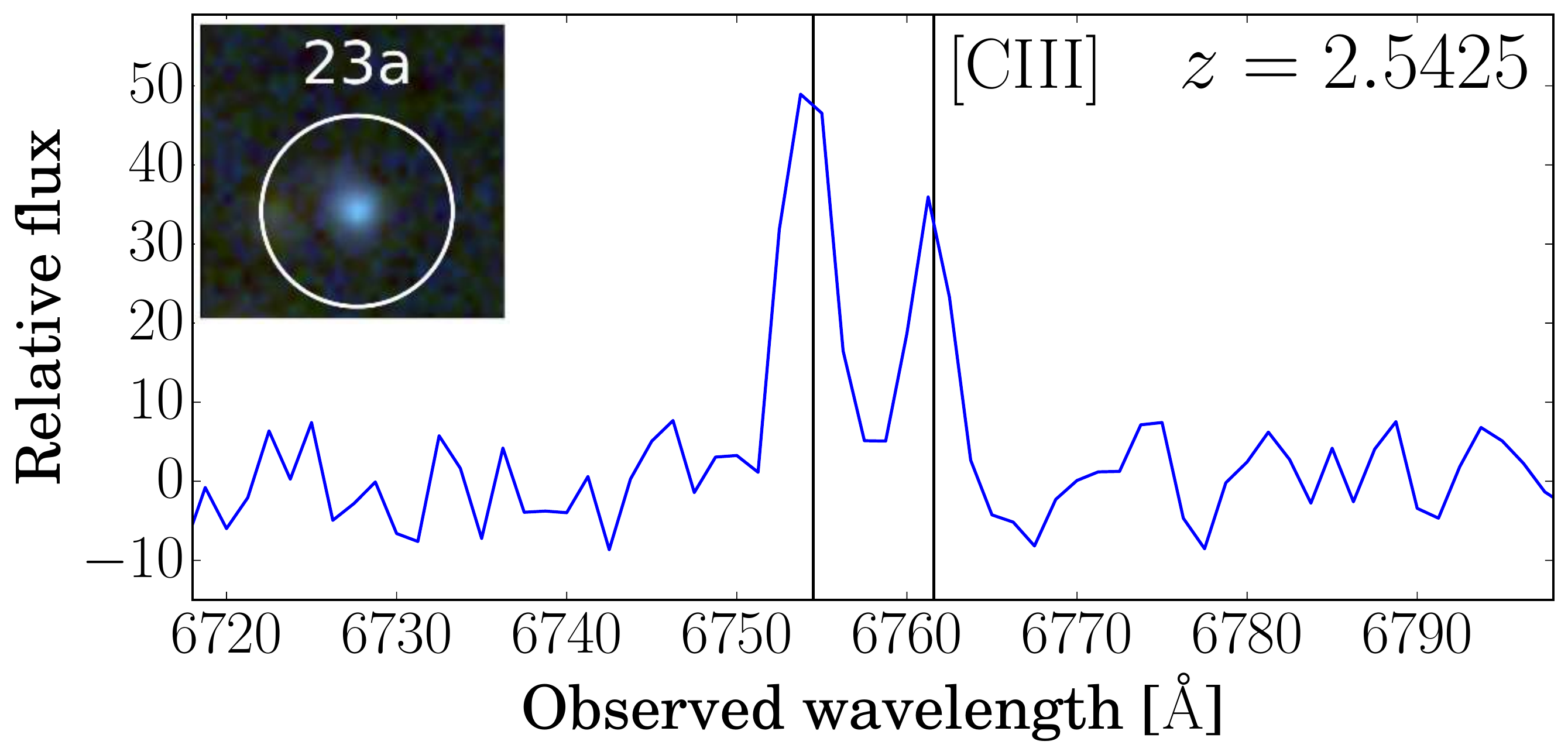}
   \includegraphics[width = 0.65\columnwidth]{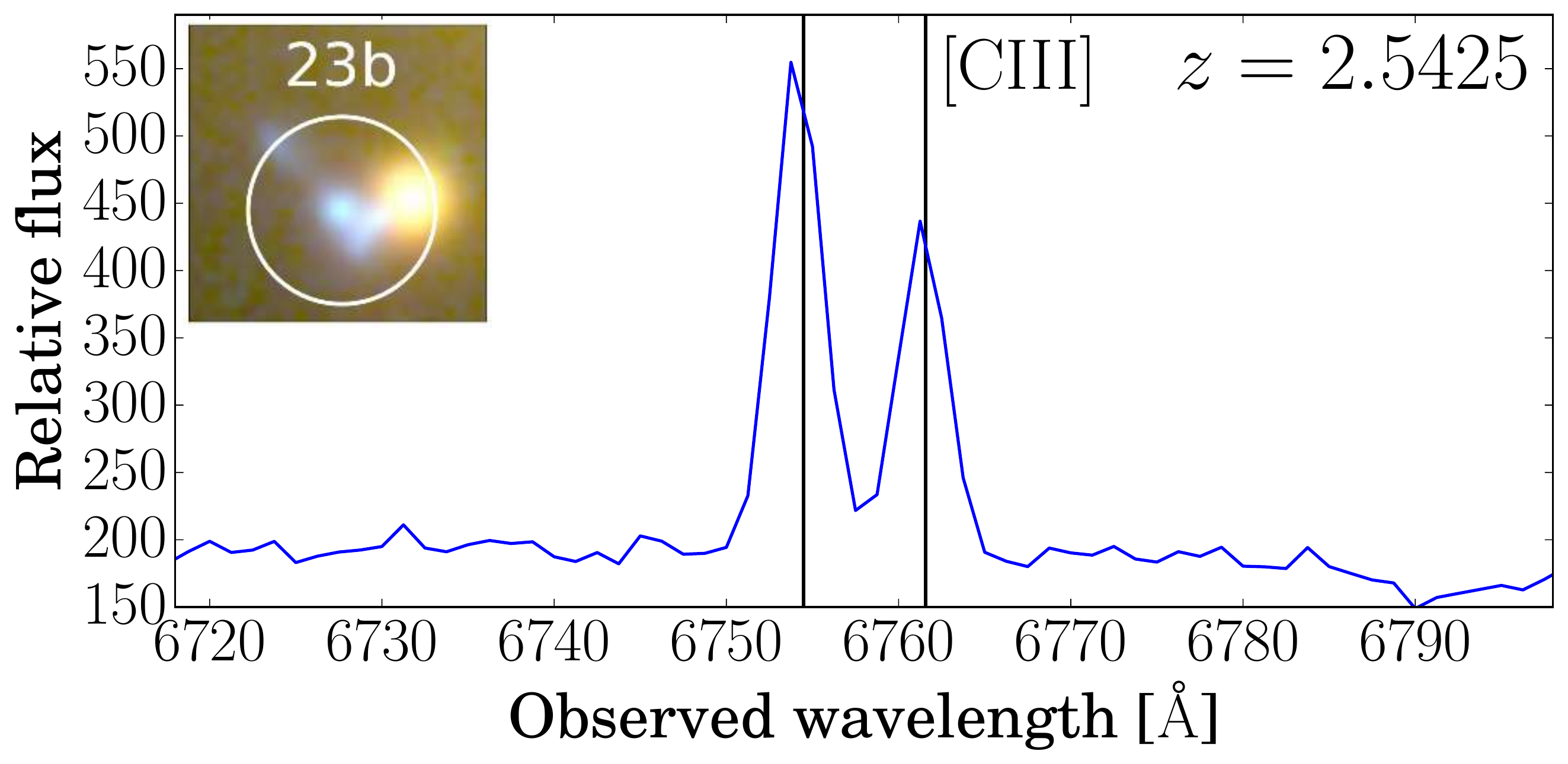}

   \includegraphics[width = 0.65\columnwidth]{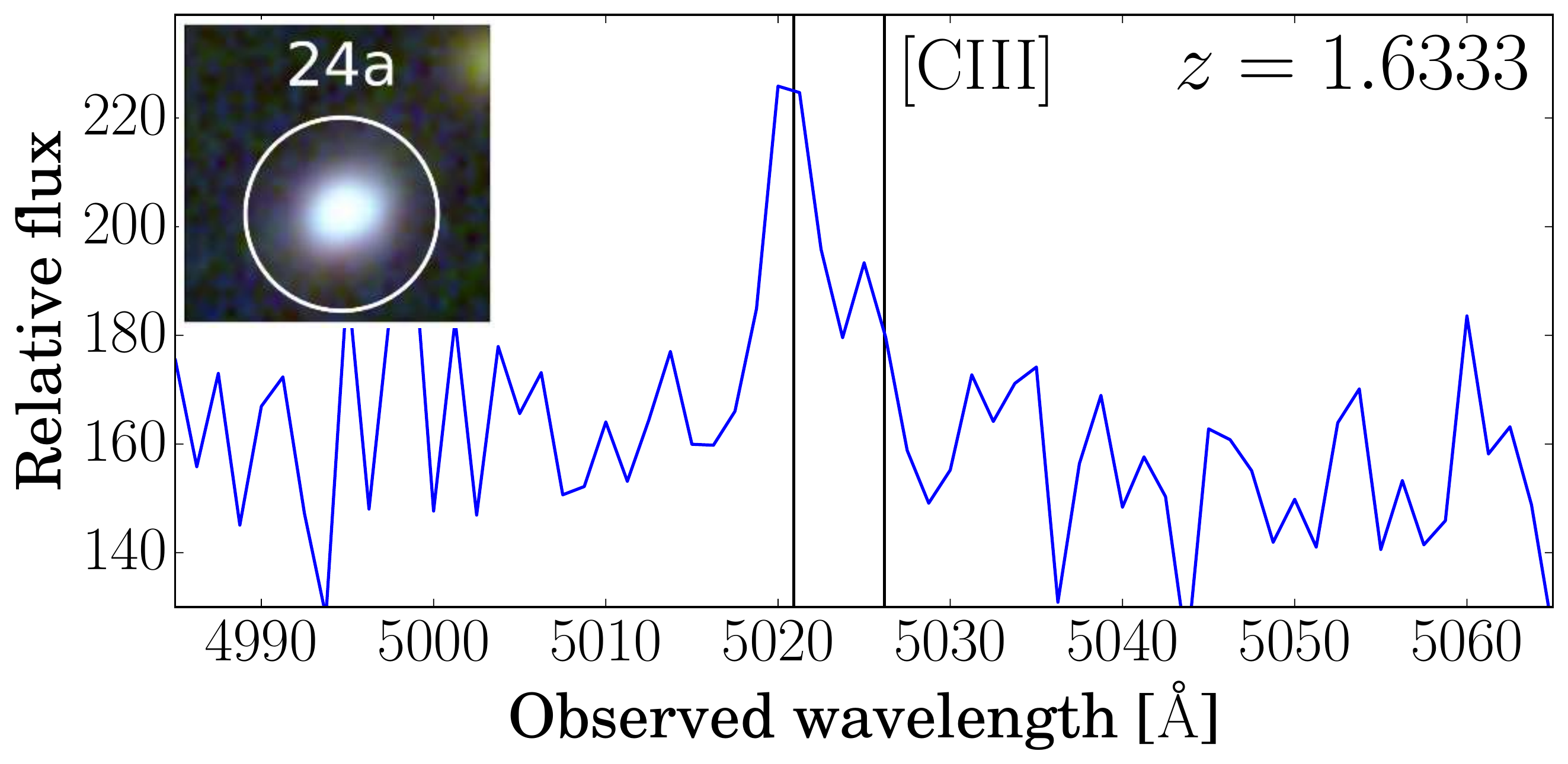}
   \includegraphics[width = 0.65\columnwidth]{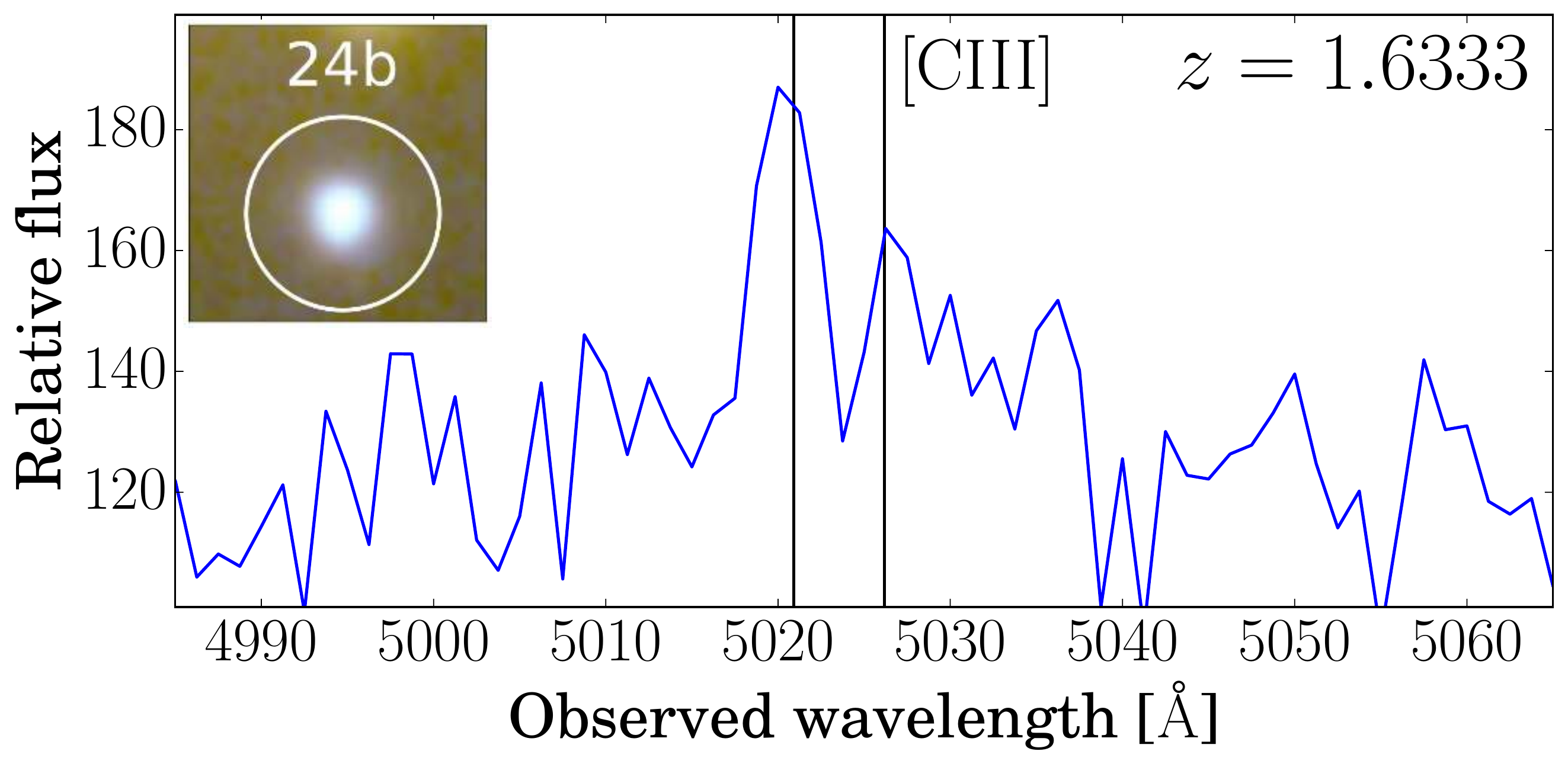}
   \includegraphics[width = 0.65\columnwidth]{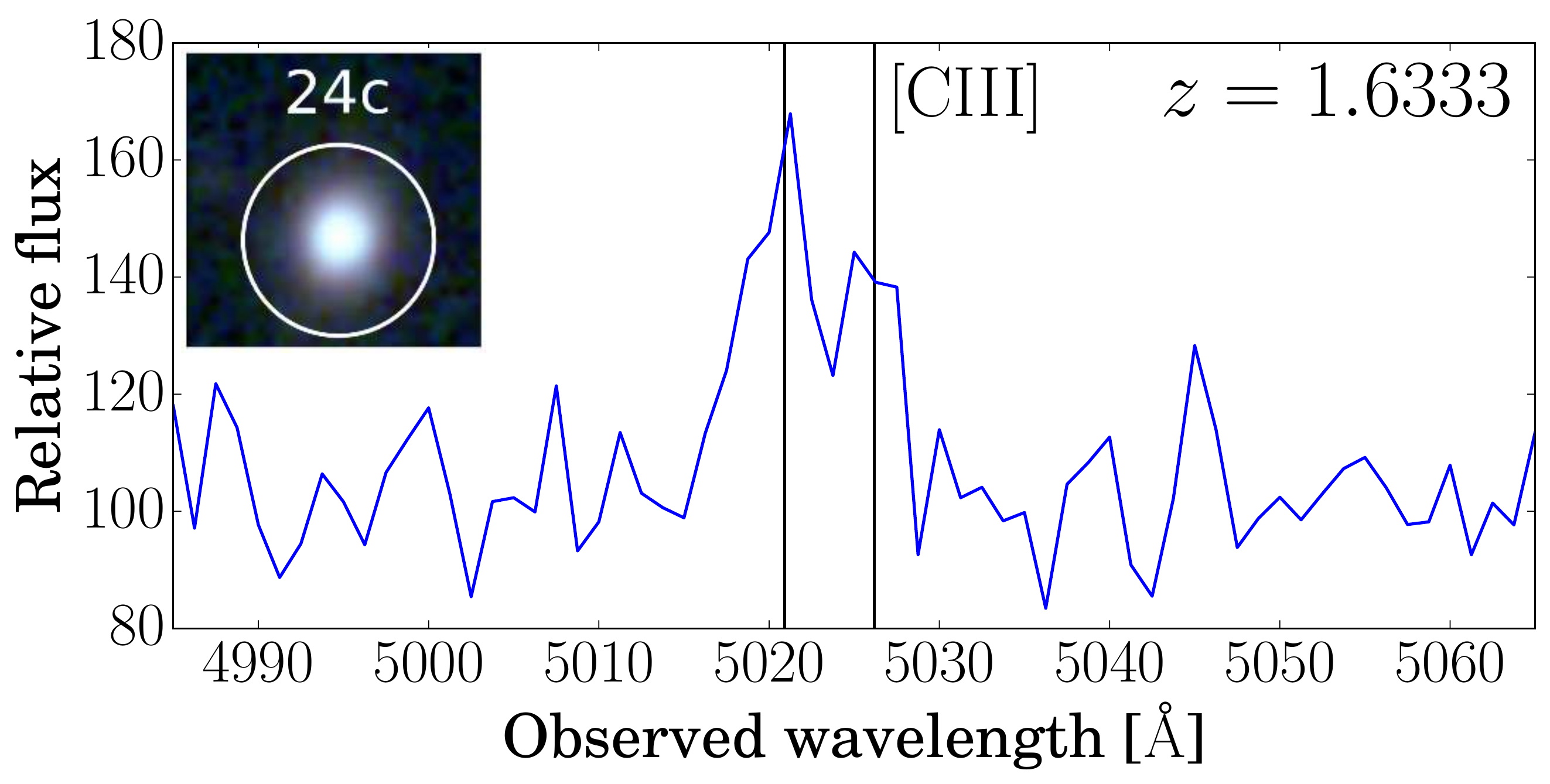}

   \includegraphics[width = 0.65\columnwidth]{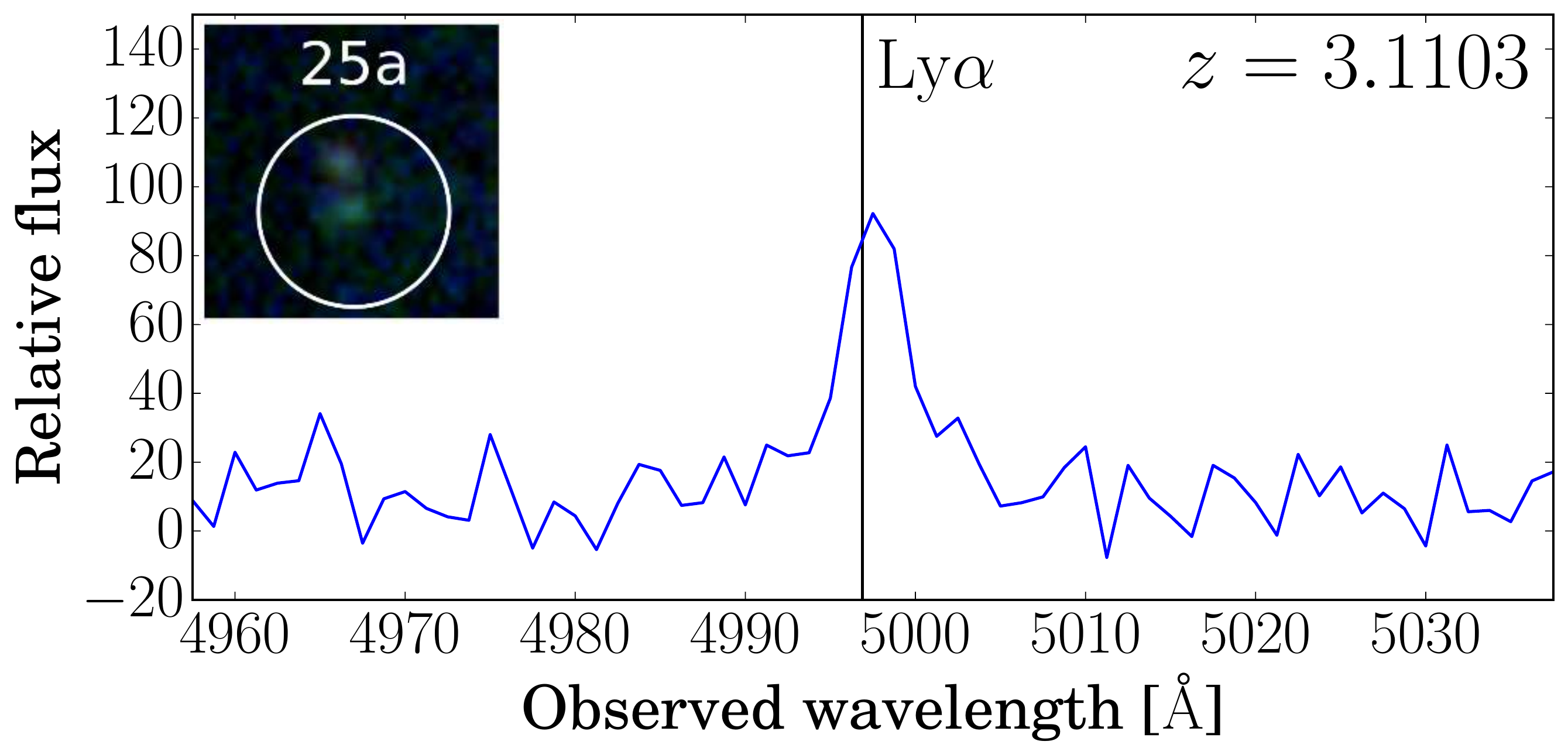}
   \includegraphics[width = 0.65\columnwidth]{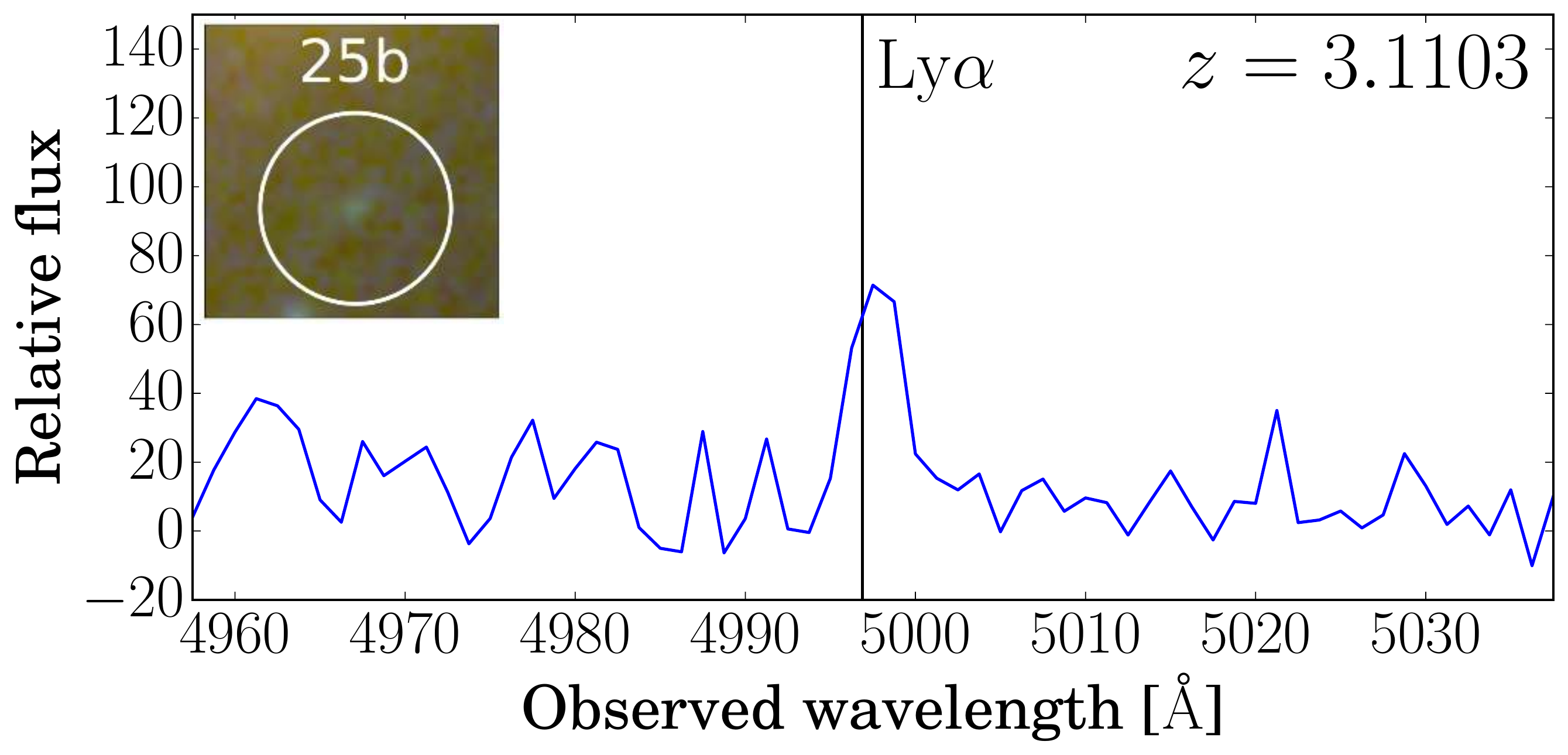}
   \includegraphics[width = 0.65\columnwidth]{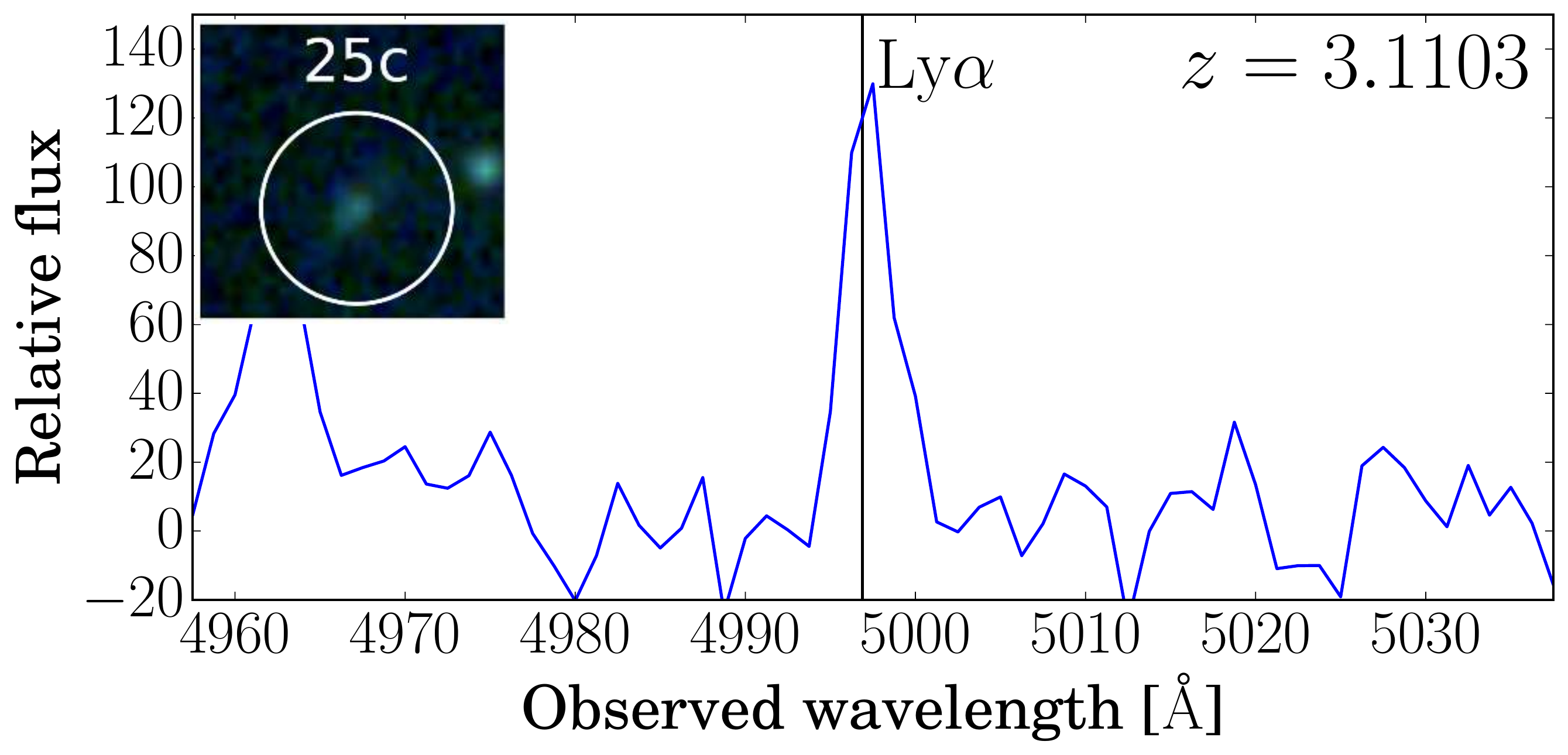}

   \includegraphics[width = 0.65\columnwidth]{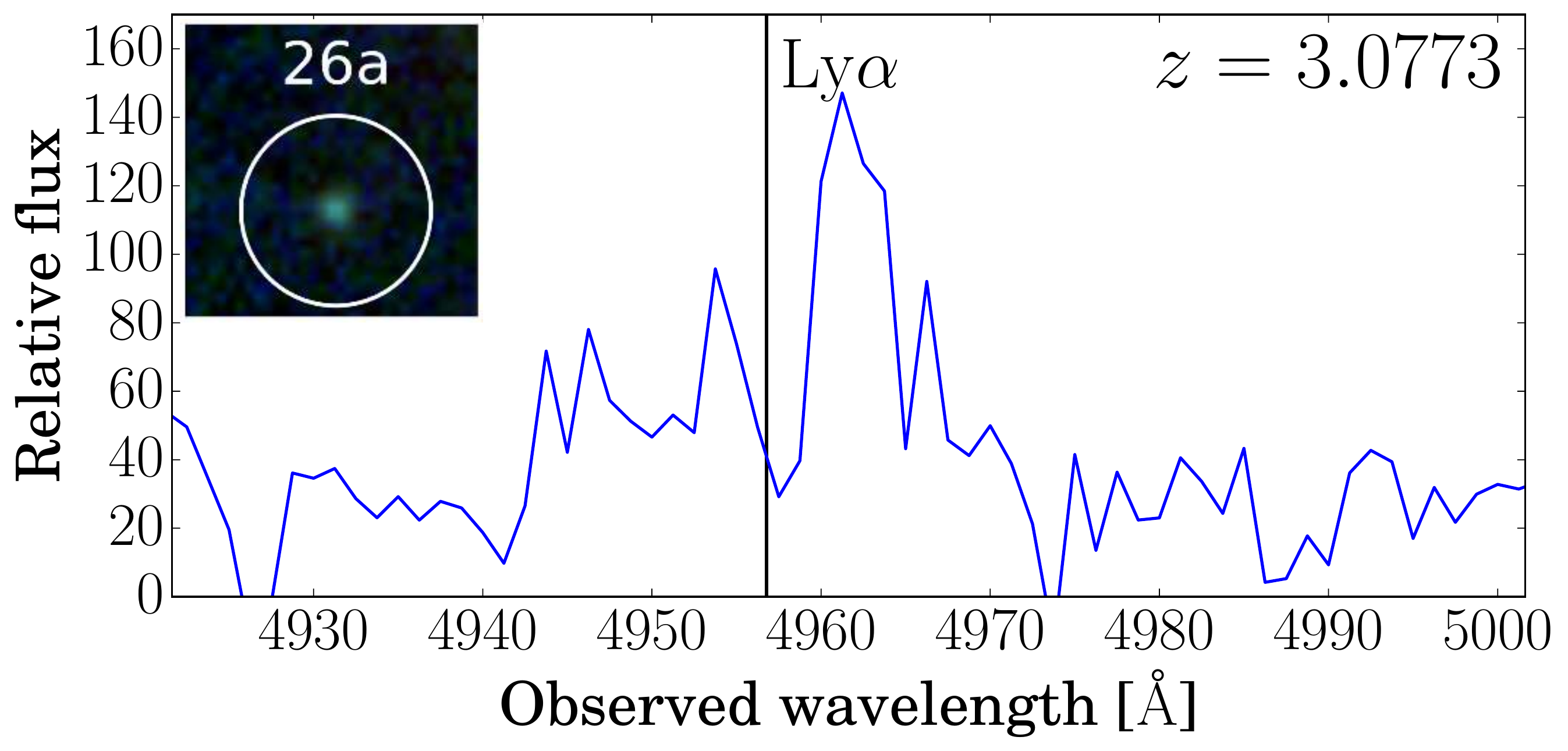}
   \includegraphics[width = 0.65\columnwidth]{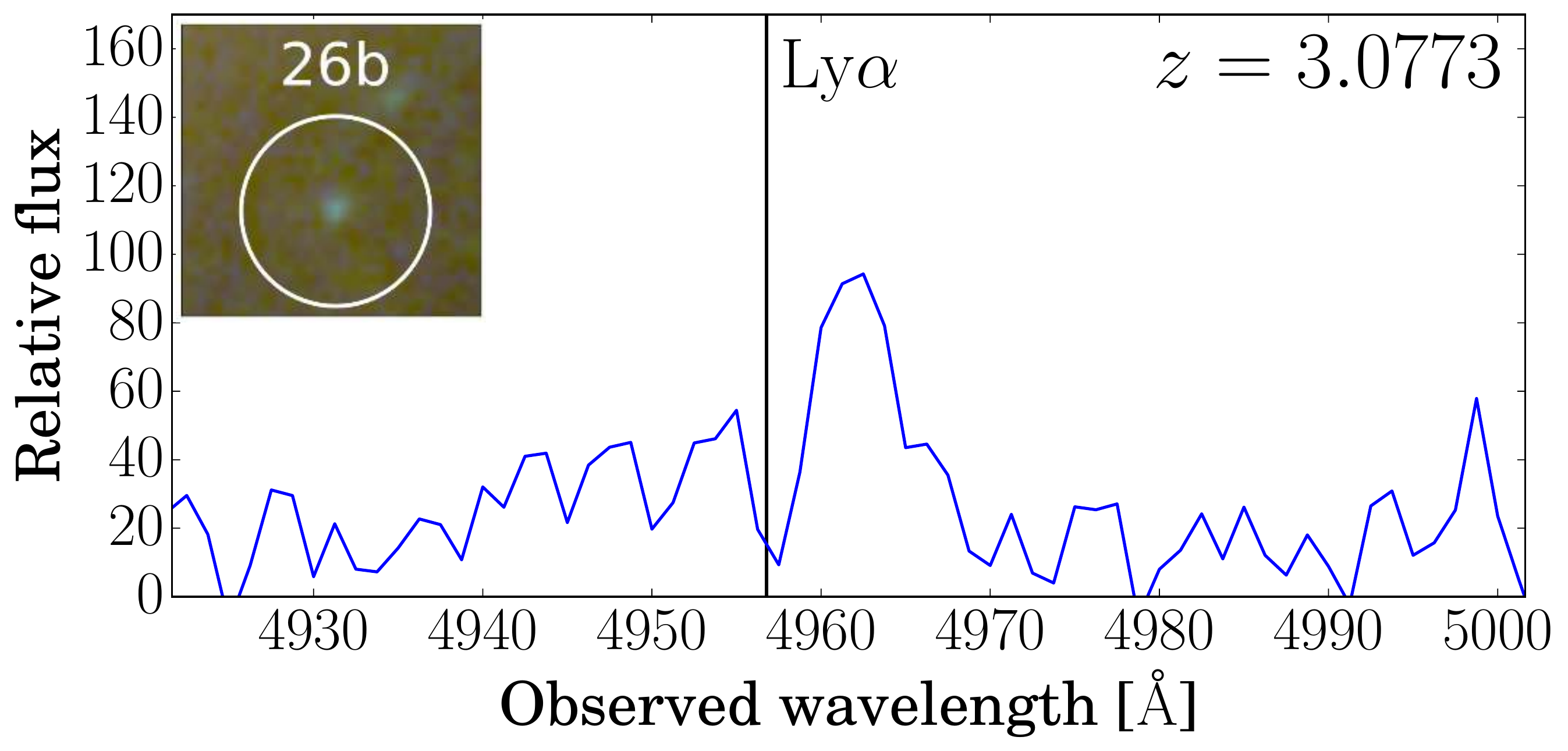}
   \includegraphics[width = 0.65\columnwidth]{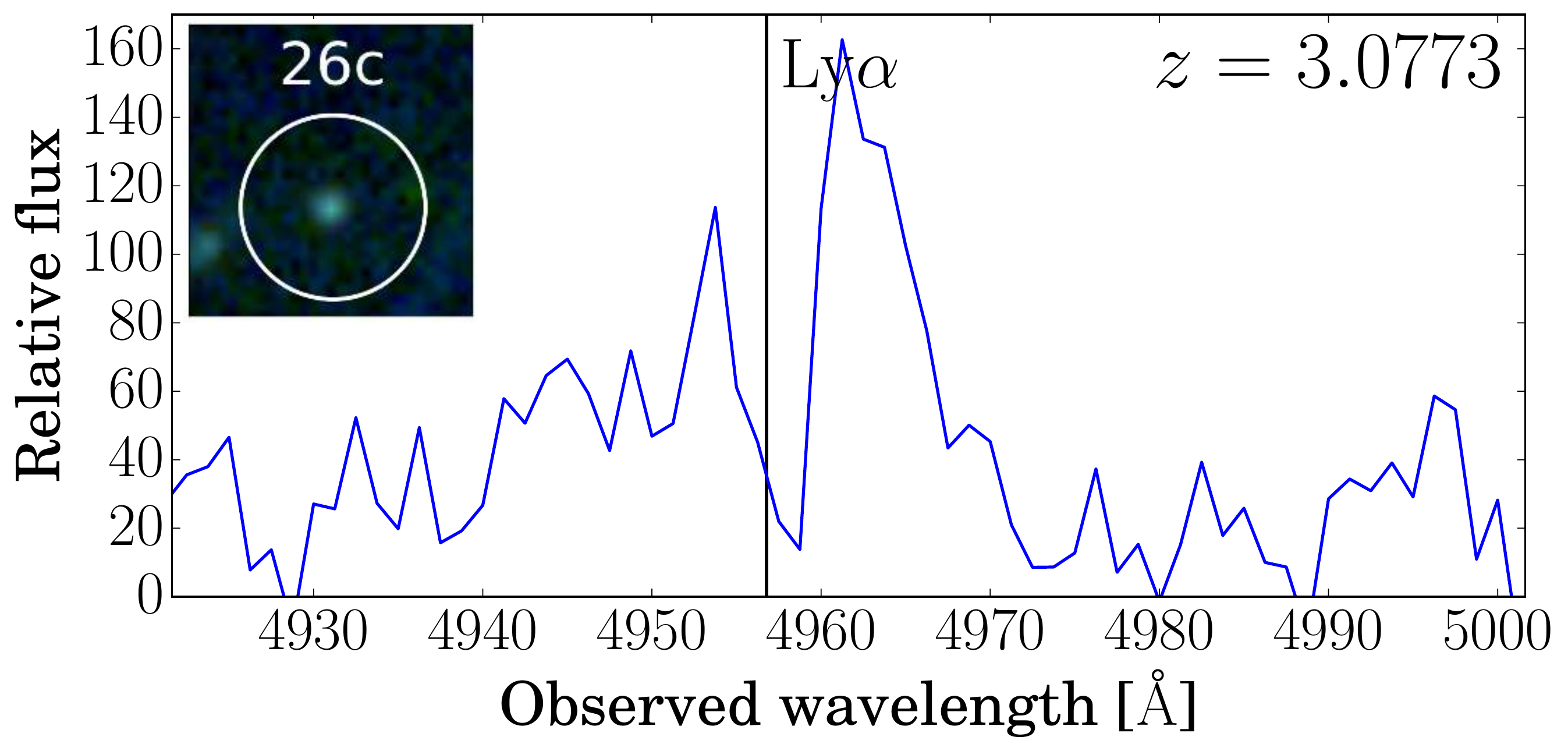}

  \caption{(Continued)}
  \label{fig:specs}
\end{figure*}

\begin{figure*}
  \ContinuedFloat
  %\centering

   \includegraphics[width = 0.65\columnwidth]{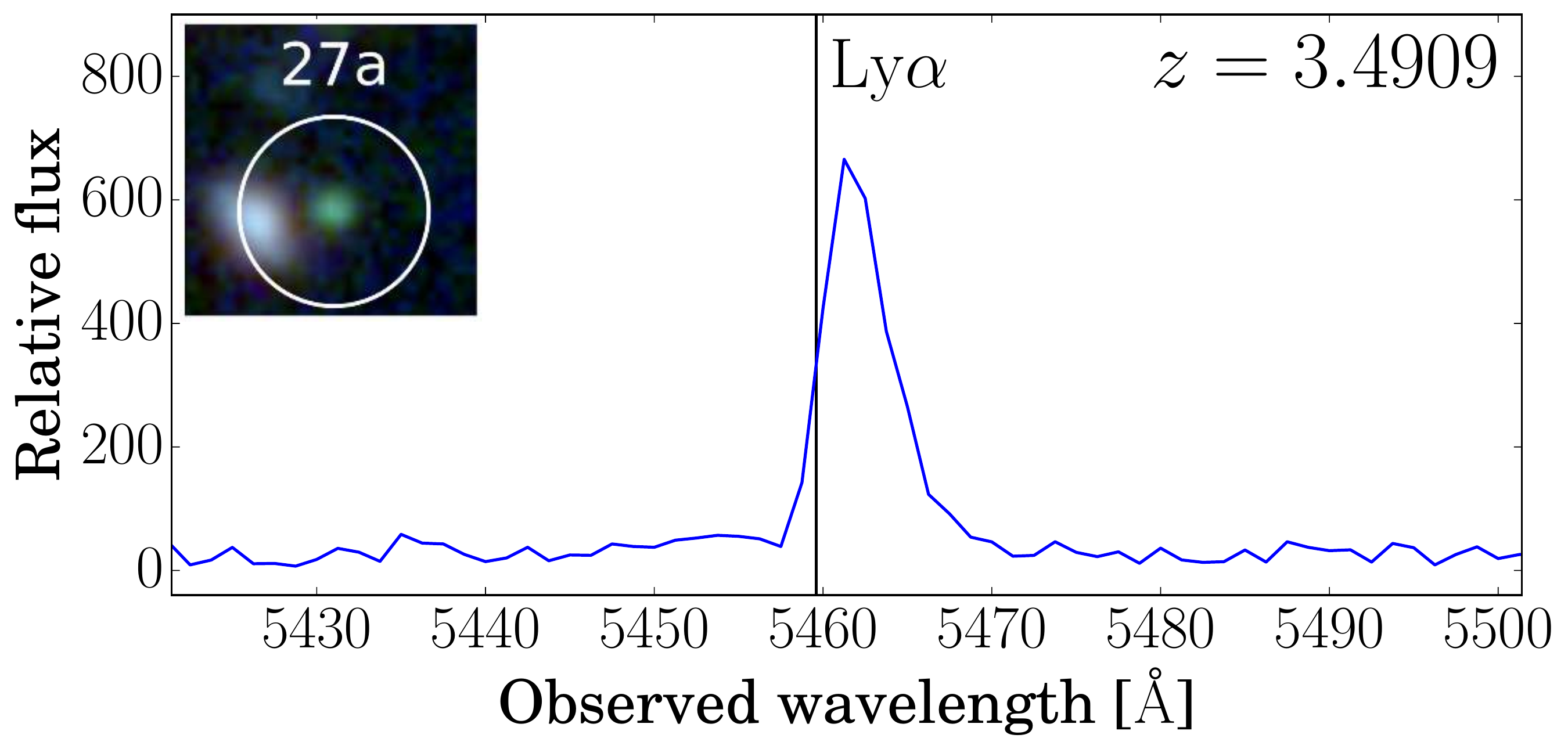}
   \includegraphics[width = 0.65\columnwidth]{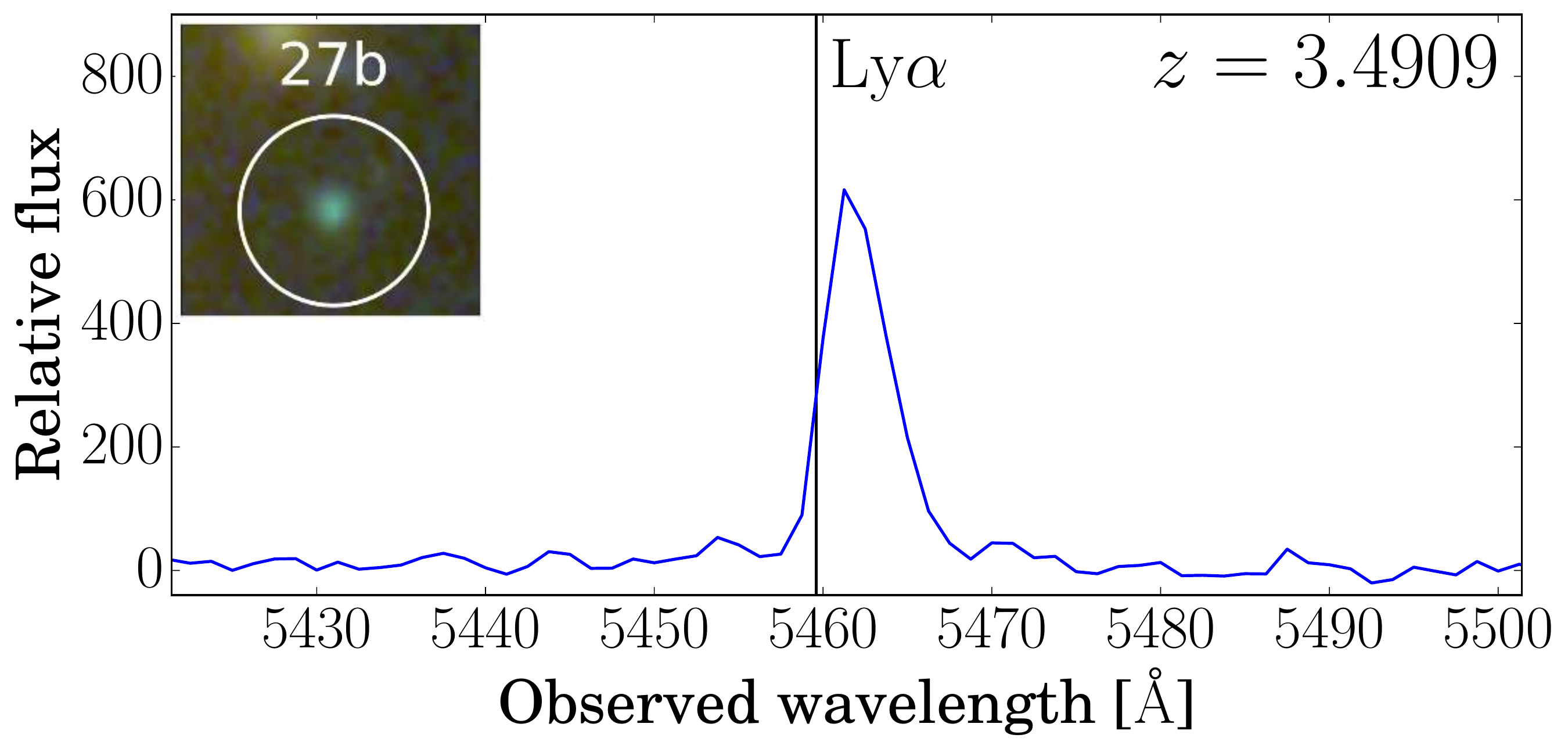}
   \includegraphics[width = 0.65\columnwidth]{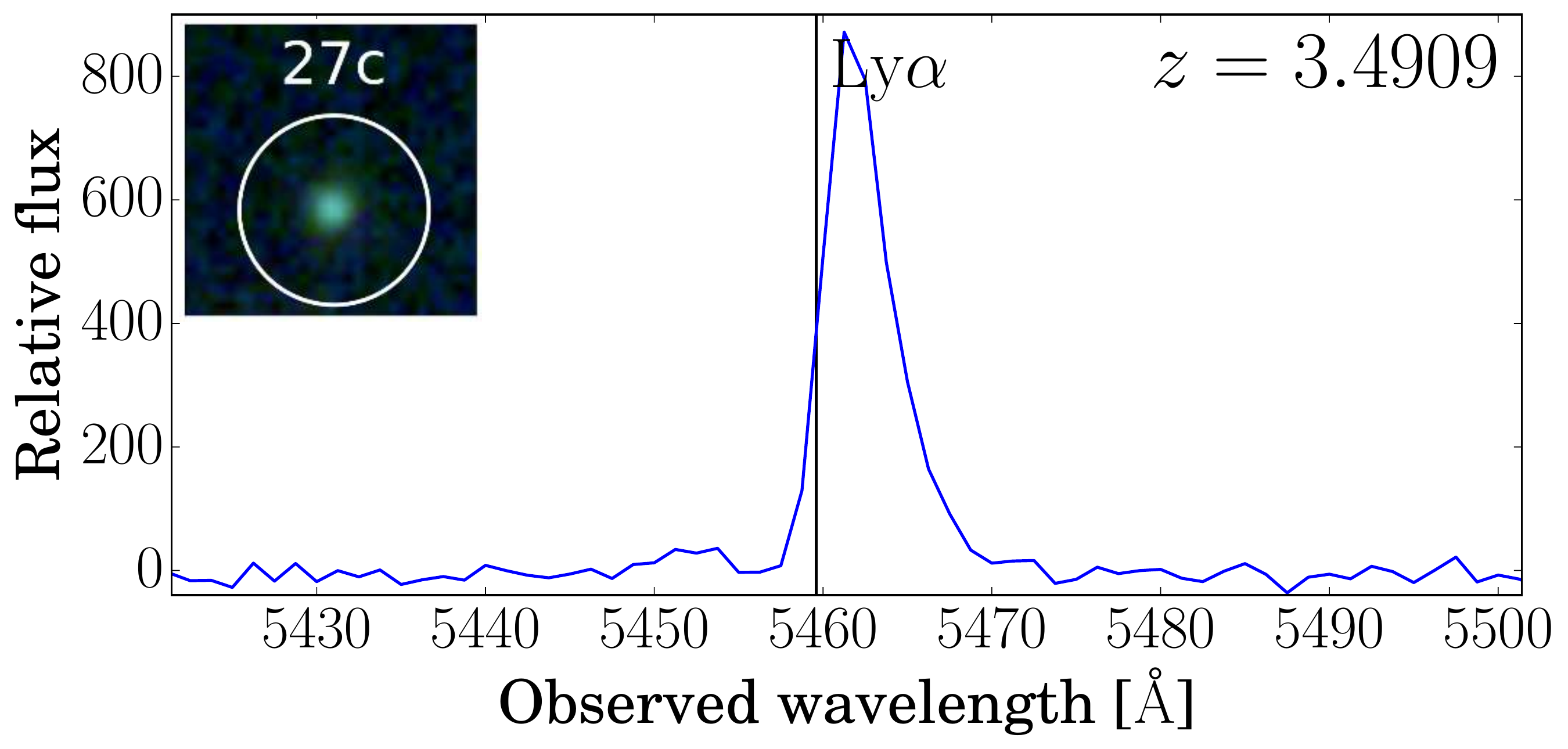}

   \includegraphics[width = 0.65\columnwidth]{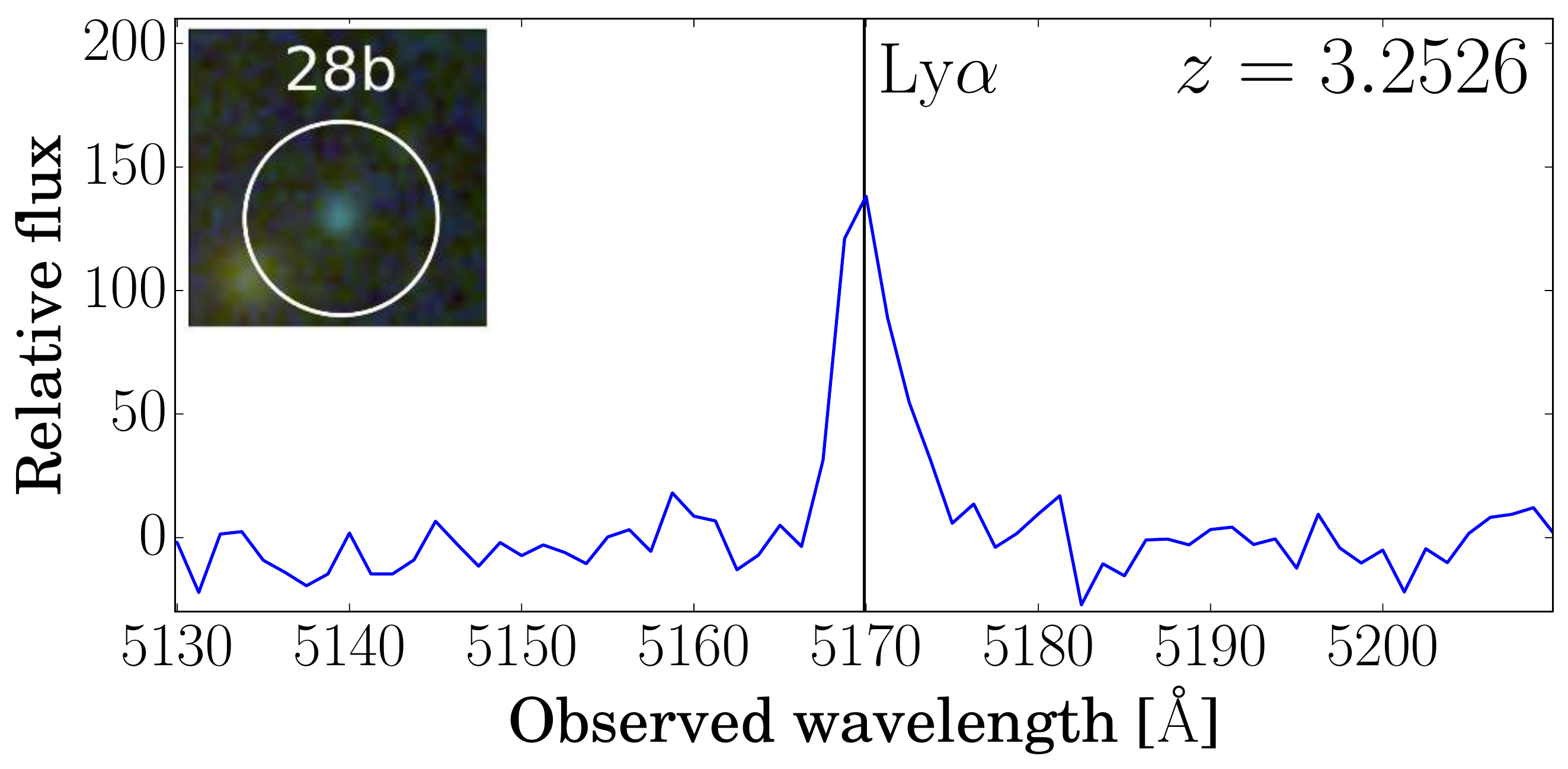}
   \includegraphics[width = 0.65\columnwidth]{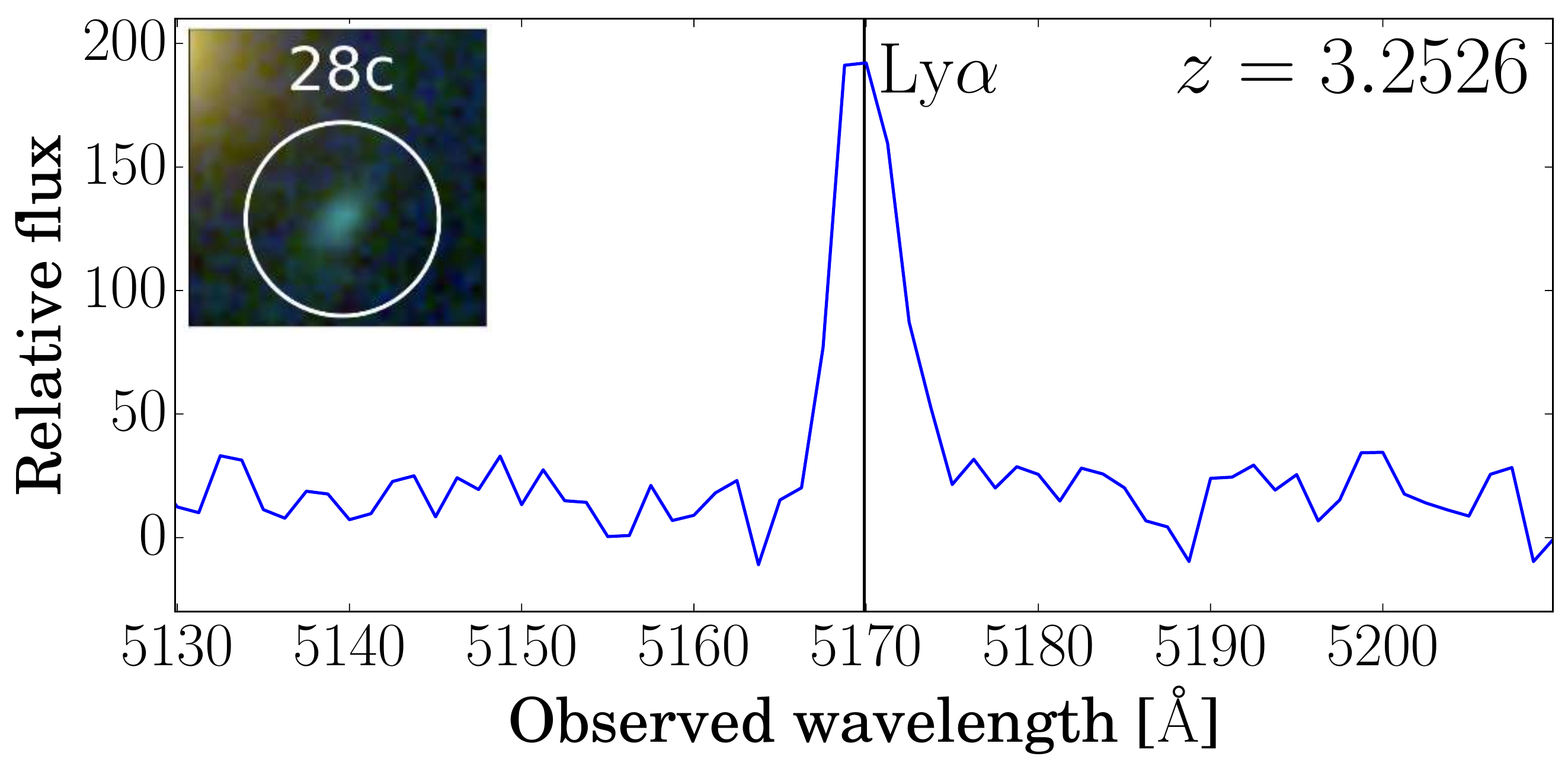}

   \includegraphics[width = 0.65\columnwidth]{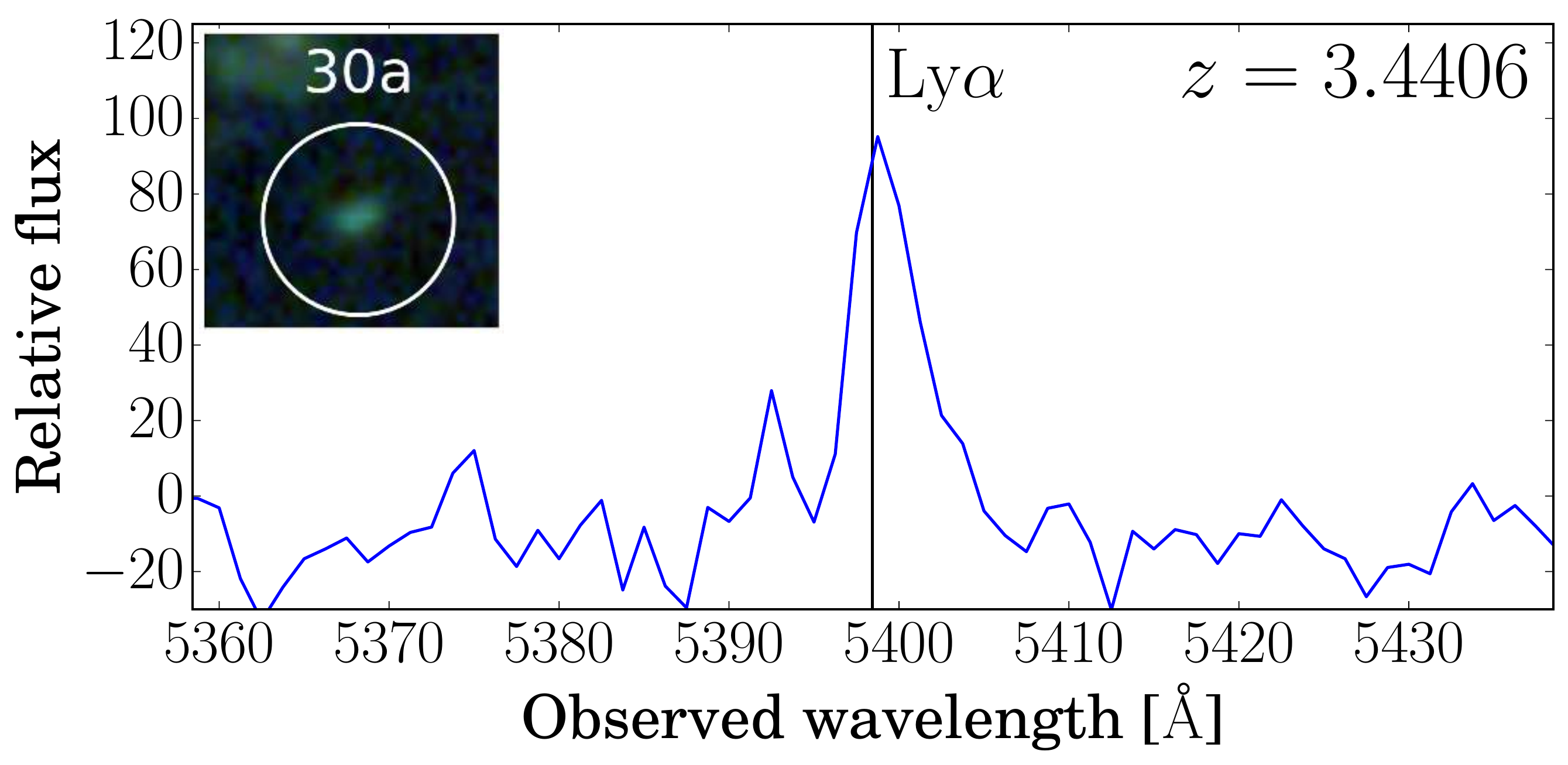}
   \includegraphics[width = 0.65\columnwidth]{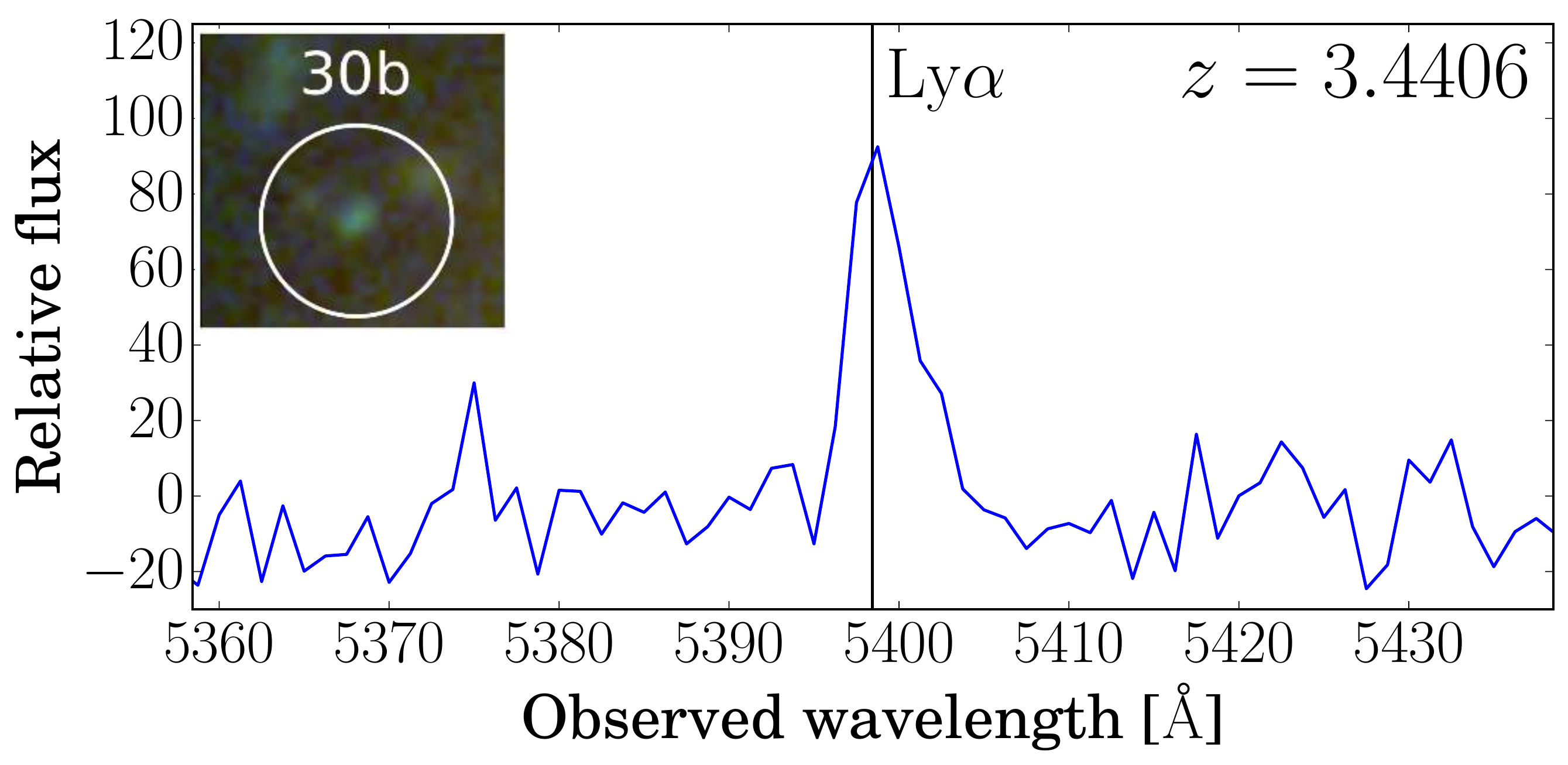}
   \includegraphics[width = 0.65\columnwidth]{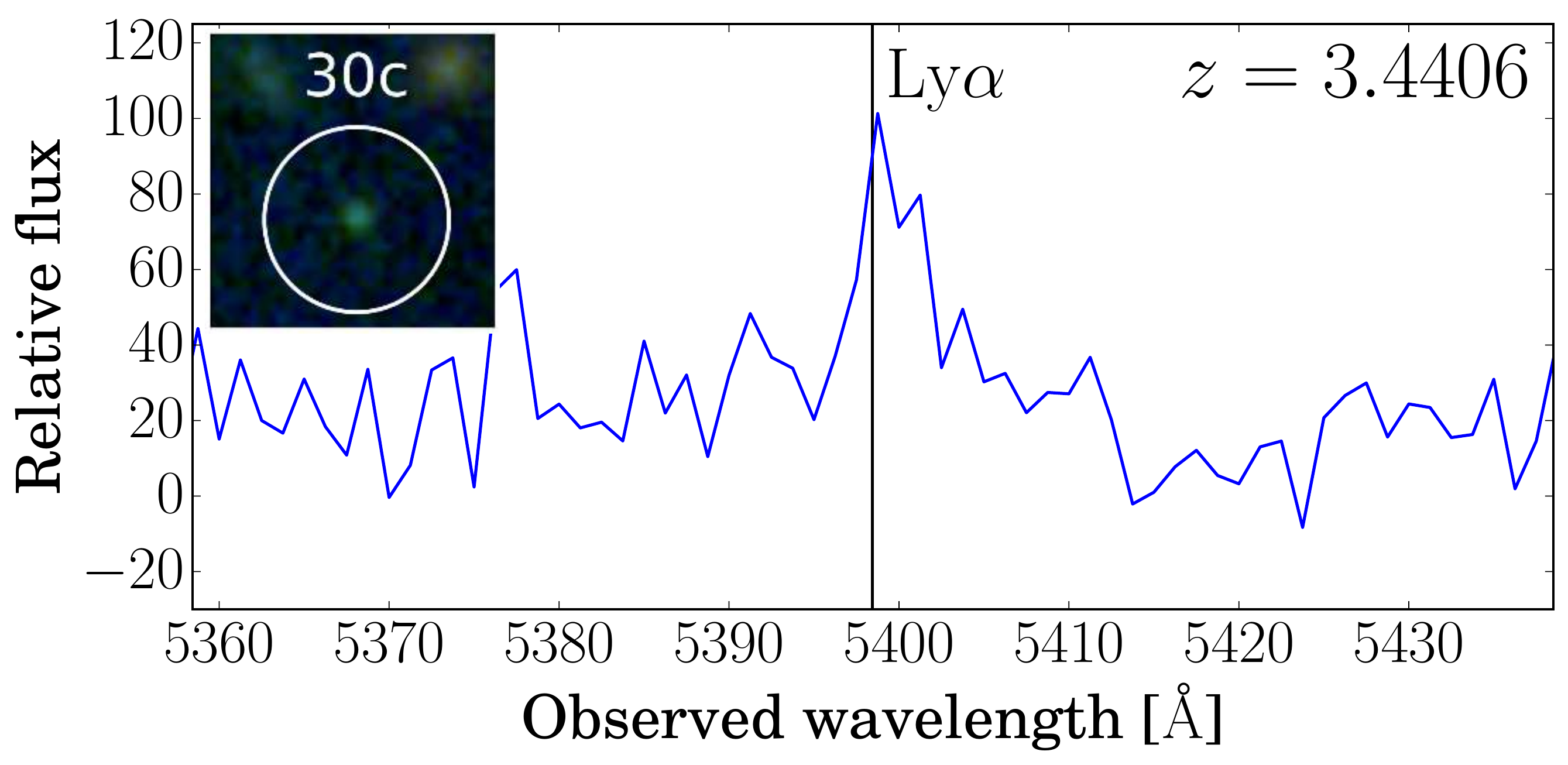}

   \includegraphics[width = 0.65\columnwidth]{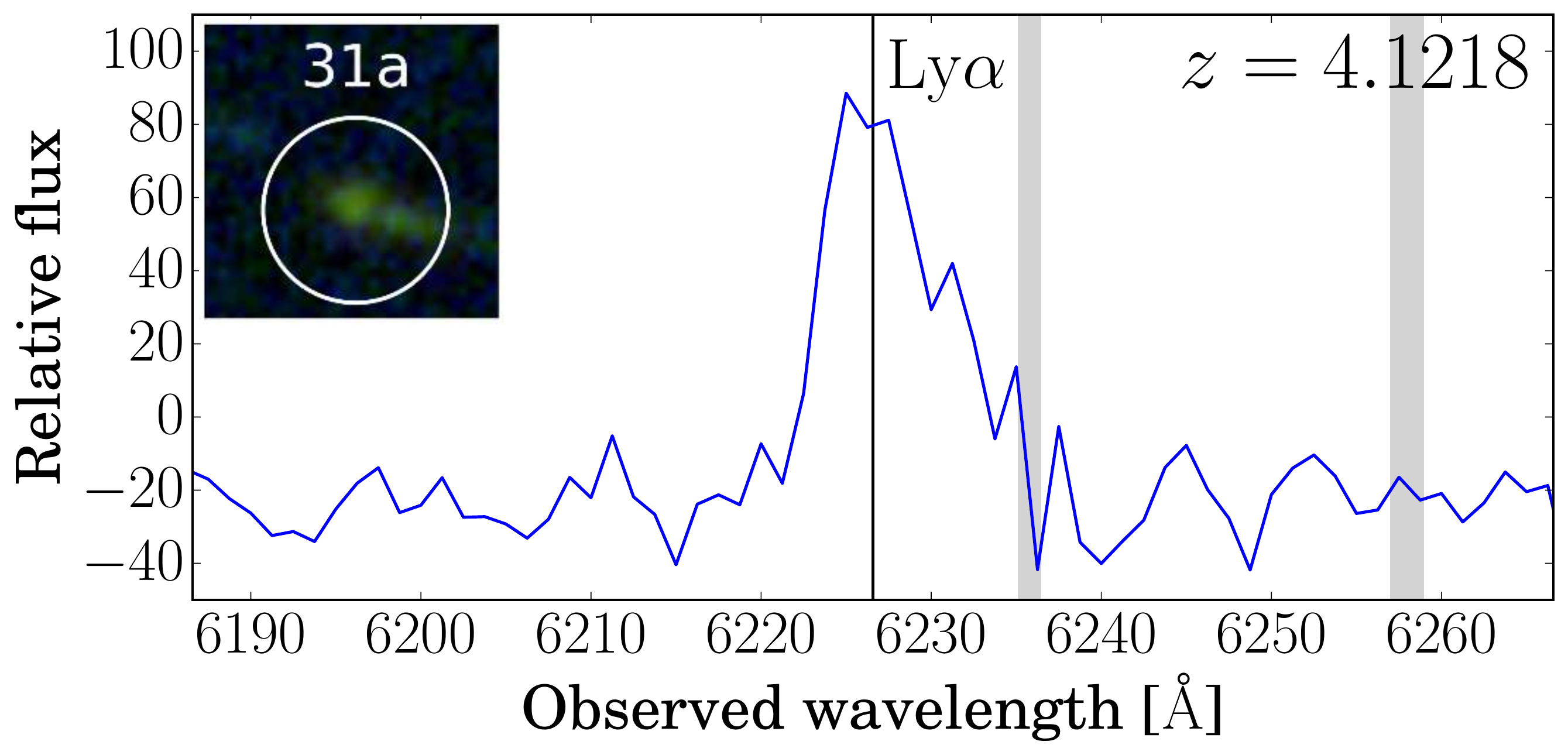}
   \includegraphics[width = 0.65\columnwidth]{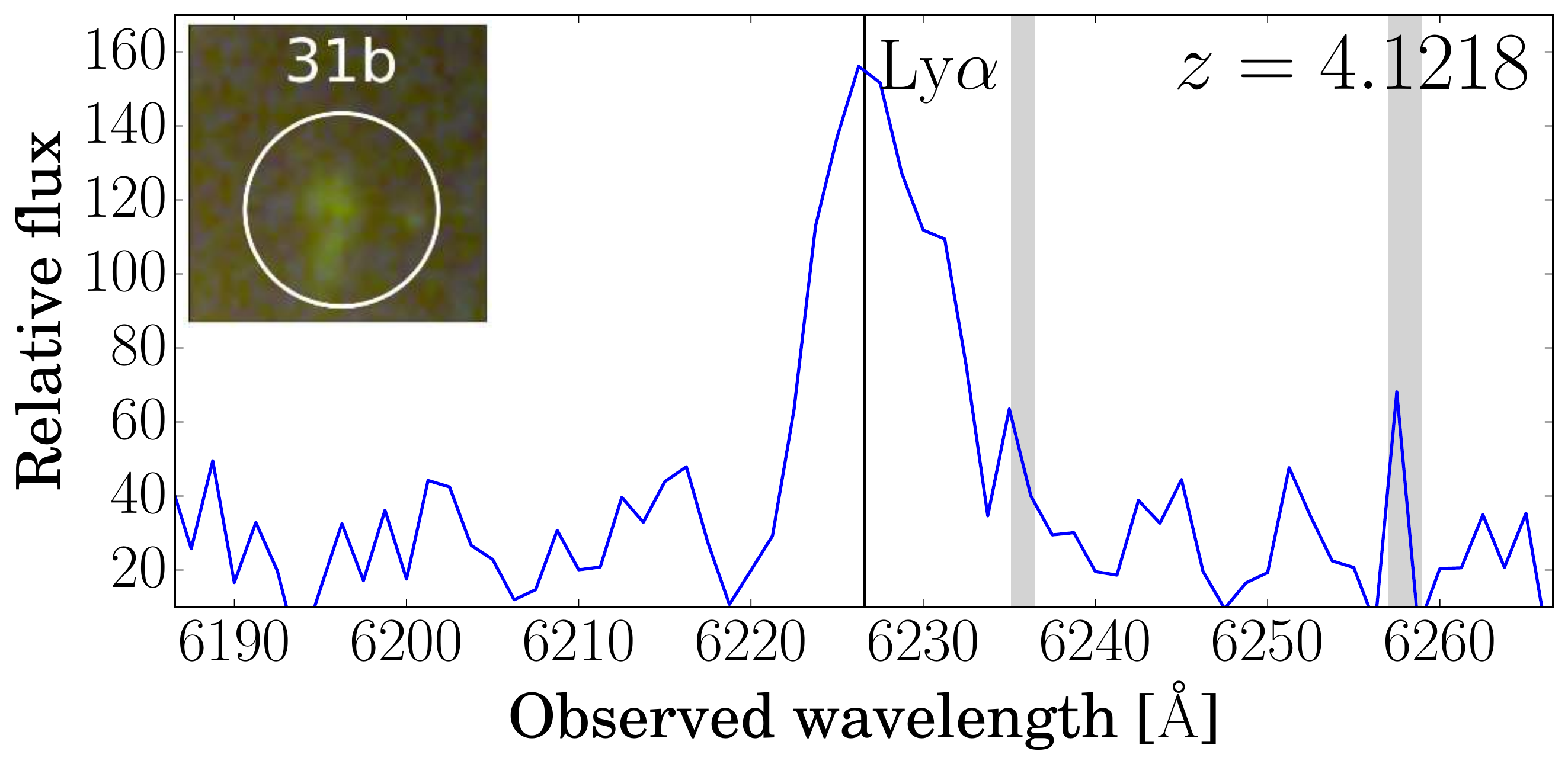}
   \includegraphics[width = 0.65\columnwidth]{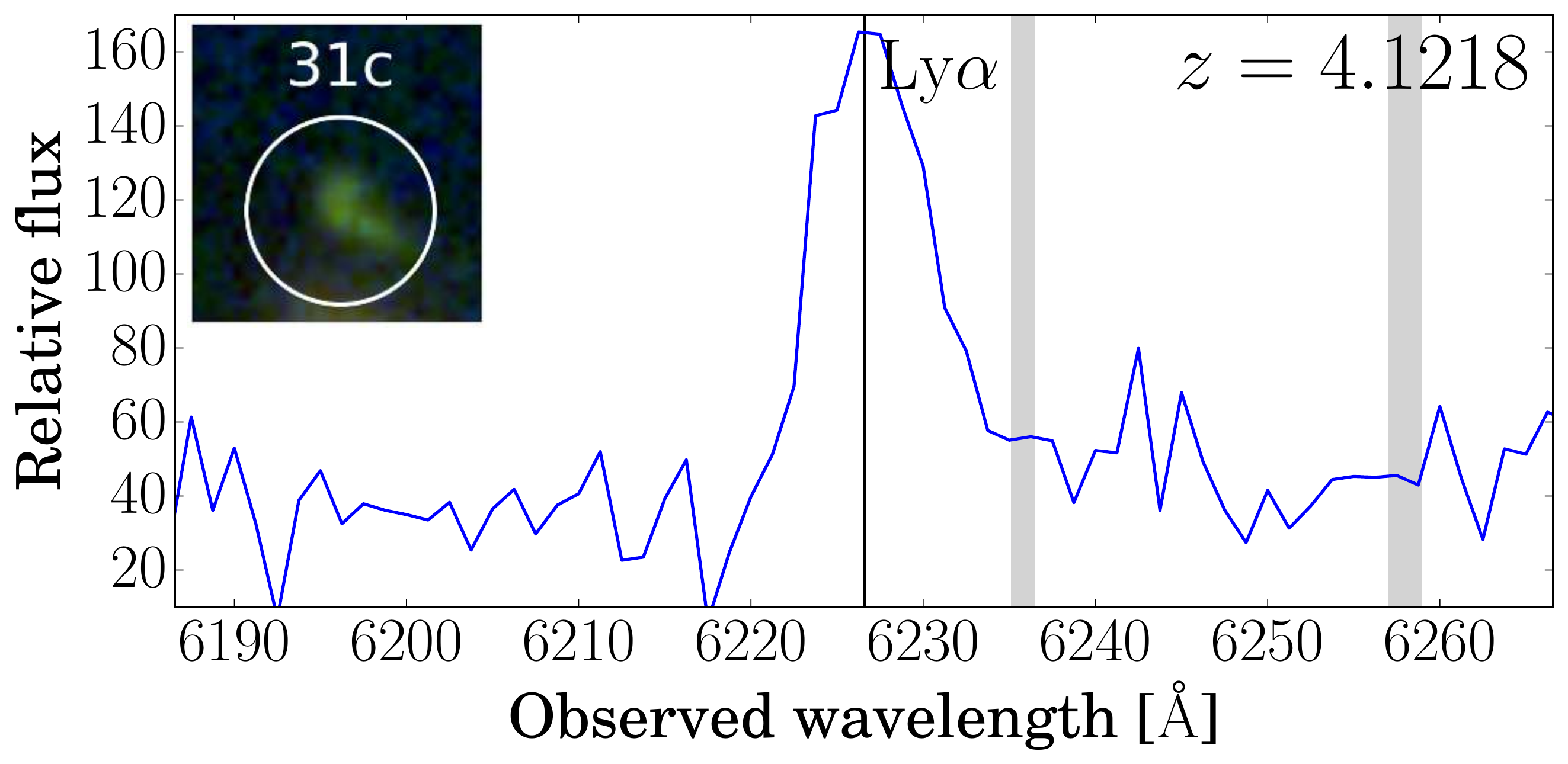}

   \includegraphics[width = 0.65\columnwidth]{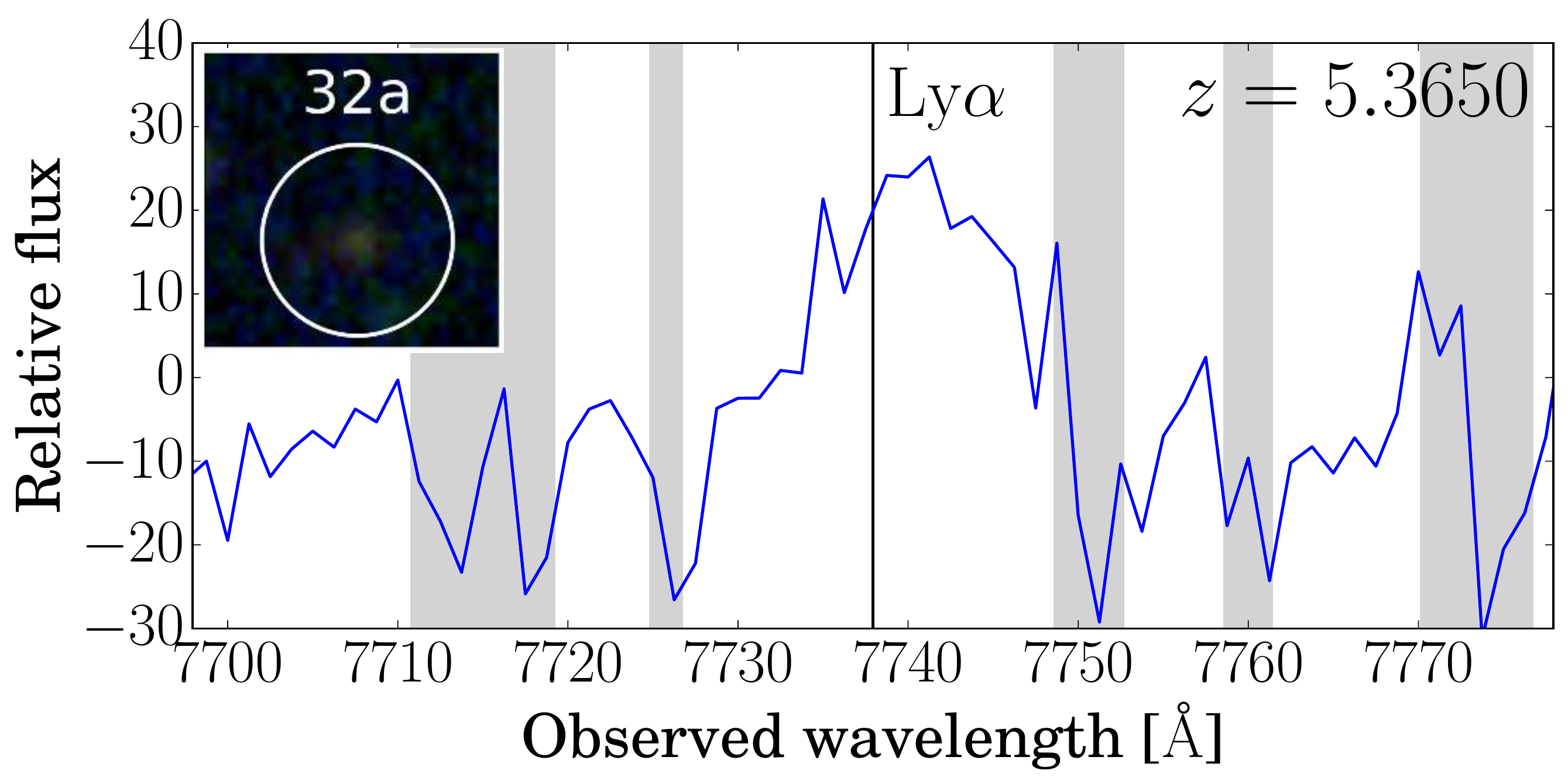}
   \includegraphics[width = 0.65\columnwidth]{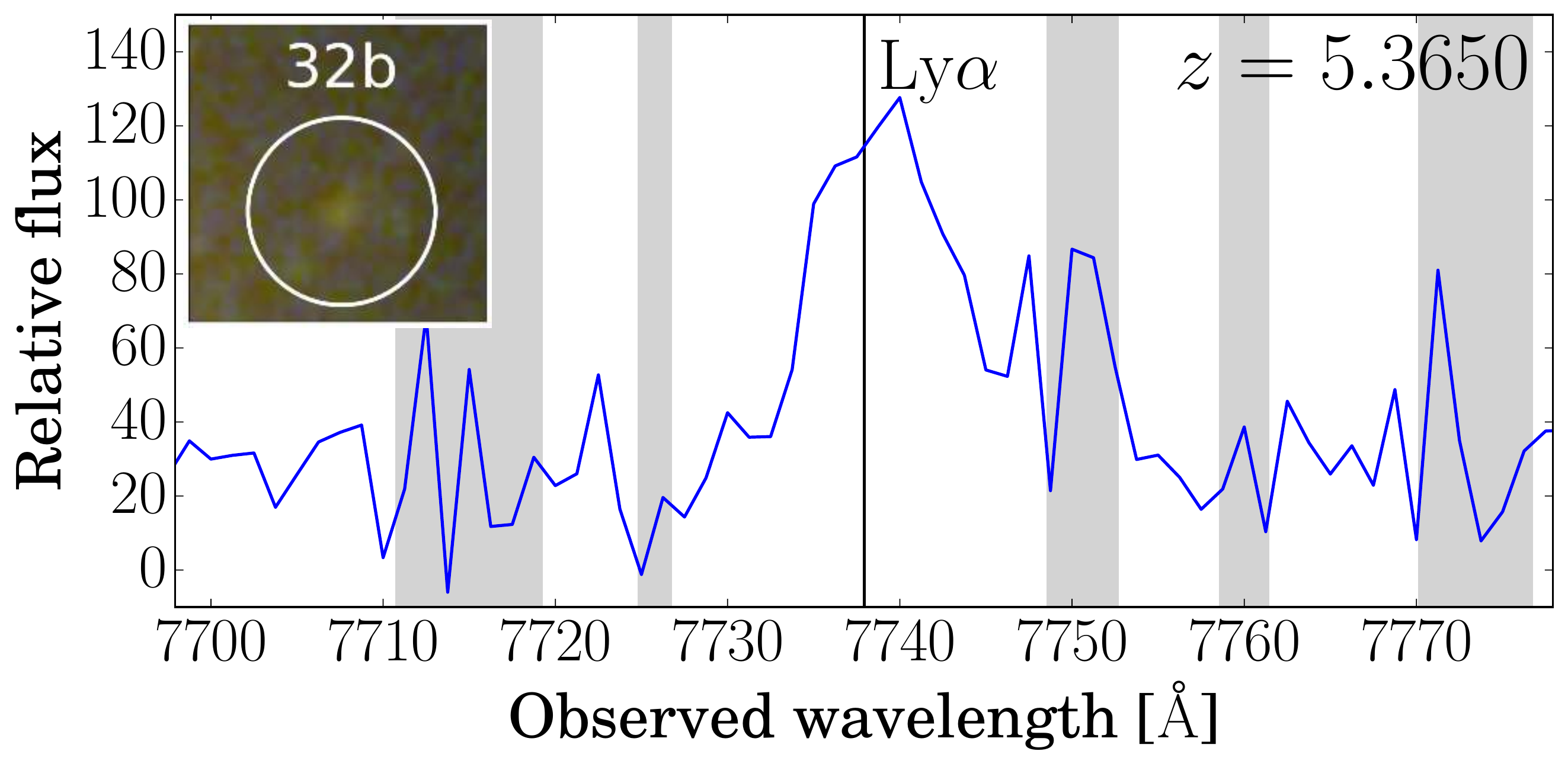}
   \includegraphics[width = 0.65\columnwidth]{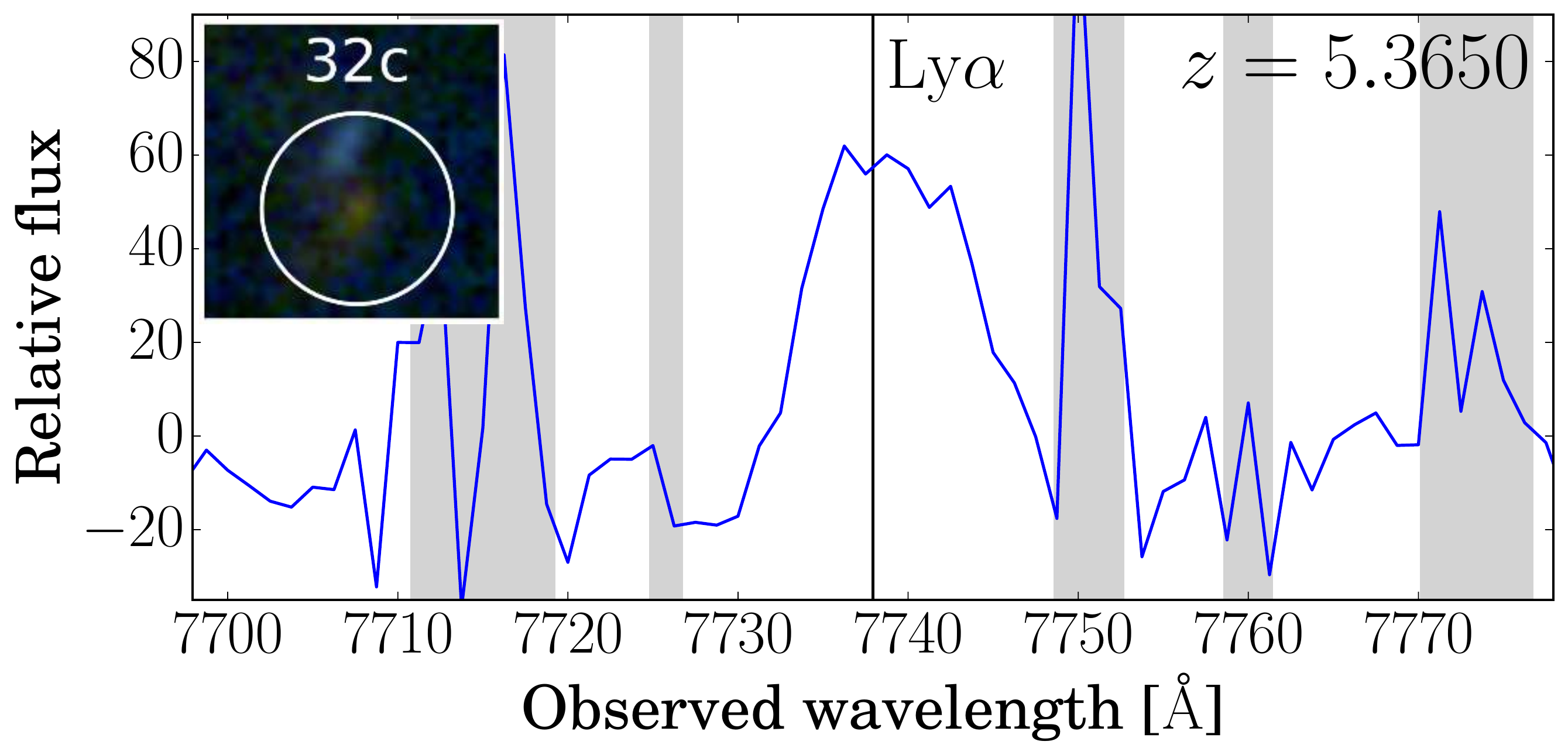}

   \includegraphics[width = 0.65\columnwidth]{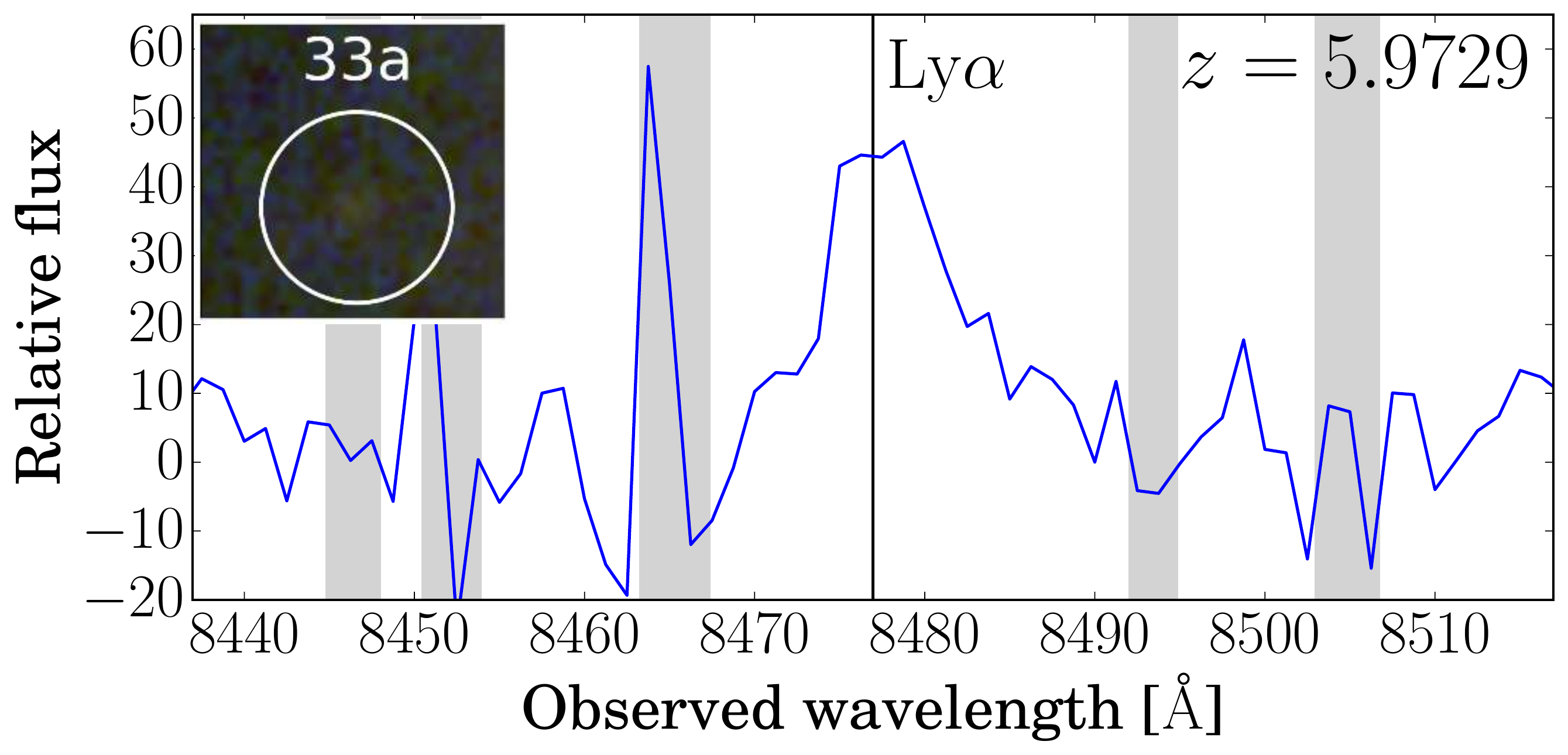}
   \includegraphics[width = 0.65\columnwidth]{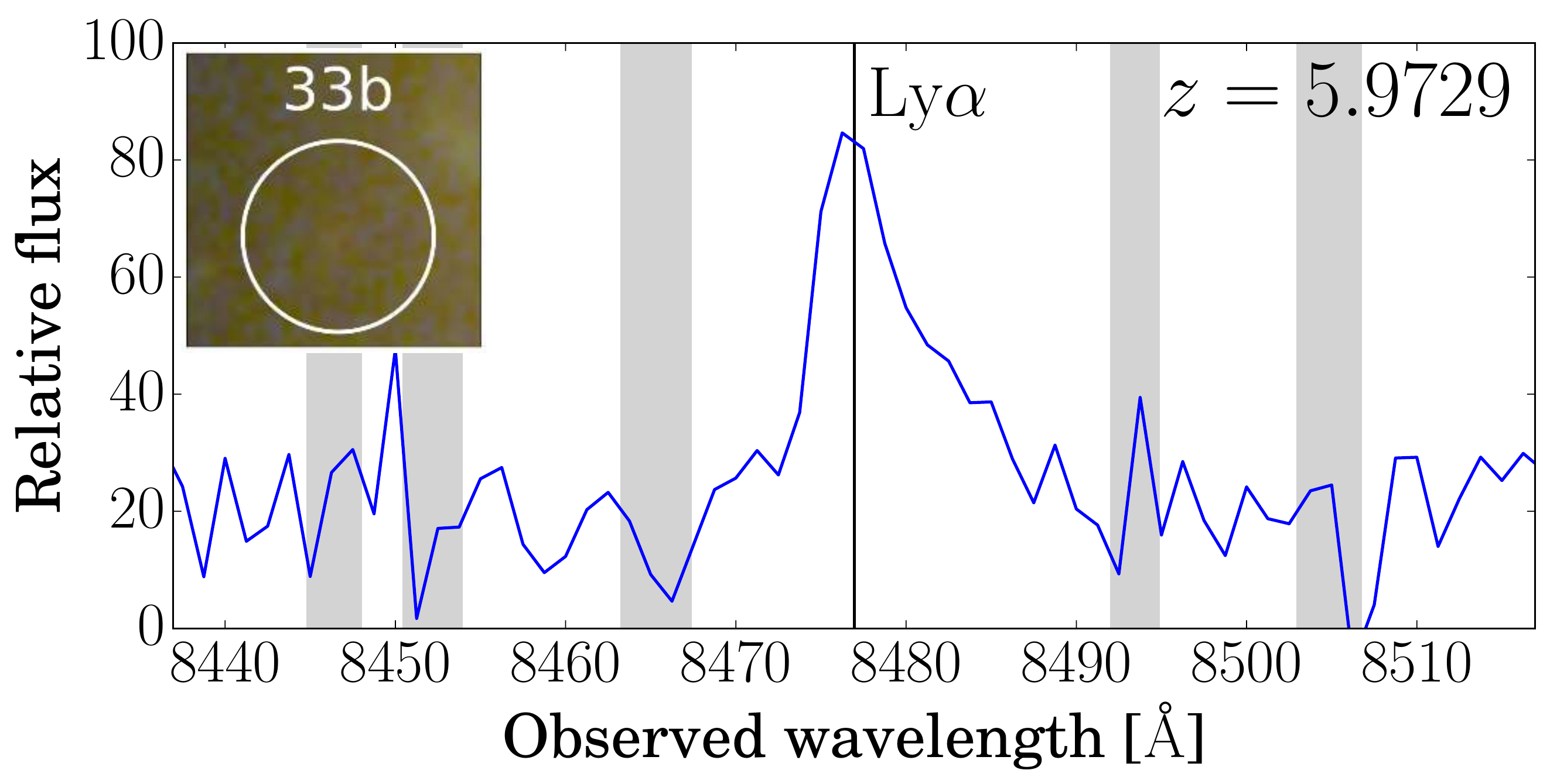}

   \includegraphics[width = 0.65\columnwidth]{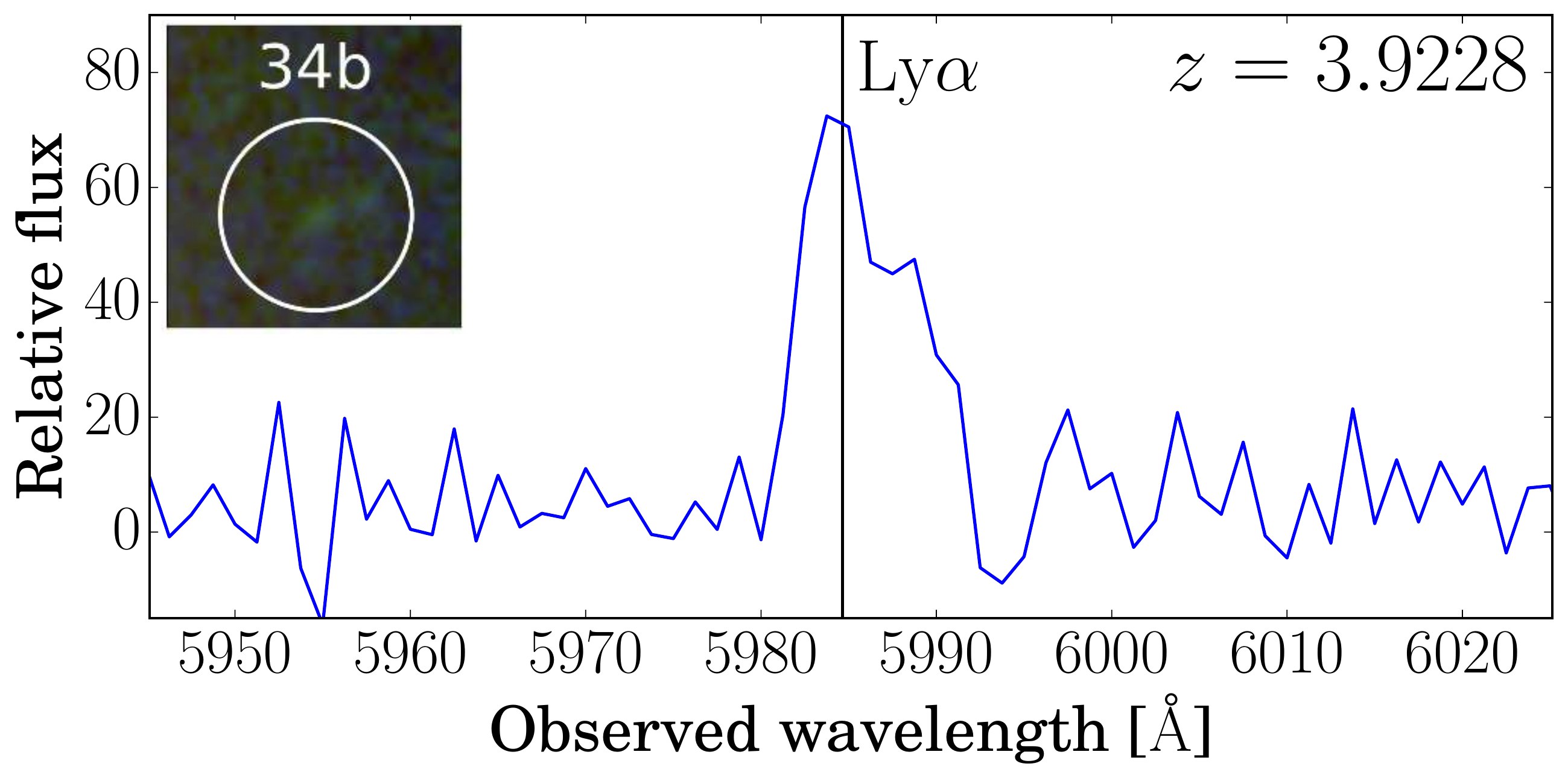}
   \includegraphics[width = 0.65\columnwidth]{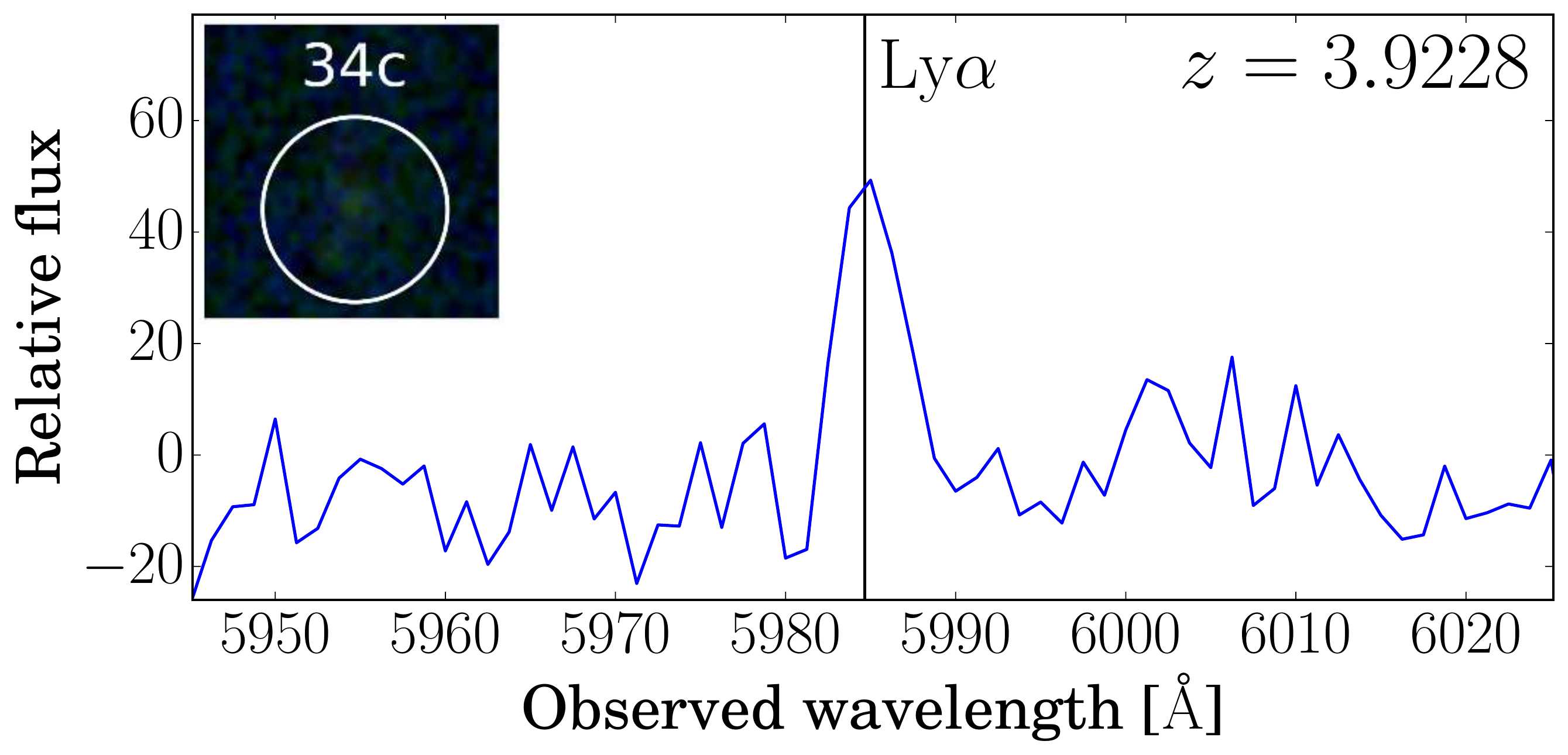}

   \includegraphics[width = 0.65\columnwidth]{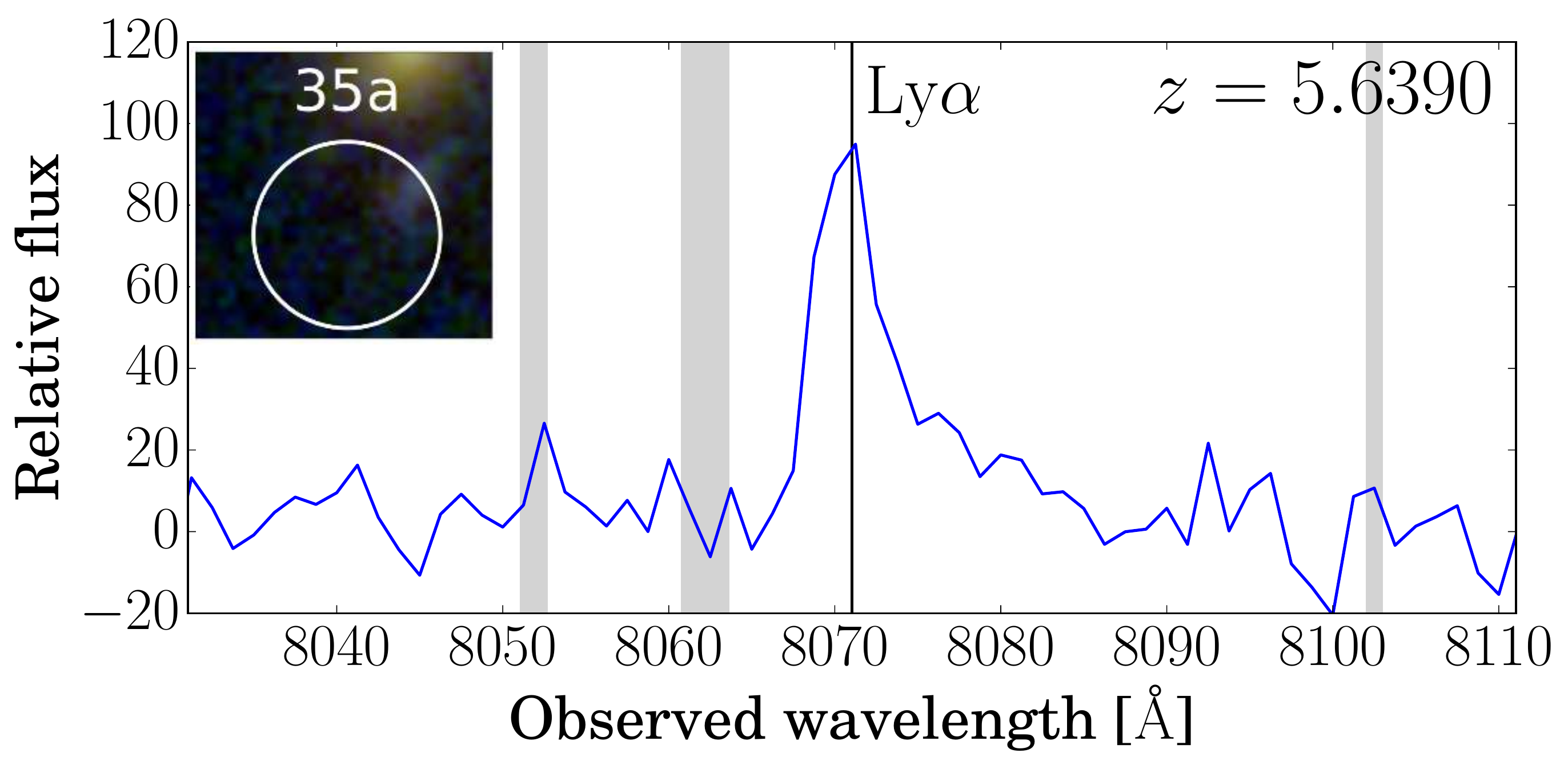}
   \includegraphics[width = 0.65\columnwidth]{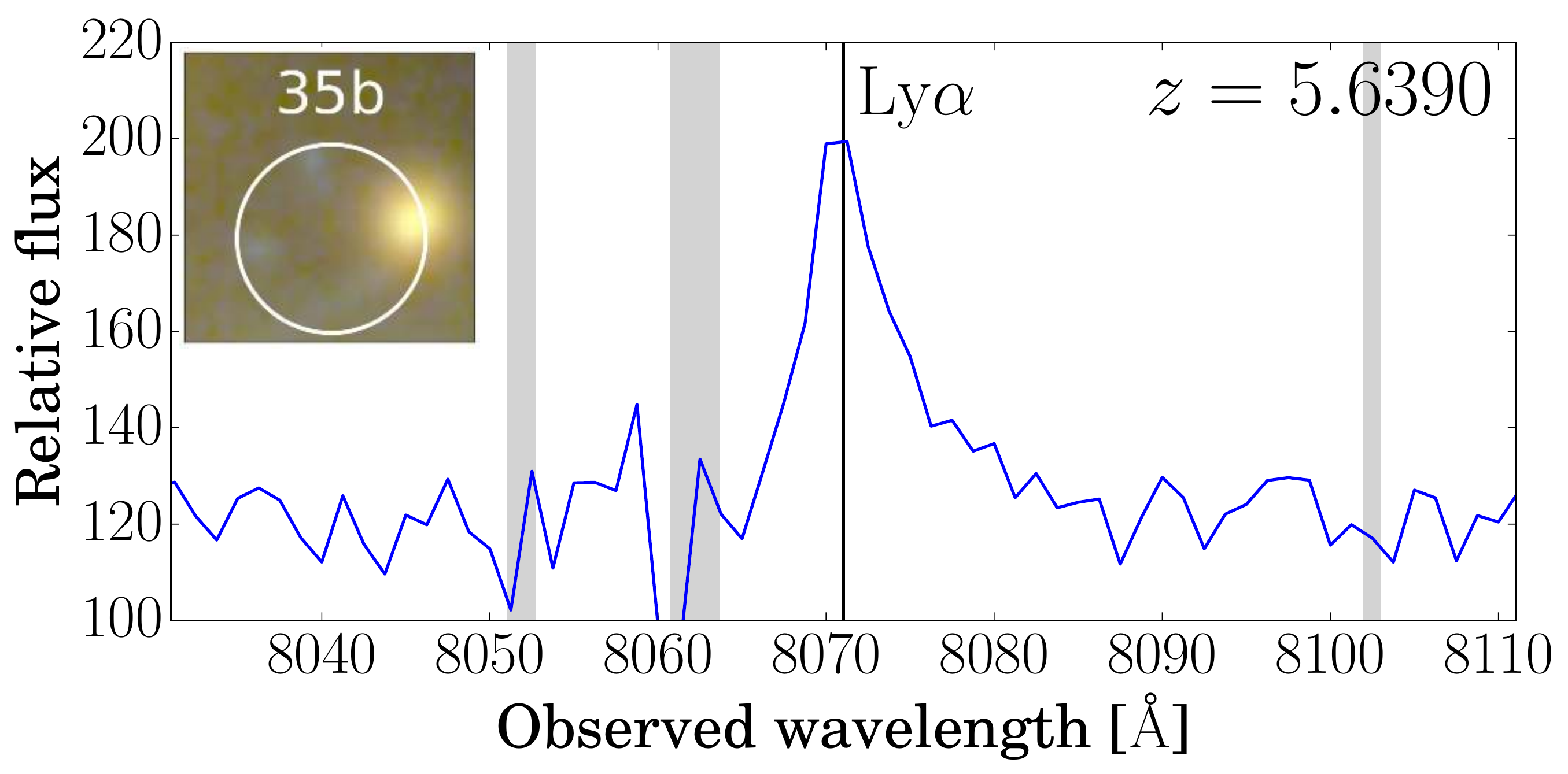}
   \includegraphics[width = 0.65\columnwidth]{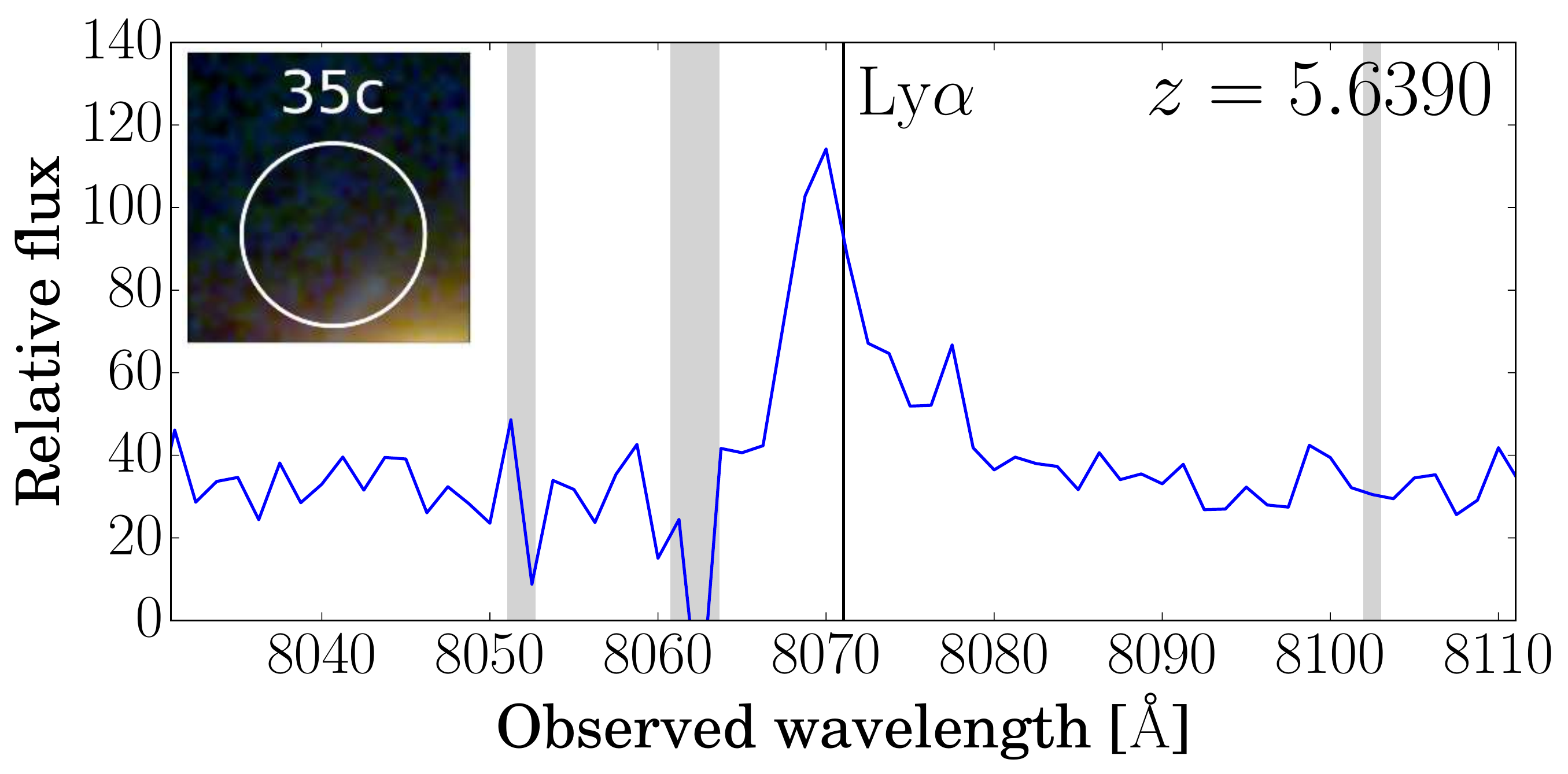}

  \caption{(Continued)}
  \label{fig:specs}
\end{figure*}

\begin{figure*}
  \ContinuedFloat

   \includegraphics[width = 0.65\columnwidth]{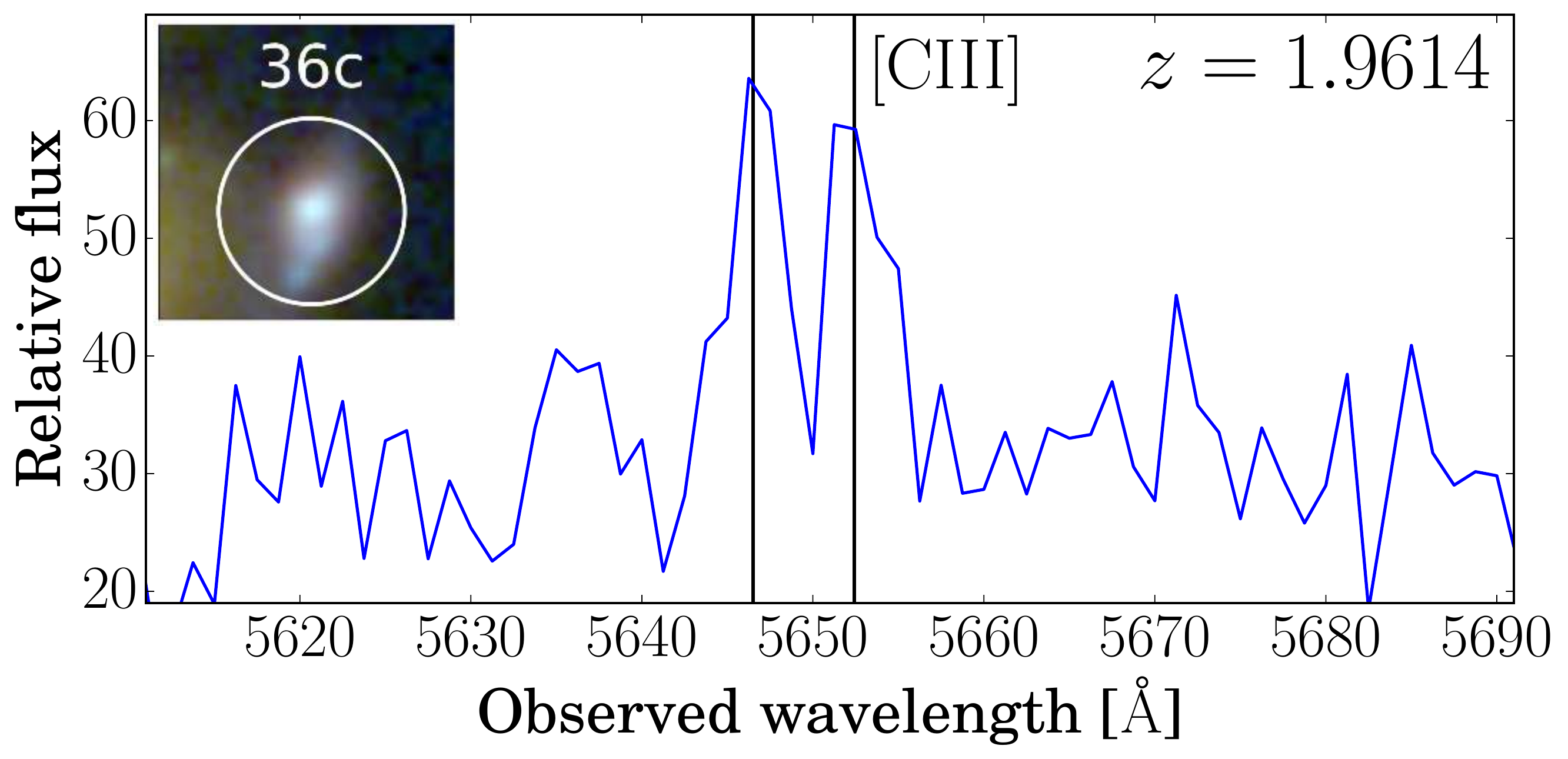}

   \includegraphics[width = 0.65\columnwidth]{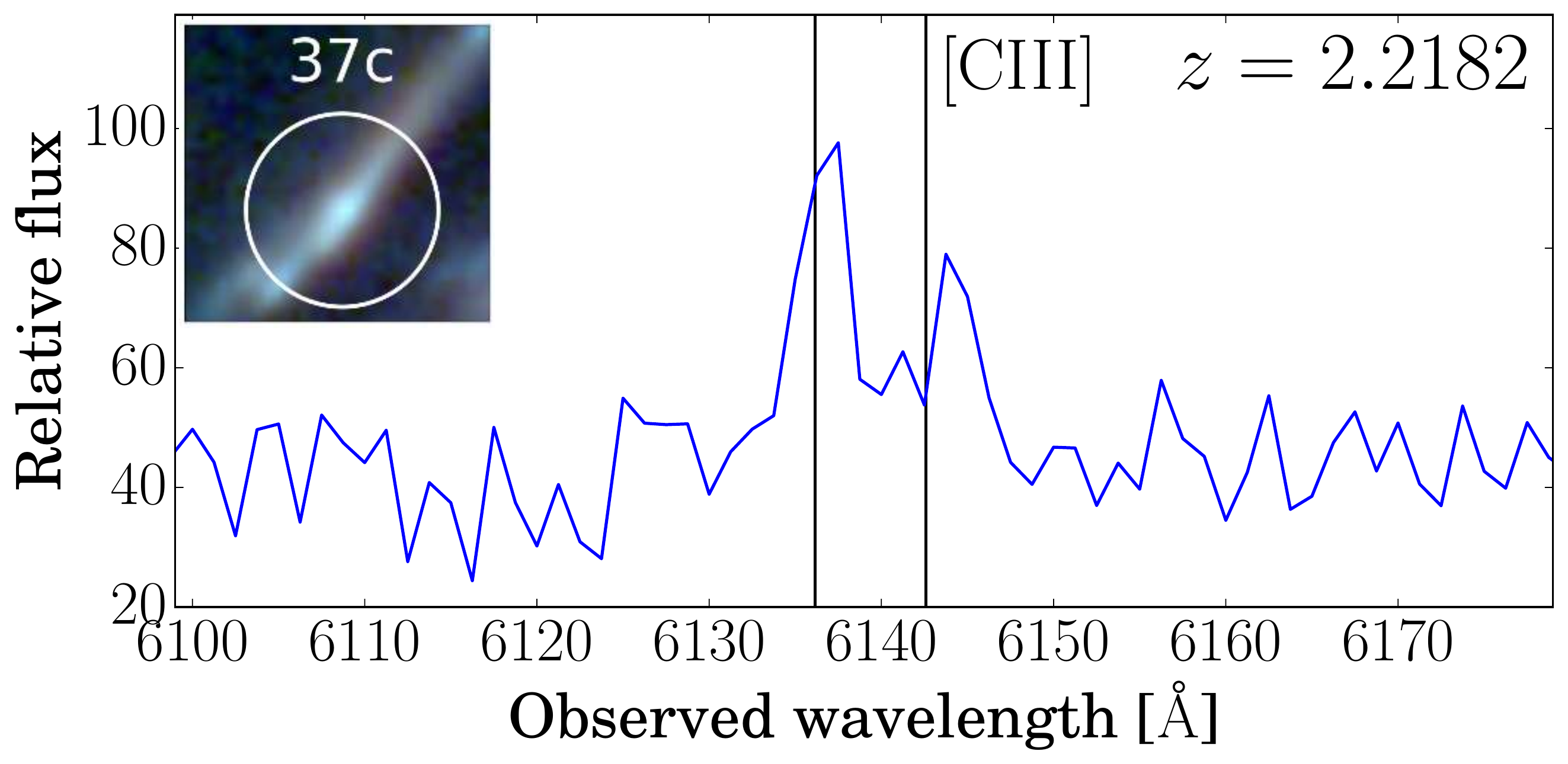}

  \caption{(Continued)}
  \label{fig:specs}
\end{figure*}

\longtab{
\setcounter{table}{1}
\begin{longtable}{ccccccc}
\caption{\label{tab:full_zcat} Information on spectroscopically identified multiple images in MACS~0416.}\\
\hline\hline
ID & RA & DEC & z & QF & Ref. & Mag \\
(1) & (2) & (3) & (4) & (5) & (6) & (7) \\
\hline
\endfirsthead
\caption{continued.}\\
\hline\hline
ID & RA & DEC & z & QF & Ref.& Mag \\
(1) & (2) & (3) & (4) & (5) & (6) & (7) \\
\hline
\endhead
\hline
\endfoot
CLASHVLTJ041512.49$-$240506.6 & 63.802059 & $-$24.085165 & 0.5772 & 3 & 1 & 22.17\\
CLASHVLTJ041512.65$-$240924.4 & 63.802703 & $-$24.156778 & 0.4043 & 3 & 1 & 22.71\\
CLASHVLTJ041512.65$-$240653.9 & 63.802720 & $-$24.114978 & 0.3032 & 3 & 1 & 18.48\\
CLASHVLTJ041512.67$-$240500.3 & 63.802779 & $-$24.083429 & 0.5767 & 3 & 1 & 21.97\\
CLASHVLTJ041512.69$-$241059.8 & 63.802859 & $-$24.183278 & 2.9782 & 9 & 1 & 23.24\\
CLASHVLTJ041512.75$-$240625.2 & 63.803134 & $-$24.107008 & 0.6372 & 3 & 1 & 23.56\\
CLASHVLTJ041512.77$-$240106.8 & 63.803218 & $-$24.018566 & 0.3102 & 3 & 1 & 20.85\\
CLASHVLTJ041512.81$-$241112.8 & 63.803373 & $-$24.186895 & 0.4004 & 3 & 1 & 21.43\\
CLASHVLTJ041512.82$-$241351.8 & 63.803418 & $-$24.231058 & 0.5361 & 3 & 1 & 20.65\\
CLASHVLTJ041512.82$-$241417.2 & 63.803430 & $-$24.238104 & 0.5788 & 3 & 1 & 23.34\\
CLASHVLTJ041512.85$-$241308.4 & 63.803547 & $-$24.219004 & 0.3007 & 9 & 1 & 24.51\\
CLASHVLTJ041512.86$-$235617.4 & 63.803602 & $-$23.938160 & 0.3436 & 3 & 1 & 21.45\\
CLASHVLTJ041513.01$-$241423.8 & 63.804188 & $-$24.239932 & 0.3398 & 3 & 1 & 22.09\\
CLASHVLTJ041513.02$-$240640.1 & 63.804258 & $-$24.111131 & 0.3046 & 3 & 1 & 20.48\\
CLASHVLTJ041513.11$-$241022.7 & 63.804605 & $-$24.172971 & 0.4204 & 2 & 1 & 22.06\\
$\vdots$ & $\vdots$ & $\vdots$ & $\vdots$ & $\vdots$ & $\vdots$ & $\vdots$ \\

\end{longtable}
\tablefoot{
The first 15 entries of the redshift table. The full catalog is available in the eletronic version of the paper and in the link \url{https://sites.google.com/site/vltclashpublic/data-release}, and contains 7 columns and 4594 rows.
The columns correspond to: (1) A unique identification reference; (2-3) the coordinates in degree; (4) the spectroscopic redshift; (5) the redshift quality flag; (6) the redshift reference (1$-$CLASH-VLT VIMOS based on LR-Blue spectra, 2$-$CLASH-VLT VIMOS based on MR spectra, 4$-$from \citealt{2014ApJS..211...21E}, 5$-$from Magellan, D. Kelson 2016 private communication, 6$-$from VLT/MUSE, this work) and (7) is the Subaru R-band magnitude.
}
}

%\end{appendix}

\end{document}